%% file: EXO-22-026_temp.tex
\pdfoutput=1
\documentclass[11pt,twoside,a4paper,cmspaper,final,collab]{cms-tdr}
\def\svnVersion{c918fd2}\def\svnDate{2025/05/12}\def\cmsCernNoTag{CERN-EP-2024-291}\def\cmsCernDate{\today}\def\cmsMessage{Published in Reports on Progress in Physics as \href{http://dx.doi.org/10.1088/1361-6633/add762}{\doi{10.1088/1361-6633/add762}.}}
\begin{document}\cmsNoteHeader{EXO-22-026}

\newcommand{\INTLUMI}{138\xspace}
\newcommand{\LSF}{\ensuremath{\mathrm{LSF_3}}\xspace}
\newcommand{\mjj}{\ensuremath{m_\mathrm{jj}}\xspace}%
\newcommand{\cwola}{\textit{CWoLa Hunting}\xspace}
\newcommand{\cathode}{\textit{CATHODE}\xspace}
\newcommand{\cathodeb}{\textit{CATHODE-b}\xspace}
\newcommand{\vae}{\textit{VAE-QR}\xspace}
\newcommand{\TNT}{\textit{TNT}\xspace}
\newcommand{\quak}{\textit{QUAK}\xspace}
\DeclareRobustCommand{\PQQst}{\HepParticle{Q}{}{*}\Xspace}
\DeclareRobustCommand{\PQTpr}{\HepParticle{\PQT}{}{\prime}\Xspace}
\DeclareRobustCommand{\PY}{\HepParticle{Y}{}{}\Xspace}
\DeclareRobustCommand{\PYpr}{\HepParticle{\PY}{}{\prime}\Xspace}
\DeclareRobustCommand{\PBpr}{\HepParticle{B}{}{\prime}\Xspace}
\newcommand{\Qstar}{$\PQQst \to \PQq \PWpr \to 3 \PQq$\xspace}
\newcommand{\XtoYY}{$\PX\to \PY\PYpr \to 4 \PQq$\xspace}
\newcommand{\Wp}{$\PWpr \to \PBpr\PQt \to \PQb \PZ \PQt$\xspace}
\newcommand{\Wkk}{$\PW_\mathrm{KK} \to \mathrm{R}\PW \to 3 \PW$\xspace}
\newcommand{\Zp}{$\PZpr \to  \PQTpr \PQTpr \to \PQt\PZ \PQt \PZ$\xspace}
\newcommand{\GtoHH}{$\mathrm{G}_\mathrm{KK} \to \PH\PH \to 4 \PQt$\xspace}
\newcommand{\cmsTable}[1]{\resizebox{\textwidth}{!}{#1}}
\ifthenelse{\boolean{cms@external}}{\newcommand{\applabel}{}}{\newcommand{\applabel}{Appendix}}

\cmsNoteHeader{EXO-22-026}
\title{Model-agnostic search for dijet resonances with anomalous jet substructure in proton-proton collisions at \texorpdfstring{$\sqrt{s}$}{sqrt(s)} = 13\TeV}
\titlerunning{Model-agnostic search for dijet resonances with anomalous jet substructure in pp collisions at 13\TeV}
\date{\today}

\abstract{
This paper presents a model-agnostic search for narrow resonances in the dijet final state in the mass range 1.8--6\TeV. The signal is assumed to produce jets with substructure atypical of jets initiated by light quarks or gluons, with minimal additional assumptions. Search regions are obtained by utilizing multivariate machine-learning methods to select jets with anomalous substructure. A collection of complementary anomaly detection methods---based on unsupervised, weakly supervised, and semisupervised algorithms---are used in order to maximize the sensitivity to unknown new physics signatures. These algorithms are applied to data corresponding to an integrated luminosity of \INTLUMI{}\fbinv, recorded by the CMS experiment at the LHC, at a center-of-mass energy of 13\TeV. No significant excesses above background expectations are seen. Exclusion limits are derived on the production cross section of benchmark signal models varying in resonance mass, jet mass, and jet substructure. Many of these signatures have not been previously sought, making several of the limits reported on the corresponding benchmark models the first ever. When compared to benchmark inclusive and substructure-based search strategies, the anomaly detection methods are found to significantly enhance the sensitivity to a variety of models.
}

\hypersetup{%
pdfauthor={CMS Collaboration},%
pdftitle={Model-agnostic search for dijet resonances with anomalous jet substructure in proton-proton collisions at sqrt(s) = 13 TeV},%
pdfsubject={CMS},%
pdfkeywords={CMS, ML, anomaly, dijet, resonance}}

\maketitle 
\section{Introduction}
Many models of physics beyond the standard model (BSM) predict the existence of new particles with hadronic decays.
One of the most generic searches for new physics at particle colliders is therefore a search for heavy resonances decaying into two jets~\cite{ua1dijet,ua2dijet,cdfdijet,d0dijet,atlas1dijet,atlas2dijet,cms1dijet,cms2dijet}.
This search is sensitive to a wide range of signals, but is dominated by an overwhelming background from quantum chromodynamics (QCD) multijet production.
To increase the sensitivity to specific decays, dedicated searches at the CERN LHC have been performed that require the jets to have a substructure and/or flavor content compatible with \PW and \PZ bosons~\cite{dibosonatlas,dibosoncms,dibosoncmslatest}, Higgs bosons~\cite{higgsatlas,higgscms}, bottom quarks~\cite{bsatlas,bsatlas2,bscms,bscms2}, or top quarks~\cite{topsatlas,topscms,topscms2}.
These searches are able to exploit the expected signature of the targeted signal to reduce the standard model (SM) background and increase the search sensitivity, but as a result they are no longer generic.
It is not possible to perform dedicated searches for every possible signature,
meaning many potential signals, whose production cross sections may be below the sensitivity of inclusive dijet searches, remain uncovered.
These gaps in coverage could result in potential discoveries being missed if new strategies are not employed.
Model-agnostic searches based on anomaly detection aim to fill in these gaps by achieving the best combination of both sensitivity and generality by using novel 
machine-learning (ML) techniques.
These searches are designed to be sensitive to a broad range of signatures, offering significant discovery potential.
Such searches have been proposed~\cite{Kasieczka:2021xcg}, and recently also performed by the ATLAS Collaboration~\cite{ATLAS_cwola,ATLAS_higgsplusX,ATLAS_twobody, ATLAS_anomaly2}.

\begin{figure}[htbp]
   \centering
   \includegraphics[width = 0.5\textwidth]{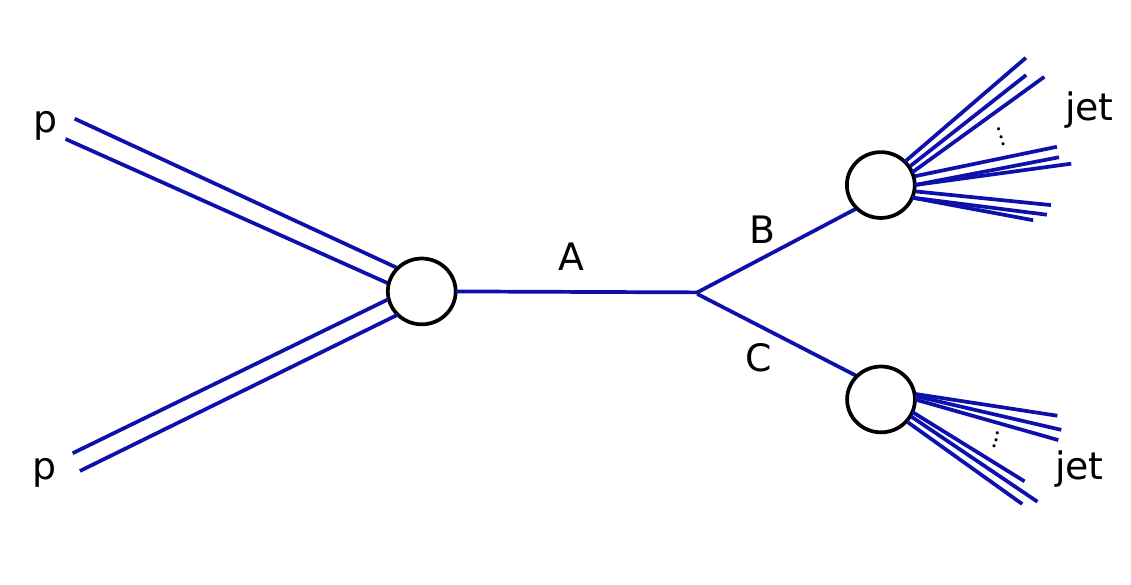}
   \caption{Production of a dijet resonance, A, in a proton-proton collision. The A resonance decays to two resonances B and C, which in turn each decay to a jet with anomalous substructure arising from multiple subjets.}
   \label{fig:diagram}
 \end{figure}

In this analysis, we present the first ever ML-based model-agnostic search for BSM physics by the CMS Collaboration. 
The search is for a narrow-width heavy resonance A with TeV-scale mass, decaying into two other resonances, B and C, in a dijet final state.
It is based on proton-proton collision data recorded with the CMS experiment~\cite{CMS:2008xjf, CMS:2023gfb} in 2016--2018 at a center-of-mass energy of 13\TeV, corresponding to a total integrated luminosity of \INTLUMI{}\fbinv \cite{CMS-LUM-17-003, CMS-LUM-17-004, CMS-LUM-18-002}.
The B and C particles could be either SM or BSM,
but are assumed to decay hadronically and have masses significantly smaller than the mass of the A particle.
The mass hierarchy results in the B and C particles being produced with high Lorentz boost, such that  their decay products are contained in large-radius jets, as illustrated in Fig.~\ref{fig:diagram}.
Masses of the A particle in the range 1.8 to 6\TeV are considered, with no specific requirements on the B and C particle masses.
The search methodology aims to maximize discovery potential for all BSM models matching these criteria.

Though the search itself is model-agnostic, a set of benchmark models representing a wide range of signatures are used to study performance and derive exclusion limits.
Signatures where the B and C jets have a varying degree of substructure, consistent with two to six subjets (prongs), are considered.
Five different methods are used to design discriminating variables that are used to identify large-radius jets with a non-QCD-like substructure.
For each method, a requirement on its corresponding anomaly discriminant is applied, enhancing the fraction of anomalous events and, therefore, of potential new physics in the signal region (SR).
A resonance is searched for in the SR by seeking a bump in the dijet mass spectrum on top of the smoothly falling QCD background.

While all methods employ ML techniques and aim to identify anomalous jets, they differ in the substructure information utilized, aspects of their learning setup, and model architecture.
Four of these methods proceed in a fully model-agnostic fashion, without relying on any signal simulation, and only make use of data events in the training of their anomaly detection model.
The fifth method is a hybrid approach and utilizes signal simulation as a loose prior on what potential anomalies may look like.
These differences lead to varying performance for signals of various cross section, mass, and substructure.
As the characteristics of the sought-after signal are unknown, all five methods are pursued as complementary approaches to provide broad coverage of potential anomalies.
All methods are found to significantly enhance the discovery sensitivity to a much wider range of signal models than traditional substructure techniques.

The first application of anomaly detection at the LHC was in a search for dijet resonances using weak supervision~\cite{ATLAS_cwola}.
However, in that prior search, only the masses of the two jets were used as input features to the anomaly detection algorithm. 
The methods employed in the present analysis use significantly larger feature sets, offering broader coverage for different types of anomalies, which could manifest in observables besides the jet mass, and significantly improving  classification performance. 
The other two uses of anomaly detection~\cite{ATLAS_higgsplusX, ATLAS_twobody} considered different final states and are not directly comparable to this search. 
The former analysis looked for resonances decaying to a Higgs boson plus a single anomalous jet, and the latter searched for two-body resonances across a variety of final states produced in association with a lepton. 
Furthermore, this analysis goes significantly beyond these prior publications by including multiple different anomaly detection methods, each varying in their input observables, architectures, and training paradigms, within a single search. 
This strategy maximizes the sensitivity to unknown physics signatures, where the most effective approach is uncertain. For several of these methods, this is their first application to LHC data, and they are shown to have complementary sensitivity to different signals.
While this work was in peer review, an additional anomaly-detection-based analysis from the ATLAS Collaboration has appeared~\cite{ATLAS_anomaly2}, emphasizing the importance and novelty of the methods developed in the present work.

The paper is organized as follows.
The CMS detector and event reconstruction procedures are described in Section \ref{sec:cms}.
Section \ref{sec:mc} describes the benchmark signal models and simulated samples used in the testing of the anomaly detection methods. 
Section \ref{sec:event_selection} describes the basic event selection that is applied prior to the anomaly detection methods.
The details of the five different anomaly detection approaches are described in Section \ref{sec:methods} and the
fit procedure is described in Section \ref{sec:fit}.
Systematic uncertainties are detailed in Section \ref{sec:sys_uncs}.
Section \ref{sec:validation} describes validation studies performed in simulation and data control regions.
The search results and interpretation are presented in Section \ref{sec:results} and 
a summary is given in Section \ref{sec:summary}.
Tabulated results are provided in the HEPData record for this analysis~\cite{HEPData}.

\section{The CMS detector and event reconstruction}
\label{sec:cms}

The central feature of the CMS apparatus is a superconducting solenoid of 6\unit{m} internal diameter, providing a magnetic field of 3.8\unit{T}.
Within the solenoid volume are a silicon pixel and strip tracker, a lead tungstate crystal electromagnetic calorimeter (ECAL), and a brass and scintillator hadron calorimeter (HCAL), each composed of a barrel and two endcap sections. Forward calorimeters extend the pseudorapidity coverage provided by the barrel and endcap detectors.
Muons are measured in gas-ionization detectors embedded in the steel flux-return yoke outside the solenoid. More detailed descriptions of the CMS detector, together with a definition of the coordinate system used and the relevant kinematic variables, can be found in Refs.~\cite{CMS:2008xjf,CMS:2023gfb}.

Events of interest are selected using a two-tiered trigger system. The first level, composed of custom hardware processors, uses information from the calorimeters and muon detectors to select events at a rate of around 100\unit{kHz} within a fixed latency of about 4\mus~\cite{CMS:2020cmk}.
The second level, known as the high-level trigger, consists of a farm of processors running a version of the full event reconstruction software optimized for fast processing, and reduces the event rate to around 1\unit{kHz} before data storage~\cite{CMS:2016ngn}.

The primary vertex (PV) is taken to be the vertex corresponding to the hardest scattering in the event, evaluated using tracking information alone, as described in Section 9.4.1 of Ref.~\cite{CMS-TDR-15-02}.
A particle-flow (PF) algorithm~\cite{CMS:2017yfk} aims to reconstruct and identify each individual particle in an event, with an optimized combination of information from the various elements of the CMS detector. The energy of photons is obtained from the ECAL measurement.
The energy of electrons is determined from a combination of the electron momentum at the PV as determined by the tracker, the energy of the corresponding ECAL cluster, and the energy sum of all bremsstrahlung photons spatially compatible with originating from the electron track.
The energy of muons is obtained from the curvature of the corresponding track.
The energy of charged hadrons is determined from a combination of their momentum measured in the tracker and the matching ECAL and HCAL energy deposits, corrected for the response function of the calorimeters to hadronic showers.
Finally, the energy of neutral hadrons is obtained from the corresponding corrected ECAL and HCAL energies.

Jets are clustered from the PF candidates in an event using the anti-\kt jet finding algorithm~\cite{Cacciari:2008gp,Cacciari:2011ma}.
In this analysis, large-radius jets with a distance parameter of $R = 0.8$ are used.
Jet momentum is determined as the vectorial sum of all particle momenta in the jet, and is found from simulation to be, on average, within 5 to 10\% of the true momentum over the entire transverse momentum (\pt) spectrum and detector acceptance \cite{Sirunyan:2020foa}.
Additional proton-proton interactions within the same or nearby bunch crossings (pileup) can contribute additional tracks and calorimetric energy depositions, increasing the apparent jet momentum.
The pileup-per-particle identification algorithm~\cite{Sirunyan:2020foa,Bertolini:2014bba} is used to mitigate the effect of pileup at the reconstructed-particle level, making use of local shape information, event pileup properties, and tracking information.
A local shape variable is defined, which distinguishes between collinear and soft diffuse distributions of other particles surrounding the particle under consideration.
The former is attributed to particles originating from the hard scattering and the latter to particles originating from pileup interactions. Charged particles identified to be originating from pileup vertices are discarded.
For each neutral particle, a local shape variable is computed using the surrounding charged particles compatible with the PV within the tracker acceptance ($\abs{\eta} < 2.5$), and using both charged and neutral particles in the region outside of the tracker coverage.
The momenta of the neutral particles are then rescaled according to their probability to originate from the PV deduced from the local shape variable, superseding the need for jet-based pileup corrections~\cite{Sirunyan:2020foa}.

Jet energy corrections are derived from simulation studies so that the average measured energy of jets becomes identical to that of particle-level jets.
In-situ measurements of the momentum balance in dijet, {$\gamma$}+jet, {\PZ}+jet, and multijet events are used to determine any residual differences between the jet energy scale in data and in simulation, and appropriate corrections are made~\cite{CMS:2016lmd}.
Additional selection criteria are applied to each jet to remove jets potentially dominated by instrumental effects or reconstruction failures~\cite{Sirunyan:2020foa}.

\section{Signal models and simulated samples}
\label{sec:mc}
Six different signal models of the $\mathrm{A} \to \mathrm{BC}$ topology are used throughout the analysis.
A set of background simulations is generated for use in the semisupervised method (Section \ref{subsec:quak}) and for testing and validation of all anomaly detection methods.
Except for the semisupervised method, these simulated samples are used only for testing and sensitivity studies, and the search itself is performed in a model-agnostic manner 
based only on the recorded data sample.

\subsection{Signal models}
\label{sec:signal_models}
A set of simplified benchmark signal models is chosen to cover a range of substructure topologies. 
The different signals are categorized based on the number of hard prongs in the B and C jets.
For all signals, the masses of the heavy resonance A are generated at 3 and 5\TeV.
The B and C particles are generated with masses of 25, 80, 170, and 400\GeV, with some values excluded for certain models due to kinematic constraints.
Jet substructure properties are not very sensitive to the spin and coupling structure of the A, B or C particles.
It is therefore expected that the anomaly detection methods will have comparable sensitivities to signal models with B and C particles 
with similar masses and decay chains to the ones chosen here.

The first signal model has a 1+2 prong topology that consists of an excited quark resonance (\PQQst), which decays into a quark and \PWpr boson \cite{Baur:1987ga,Baur:1989kv}.
The \PWpr boson decays into two light quarks with the same flavor composition as the SM \PW boson.
The second signal model has a 2+2 prong topology and consists of a heavy resonance \PX decaying to two resonances, \PY and \PYpr, each of which then decays into two light quarks.
The third signal model has a 3+3 prong topology and consists of a \PWpr boson decaying into a vector-like quark (\PBpr) and a top quark \cite{Barducci:2012kk}.
The \PBpr then decays into a bottom quark and a \PZ boson.
The fourth signal model has a 2+4 topology and consists of a $\PWpr_\mathrm{KK}$ boson decaying into a radion ($\mathrm{R}$) and a \PW boson \cite{Agashe:2016rle,Agashe:2017wss}.
The radion then decays into two W bosons.
The fifth signal model has a 5+5 topology and consists of a \PZpr boson decaying into two vector-like quarks, \PQTpr \cite{Okada:2012gy, Buchkremer:2013bha}.
The vector-like quarks then decay into a top quark and a \PZ boson.
The last signal model has a 6+6 topology and consists of a heavy spin-2 Randall--Sundrum graviton ($\mathrm{G}_\mathrm{KK}$) which decays into to two lighter Higgs-like scalars (\PH)~\cite{Carvalho:2014lsg}.
The lighter Higgs-like scalar then decays into a top quark-antiquark pair.
In the rest of this paper, these signal models are denoted \Qstar, \XtoYY, \Wp, \Wkk, \Zp, and \GtoHH, respectively.

The \Wp process has been previously searched for by the CMS Collaboration \cite{CMS_Wp_Bp}.
However, the previous search did not consider the parameter space probed in this analysis, where the \PBpr is sufficiently light compared to the \PWpr so that its decay products are merged into a single jet, and therefore did not have sensitivity for the mass points chosen here for our benchmarks.
Diboson resonance searches performed by the ATLAS \cite{dibosonatlas} and CMS \cite{dibosoncms, dibosoncmslatest} Collaborations have sensitivity to the 2+2 prong mode, \XtoYY, for the case where $M_{\PY} = M_{\PYpr} = 80\GeV$, but do not have sensitivity for other \PY and \PYpr mass combinations.
The \Wkk process was previously searched for by the CMS Collaboration \cite{CMS_triboson, CMS_triboson_hadronic} in the same boosted regime as this analysis.
Therefore, the inclusion of this signal in our benchmarks allows a comparison of the sensitivity of anomaly detection methods to a previous optimized dedicated search.
We note that the specific 3+3 (\Wp) and 5+5 (\Zp) models chosen are excluded by searches for the direct production of the vector-like quarks \PBpr and \PQTpr.
These searches have set lower limits on the \PBpr and \PQTpr masses above a \TeV \cite{CMS:VLQ, ATLAS:VLQ}.
However, the inclusion of such models is still useful, as a demonstration of the sensitivity of the methods to the 3+3 and 5+5 prong topologies, which encompass many other potential signal models.

\subsection{Simulated samples}
All signals are generated at leading order with \MGvATNLO version 2.6.5~\cite{Alwall:2014hca}, whereas a variety of generators are used for different backgrounds.
All samples used the NNLO NNPDF 3.1 parton distribution functions \cite{Ball:2010de,NNPDF:2014otw,NNPDF:2017mvq} and interfaced to \PYTHIA version 8.240~\cite{Sjostrand:2014zea}, with the underlying event tune CP5~\cite{CMS:2019csb} to simulate the parton shower and hadronization.

Although this analysis targets hadronic final states, we did not enforce hadronic-only decays of SM particles in the signal simulation.
Some sensitivity to semileptonic decays is retained when the daughter particles are boosted such that the leptonic decay products end up inside the jet along with hadronic decay products.

Simulations of the QCD multijet background are generated using \PYTHIA description of $2 \to 2$ scattering at leading order (LO) accuracy.
Additional jets in the simulated events arise from initial- and final-state radiation, within a full parton shower provided by the generator.
The \ttbar, $\PQt\PW$, and single \PQt production processes are generated with the next-to-leading order (NLO) generator \POWHEG v2.0 \cite{Nason:2004rx,Frixione:2007nw,Alioli:2010xd}.
Simulated events originating from {\PW}+jets and {\PZ}+jets are generated using \MGvATNLO at LO accuracy.
Production of up to 3 (4) extra partons in the hard process are considered for the {\PW}+jets ({\PZ}+jets) simulation.
Double counting, which occurs between the partons generated by \PYTHIA and those of \MGvATNLO,
is eliminated using the MLM method~\cite{Alwall:2007fs}.

\section{Basic event selection}
\label{sec:event_selection}
For all anomaly detection methods, events were selected online using a variety of different jet triggers, based on either the highest (leading) jet \pt or the scalar \pt sum of all the jets in the event (\HT).
The thresholds of these triggers increased slightly in later data taking years, so that the same trigger rate was maintained under higher instantaneous luminosity.
The thresholds of these leading-jet \pt triggers (\HT triggers) varied from 450 (800)\GeV in 2016 to 500 (1050)\GeV in 2018.

Offline, events are selected requiring at least two jets with $\pt > 300\GeV$ and pseudorapidity $\abs{\eta} < 2.5$.
The two jets with the highest \pt in each event are selected as potential B and C candidates, and are further required to have an invariant mass $\mjj > 1455\GeV$ in order to ensure the trigger is fully efficient.
Additional jets are ignored.

In the sought-after signal topology, the resonance A is produced via the $s$ channel. The dominant QCD background proceeds via the $t$ channel, which has an angular distribution peaking towards large rapidity separation between the two jets.
The amount of QCD background is therefore reduced relative to that of the signal by requiring the two jets to have a pseudorapidity difference of ${\Delta\eta_\mathrm{jj}} < 1.3$.
A signal-depleted control region (CR) is constructed from events with $2.0 < \abs{\Delta\eta_\mathrm{jj}} < 2.5$ that also meet additional requirements on the \pt balance of the two jets. This region is used to check that the analysis procedure does not produce any spurious excesses of events.

This event selection is also used to define an \textit{inclusive} search strategy, which does not make any further selections, and therefore does not use substructure information.
This inclusive strategy is used as a reference when evaluating the performance of the anomaly detection methods.

\section{Anomaly detection methods}
\label{sec:methods}
The five anomaly detection methods are based on three different training paradigms for ML-based anomaly detection: unsupervised, weakly supervised, and semisupervised learning.
Each method is described briefly here, with further details reserved for \applabel~\ref{sec:method_details}.

\subsection{Unsupervised method}
\label{subsec:vae}
The unsupervised learning algorithm attempts to construct a model to identify anomalous jets without using any labeled references.
The method employed here consists of a variational autoencoder (VAE) \cite{kingma2014autoencoding} trained on a data sample dominated by QCD jets.
A quantile regression (QR) network is then used to decorrelate the anomaly score of the VAE from the dijet mass.
This method is referred to as \vae.

Autoencoders are a type of neural network which are trained to compress inputs into a smaller representation and decompress it to recover the original inputs.
The VAE employed here takes as input the 100 highest $\pt$ particles of a jet, with the ordering obtained from a Cambridge--Aachen \cite{Dokshitzer:1997in} reclustering of the jet constituents.
Each particle is represented as a set of three values, which are the $x$, $y$, and $z$ component of its momentum $\vec{p}$.
The VAE is trained using jets from the CR.
It therefore learns how to perform this compression and decompression on QCD background jets, but might fail in performing this task
as well on anomalous jets not present in the training sample.
Therefore, the difference between the original and autoencoded jets can be used as an effective anomaly score, with higher values corresponding to more anomalous events.

High invariant mass events are rare, being on the tail of a steeply falling spectrum, and therefore constitute a small fraction of the VAE's  training set. 
This would naturally cause the VAE to find high invariant mass events anomalous, which would distort the dijet mass distribution, making it difficult to estimate 
the background and any potential signal.
To decorrelate the VAE's anomaly score from the dijet invariant mass a QR method~\cite{refId0} is used.
The QR model is trained to find the threshold on the anomaly score, as a function of \mjj, which corresponds to a fixed data efficiency in the SR.
Using a selection criterion defined in this way automatically preserves the \mjj shape of the inclusive sample.
A selection corresponding to the 10\% most anomalous data is used for the generic model-independent search.
For the limit setting, where the signal is known, three orthogonal categories are used to enhance sensitivity.
They are defined as follows: the most anomalous 1\% of data, the data between the 5\% and 1\% percentile range, and the data between the 10\% and 5\% percentile range.
This multicategory approach cannot be used in the model-independent search, because the relative signal yield in each category is signal dependent.
To ensure that the full data set is used, and that the VAE does not evaluate anomaly scores on events on which the QR is trained, we use $k$-fold cross-validation, discussed in \applabel~\ref{subsec:cross_val}.
This is a method where the data set is divided into $k$ equal parts, and the model is trained $k$ times, each time using a different part as the test set and the remaining parts as the training set.

\subsection{Weakly supervised methods}
\label{subsec:weak_sup}
In weakly supervised training, a signal versus background classifier is trained exclusively on data, without the use of any Monte Carlo simulation. 
This can be done using the classification without labels (CWoLa)~\cite{cwola} paradigm, where classification is performed between two groups of data events rather than relying on individual event labels. 
This paradigm requires two mixed samples of data events, chosen such that one is a mixture of potential signal events and background events, and the other is nearly pure background. For example, our methods aim to construct the first sample so that it consists of data events within the mass peak of a hypothetical resonance, and the second sample consists of background events from the sidebands.

The classifier is trained to distinguish between events in these two samples. If sufficient signal is present in the data set, the classifier will learn to distinguish the signal from background, provided that the background composition in the two samples is the same. These methods train directly on data events in the SR of the analysis, and thus learn the specific characteristics of the signal if it is present in the data set.
A k-fold cross-validation is used to ensure each weakly supervised classifier does not evaluate anomaly scores of events on which it is trained.

Three different methods based on weak supervision are employed. These are \cwola~\cite{cwolabumphunt}, \textit{Tag N' Train} (\TNT)~\cite{Amram:2020ykb}, and \textit{classifying anomalies through outer density estimation} (\cathode)~\cite{Hallin:2021wme}.

All of the weakly supervised methods pursued in this analysis assume the signal is a narrow resonance, and use the invariant mass of the two
jets in their definition of the mixed samples.
These methods therefore assume a particular hypothesis for the mass of the new resonance in order to construct the samples for training.
If a narrow resonance exists inside the hypothesized region in sufficient abundance, the weakly supervised training procedure will produce a classifier able to
discriminate between it and the QCD background.
In the absence of a signal, the two samples will both consist solely of background events, and therefore be indistinguishable in the training procedure.
The resulting classifier will therefore produce random results driven by the statistical fluctuations between the two samples.

For all three weakly supervised methods, the training procedure is repeated for multiple hypotheses for the signal mass in order to scan over the full dijet mass spectrum.
The SR is split into 8 nonoverlapping \mjj windows to be used for the training procedure.
The mass resolution for a narrow fully merged resonance varies from $\sim$80 to $\sim$200\GeV over the considered mass range. 
The size of the signal windows is chosen to be significantly larger than this resolution, such that a narrow resonance would be expected to be nearly fully contained in a single bin.
Two sets of windows with different centers are defined in order to ensure no signals close to a window boundary are missed.
The two sets of window boundaries are  [1350, 1650, 2017, 2465, 3013, 3682, 4500, 5500, 8000]\GeV and [1492, 1824, 2230, 2725, 3331, 4071, 4975, 6081, 8000]\GeV, which are referred to as mass windows $\alpha$ and $\beta$, respectively.
Each SR is required to have both an upper and lower sideband to be used as control regions. 
Therefore, the highest and lowest mass regions in each set of windows are used only as sidebands, resulting in 12 total SRs in which the training procedure is performed.
The \cathode method does not use the highest mass $\beta$ SR due to the limited size of the data sample available for training.
The 1650--2017\GeV region is the lowest mass SR for all methods and the region from 4975--6081\GeV (4500--5500\GeV) the highest mass SR for the \cwola and \TNT methods (\cathode method).

For \cwola, the two mixed samples are obtained directly from windows in dijet invariant mass.
A potentially signal-rich sample is defined as all events falling in a particular signal window,
and the background-rich sample is defined as events in the neighboring sideband windows.
To account for the statistical imbalance, events from the lower sideband are reweighted in the training to match the total weight of the upper sideband.
Separate classifiers are trained for the heavier and lighter jet in each event, as defined by each jet's soft-drop mass $m_{\mathrm{SD}}$~\cite{softdrop},
rather than a single classifier for the full dijet system, as was done in Refs.~\cite{cwolabumphunt,ATLAS_cwola}.
This allows an additional reweighting procedure during the training:
jets in the background-rich sample are reweighted to match the \pt distribution of jets in the SR.
This procedure reduces the correlation between the final anomaly score and \mjj.

The classifiers take as input the $m_{\mathrm{SD}}$ of the jet, the $N$-subjettiness variables $\tau_{21}$, $\tau_{32}$, and $\tau_{43}$~\cite{nsubjettiness},
the number of PF candidates inside the jet $n_\mathrm{PF}$,
the maximum b tagging score from the \textsc{DeepCSV} algorithm~\cite{deepcsv} of the two leading subjets of the large-radius jet, and
a measure of the energy fraction carried by leptons, the lepton subjet fraction (\LSF)~\cite{lsf}.
The inclusion of soft-drop mass and $N$-subjettiness variables helps to differentiate between nearly massless QCD jets and significantly boosted resonances, and determines the number of subjets in a jet, respectively.
The count of PF candidates in a jet detects jets with higher numbers of constituent particles, while the \LSF and b tagging score pinpoint jets containing enhanced contributions from leptons and heavy-flavor quarks.

In \TNT, the two mixed samples are defined in a similar way to \cwola, with one modification: an additional unsupervised classifier, a jet-based autoencoder, is used to increase the purity of the signal-rich sample.
The method assumes that for a true signal, both the B and C jets would be anomalous.
This implies a correlation between the B and C jets for signal events present in the data:
an anomaly in one jet implies an anomaly in the other.
This relationship is not present for background QCD events in which the anomaly scores of the two jets are uncorrelated.
The procedure therefore constructs a signal-enriched sample of C (B) jets, by utilizing the anomaly scores of jet B (C).
The criteria for constructing the mixed sample now encompass not only an event's position within the \mjj SR, but also the unsupervised anomaly score of the jet not involved in the classifier's training.
This dual-criterion approach markedly enhances the signal purity within the signal-rich sample.

For each SR, a separate autoencoder is trained using events from the corresponding sidebands.
The autoencoder takes as input an image representation of the jet~\cite{Farina:2018fyg,Heimel:2018mkt}.
Internally, the autoencoder works by compressing the image into a six-dimensional latent space, before attempting to reconstruct the original image.
The difference between the original and reconstructed images is used as the anomaly score.
The anomaly score of the autoencoder is evaluated on each jet in the event, one at a time.
This anomaly score is used in addition to the dijet mass information to sort events into the signal-rich or background-rich categories.

To construct the signal-rich and background-rich samples, the two jets in each event are first randomly sorted into two groups.
The dijet mass of the event and the autoencoder score of the jet from one group, are used to categorize jets of the other group into mixed samples, and vice versa.
The signal-rich mixed sample of jets is defined as jets from the second group, for which the jet in the first group is in the top 20\% of the autoencoder anomaly scores, and in an event that has \mjj within the SR.
The background-rich mixed sample of jets is defined as jets of the second group, coming from events in the dijet mass sidebands, or which have jets from the first group in the bottom 40\% of the autoencoder anomaly scores.
The samples of categorized jets from both jet orderings are then merged together to train a single weakly supervised jet classifier, using identical network architecture, reweighting schemes, and input variables as used for the \cwola~method.
Fig. \ref{fig:TNT_diagram} in \applabel~\ref{subsec:weaksup_details} shows a graphical representation of this procedure.

For \cathode, a dijet mass window is also used to define the potentially signal-rich sample, but a different approach is used for the background-rich sample.
First, the conditional probability density of background events as a function of invariant mass is learned, using a normalizing flow generative model~\cite{germain2015made, dinh2015nice,rezende2016variational,dinh2017realnvp,kingma2018glow,papamak2018flows,durkan2019splineflows,vdberg2019sylvesterflows,cranmer2020mflows} trained on all events outside the signal window.
This probability density is then interpolated into the signal window, and used to generate a sample of synthetic background events.
An event-level weakly supervised classifier is then trained to distinguish between data events from the SR and the synthetic background sample.

This approach fully learns correlations between the input features and \mjj, allowing the use of variables that are significantly correlated with \mjj.
The input variables used in \cathode~are the mass of the heavier jet $m_\mathrm{j_1}$, the mass difference between the two jets $\Delta m_{\mathrm{j_1j_2}} = m_\mathrm{j_1} - m_\mathrm{j_2}$, and the $N$-subjettiness variable $\tau_{41}$ for each jet. This $N$-subjettiness variable is found to perform the best for the signals under study.
An additional method, \cathodeb, uses the same input variables as \cathode, but also includes the b tagging \textsc{DeepCSV} score of each leading jet. This variant is expected to have higher sensitivity for processes yielding b jets, but worse sensitivity otherwise.

\subsection{Semisupervised method}
\label{subsec:quak}
Finally, a semisupervised algorithm referred to as quasi-anomalous knowledge (\quak)~\cite{Park:2020pak} is used.
This method seeks a middle ground between the fully model-agnostic approach of the previous methods and a standard dedicated search.
In \quak, density estimators are used to encode a `prior' for the likely signature of a new physics signal, based on labeled signal Monte Carlo samples.
To help reject QCD events, an additional density estimator trained on simulated background events is used.

A normalizing flow is trained to estimate the probability density of each event, based on the substructure information of both jets.
The substructure variables used are the same as those used in the \cwola~method except a
modified $N$-subjettiness metric, $\tau_s = {\sqrt{\tau_{21}}}/{\tau_1}$, and the ratio of the jet mass and transverse momentum, $\rho = {m_\mathrm{SD}}/{\pt}$, are used instead of the \LSF variable and soft-drop mass variables, respectively.

To maintain sensitivity to a broad class of signals, six distinct flows are trained on combinations of signal samples grouped according to the B and C particle masses.
The output of these six flows is then combined into one signal-like score by adding together their scores raised to the fifth power, weighted by the sign of each score.
This relatively high power is chosen so that the combined score is sensitive to very signal-like events in any of the six flows.
A single flow is also trained for background events in simulation.
A two-dimensional (2D) \quak~space is formed, in which each event's position is defined by its background-like and signal-like scores.
The most anomalous events are expected to populate the region with low background-like score and high signal-like score.
For a given mass hypothesis of the intermediate resonance, $m_\mathrm{A}$, a template of background
events is created by considering events with dijet mass in the sidebands, $m_\mathrm{A} - 900\GeV < m_\mathrm{jj} < m_\mathrm{A} - 400\GeV$ and $m_\mathrm{A} + 200\GeV < m_\mathrm{jj} < m_\mathrm{A} + 700\GeV$.
The SR is taken to be $m_\mathrm{A} - 400\GeV < m_\mathrm{jj} < m_\mathrm{A} + 200\GeV$.
The template itself is a binned 2D histogram of the \quak~space. The
\quak~space bins with an excess of events in the SR, as opposed to the sidebands, are chosen for the selection.
Events from the full dijet mass spectrum which fall into these selected \quak~bins are then utilized in the fit for a resonant signal.

A general search, referred to as 'generic \quak', is performed using a combination of all the benchmark signal samples as a signal prior.
Additionally, model-specific versions of the searches targeting each benchmark signal are performed by using \quak~with a signal prior consisting solely of the targeted signal.
Because it specifically targets each signal separately, this latter method is expected to yield better performance for the benchmark signals, but have worse generalization to unknown signals.
It is used as a point of comparison in evaluating performance of the anomaly detection methods.

\section{Fit procedure}
\label{sec:fit}

After the set of potentially anomalous events are selected for each method, the resulting  dijet invariant mass spectra are scrutinized for the presence of a potential signal.
In each spectrum, the SM background is expected to be smoothly falling, while a signal with a narrow width is expected to peak at its resonance mass.

The shape of the signal is modeled with a double Crystal Ball function~\cite{Oreglia:1980cs, Gaiser:1982yw} from fits to simulated narrow-width resonances.
When the model-agnostic search is performed, the shape of the \XtoYY signal model is used for the signal hypothesis.
For limits on a particular signal model, a signal shape derived from dedicated simulations of that model is used.
The search is performed considering resonance masses in the range 1.8 to 6\TeV in intervals of 100 GeV.
Simulations of the \XtoYY signal for $m_\mathrm{A}$ values of 2, 3, and 5\TeV are interpolated and extrapolated to produce the signal shapes covering the full range of mass hypotheses.

For the model-agnostic search, we test the effect of using only the signal shape from the \XtoYY model, when extracting signals produced by other models with different resonance shapes.
It is found that the use of the \XtoYY signal shape in the extraction of a true \Wp, \Wkk, or \GtoHH signal, each of which produces a different signal shape, reduces the statistical significance by less than 1 standard deviation ($\sigma$)
as compared to using the true resonance shape, which is deemed acceptable.

The background is modeled via a fit to data with the parameterization:
\begin{equation}
    \frac{\rd{N}}{\rd{\mjj}} = \frac{ P_0(1-x)^{P_1} } {(x)^{P_2+P_3 \log(x)+P_4 \log^2(x)}},
    \label{eq:dijet_fn}
\end{equation}
where $x=\mjj/\sqrt{s}$.
This is from the same family of functional forms as deployed in previous dijet searches~\cite{atlas1dijet,atlas2dijet,cms1dijet,cms2dijet}.
The parameters $P_3$ and $P_4$ are initially fixed to zero, but allowed to be nonzero if including them significantly improves the fit quality
as quantified through a Fisher F-test~\cite{FisherTest}.
All the $P_i$ selected by the F-test are allowed to freely float in the final fit.
As the selection is done separately for each SR, the \cwola, \TNT, \cathode, and \quak~methods, this approach results in a distinct mass spectrum for the search in each SR.

The F-test is repeated separately for each of these spectra.
To assess the quality of the fit, the $\chi^2$ per degree of freedom is computed, and checked that it corresponds to a probability value $p > 0.05$ of the data arising from the background-only hypothesis.
If the fit fails this quality check, the \mjj range of the fit is reduced and the fitting procedure is repeated until the quality criterion is reached.
Alternate functional forms of the background parameterization were tried and used to test for biases due to the parameterization choice.
It was found that the resulting bias on the inferred signal strength was 1 $\sigma$ or less, which was deemed acceptable.
More details on this study of potential fit bias are given in Appendix \ref{sec:fit_bias}.

Evidence of a signal is determined from a likelihood ratio test \cite{CMS-NOTE-2011-005}, comparing a background-only fit to the data and a combined signal plus background fit.
Upper limits are set using the profile likelihood ratio as the test statistic with the CL$_s$ criterion \cite{CLS1,CLS2}.
Asymptotic formulae are used to simplify calculations \cite{Cowan:2010js}.
The CMS analysis tool \textsc{Combine}~\cite{CAT-23-001}, which is based on the \textsc{RooFit}~\cite{Verkerke:2003ir} and \textsc{RooStats}~\cite{Moneta:2010pm} frameworks, is used for all statistical results.

\section{Systematic uncertainties}
\label{sec:sys_uncs}

Systematic uncertainties in the estimation of the background, the modeling of the shape of the signal resonance, 
and the signal selection efficiency are considered.
The former two are intrinsic to the model-agnostic search, and therefore affect all reported results.
In contrast, the latter uncertainties are specific to the interpretation of the search under a particular signal model, and therefore only affect the derived exclusion limits;
they have no impact on the significances of reported excesses.

As discussed in Section \ref{sec:fit}, the parameters describing the background shape are freely floating and constrained
only by the observed data in the SR.
This allows for the uncertainty in the background shape to be correctly accounted for in all reported results.
Uncertainties in the signal shapes arise from uncertainties in the energy scale and resolution of jets.
These uncertainties are assessed by systematically varying jets by the uncertainties in their energy scale and resolution.
For each variation, the signal shape is re-derived and the resulting shifts in the parameters of the double Crystal Ball function are then taken as uncertainties.
The uncertainty in the location (width) of the signal mass peak is found to be $\sim$2\% ($\sim$4\%).
These uncertainties are then included as Gaussian-constrained nuisances in the likelihood. 

Systematic uncertainties in the selection efficiency are assessed for each benchmark signal model.
The dominant affect comes from the uncertainty in the jet substructure modeling, which affects the estimated efficiency of the anomaly tag.
A correction to, and uncertainty in, the jet substructure modeling of each signal is applied based on the method of Ref. \cite{Lund_PAS}.
The method is based on reclustering the jet's constituents into several subjets, such that each subjet captures a single prong.
Each subjet's simulated radiation pattern in the Lund jet plane \cite{Dreyer:2018nbf} is corrected to match that of subjets from a data control region.
Uncertainties in the correction originate from statistical and systematic components of the splitting frequency of the subjets in data, and limitations of the reclustering procedure for high-prong jets.
The resulting uncertainties in the selection efficiency range from 10--25\%, increasing for signal models with larger numbers of prongs. 

The simulation of subjet b tagging scores, used as inputs to some of the anomaly detection methods, are corrected to match the distribution 
observed in data control regions \cite{deepcsv}.
An uncertainty in the modeling of this feature is assessed by varying the simulated distribution according to the uncertainties in the correction. 
This results in an uncertainty of up to $\sim$5\% in the tagging efficiency of signals which produce b jets.
Other uncertainties in the signal selection efficiency, including the modeling of pileup, parton distribution functions, and renormalization and factorization scales,
are assessed, but found to be sub-dominant.
The weakly supervised methods receive an additional uncertainty in the selection efficiency of up to $\sim$10\% due to stochasticity of their training performance in data.
The estimation of the signal selection efficiencies and incorporation of the systematic uncertainties for these methods is discussed further in \applabel~\ref{sec:limits}.

\section{Performance and validation}
\label{sec:validation}

The performance of the anomaly detection methods is verified in a simulated pseudo-data set.
The pseudo-data set is constructed by selectively sampling events from simulations of different background processes, in proportion to their cross sections, instead of the commonly used practice of applying event weights.
This sampling procedure better captures the data set size and statistical fluctuations that would be present when the methods are being applied to data, both of which affect the performance achieved when training neural networks.
The majority of the pseudo-data set consists of QCD multijet events, with minor contributions from other background processes.
The combined fraction of {\PW}+jets and {\PZ}+jets (\ttbar, $\PQt\PW$, and single \PQt production) events varies from $\sim$2\% ($\sim$0.3\%) at a dijet mass of 2\TeV to $\sim$4\% ($\sim$0.5\%) at a dijet of mass of 5\TeV.
Due to the limited event count of the QCD multijet simulations, the pseudo-data sets corresponded to an equivalent integrated luminosity of only 26.8\fbinv.
Versions of the pseudo-data set with different amounts of injected signal events are constructed, and the search procedure is repeated on each version.
A background-only version of the pseudo-data set is used to verify that no method produced artificial excesses of events.

Pseudo-data sets with injected signals are used to test the sensitivities of the anomaly detection methods.
These data sets are used to determine the expected statistical significance of
the signal as a function of the size of the injected signal.

The sensitivities of the anomaly detection methods are compared to several standard methods to better contextualize their performance.
These comparison methods utilize the same basic selection criteria, fitting procedure and statistical analysis as employed by the anomaly detection methods.
The only difference is slightly modified event selections.
The inclusive search (defined in Section \ref{sec:event_selection}) is used as a comparison model-agnostic approach.
The first (second) model-specific event selection is a typical substructure selection tailored to 2-prong (3-prong) signals and requires $\tau_{21} < 0.4$ ($\tau_{32} < 0.65$), and $m_\mathrm{SD}> 50\GeV$ for both jets in the event.
The final model-specific event selection is intended to maximally exploit signal information to achieve superior sensitivity.
For this, a version of the \quak{} procedure, which had a signal prior exactly matching the injected signal is used.

\begin{figure*}[htbp]
   \centering
   \includegraphics[width = 0.9\textwidth]{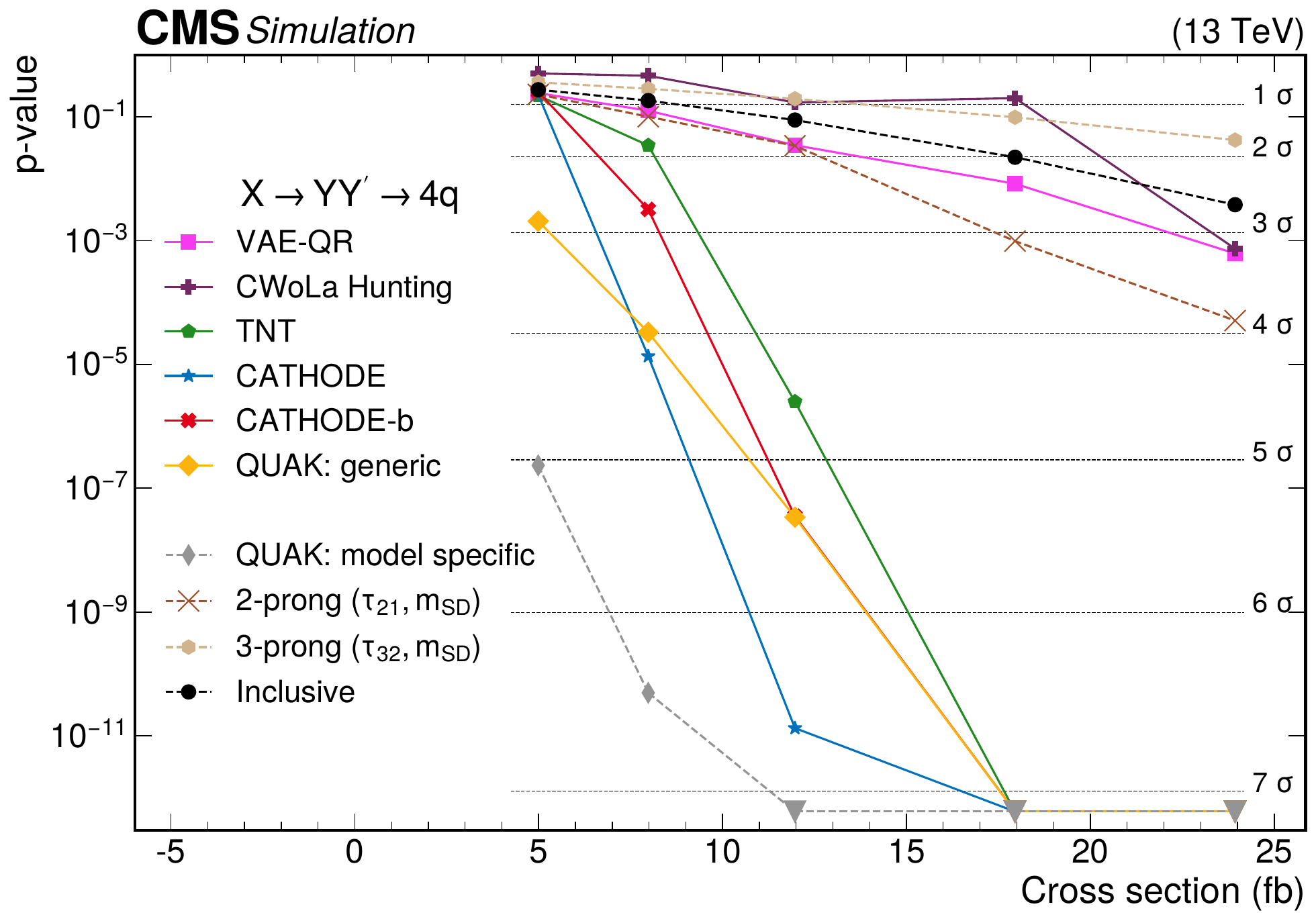}
   \includegraphics[width = 0.9\textwidth]{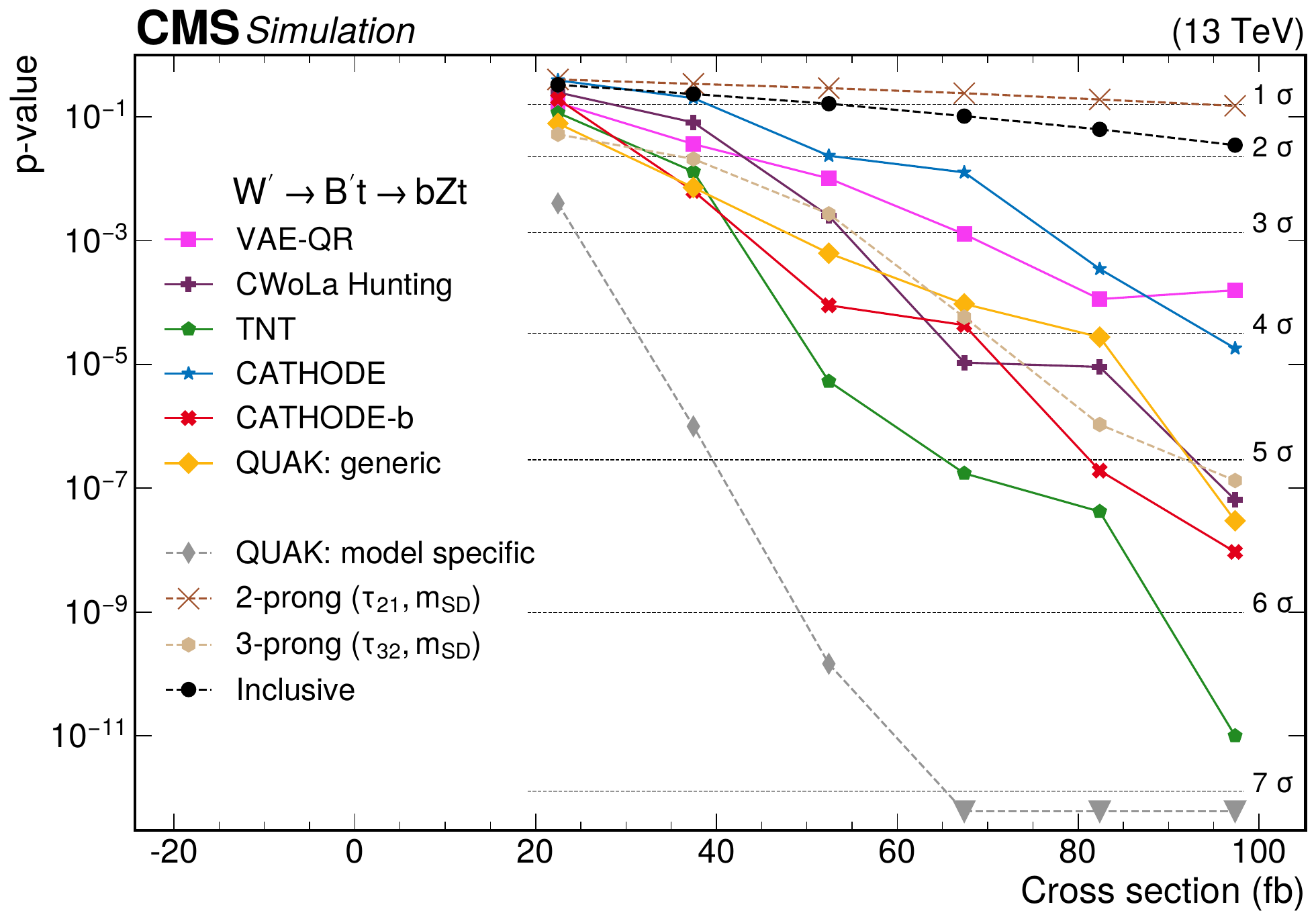}
   \caption{The $p$-values as a function of the injected signal cross sections for the different analysis procedures for two different signals: (upper) the 2-prong \XtoYY signal with $m_\mathrm{X}=3\TeV$, $m_\mathrm{Y}=170\GeV$, and $M_\mathrm{Y'}=170\GeV$, and (lower) 3-prong \Wp signal with $M_\mathrm{W'}=3\TeV$ and $M_\mathrm{B'}=400\GeV$. Significance values larger than 7$\sigma$ are denoted with downwards facing triangles.}
   \label{fig:pvals}
 \end{figure*}

Figure~\ref{fig:pvals} shows simulations of the sensitivity of the search methods, comparing the extracted p-value as a function of the signal cross section for two benchmark signals, the 2+2 prong \XtoYY and the 3+3 prong \Wp.

As expected, the inclusive search is sensitive to both models, but is unable to reach evidence or discovery-level significances at the considered signal cross sections because of the large QCD background.
The two-prong (3-prong) targeted selection is found to improve sensitivity beyond the inclusive search for the 2-prong \XtoYY signal (3-prong \Wp signal), but is found to be significantly worse than the inclusive selection on the 3-prong (2-prong) signal.
In contrast, all anomaly detection methods are able to demonstrate increased sensitivity above an inclusive search for both signals.
The relative performance of the anomaly detection methods are seen to vary between the two signals and no single method is seen to be optimal for both.
The model-specific \quak search is found to yield the best sensitivity for both signals as expected, reaching the discovery level at significantly lower cross sections than other methods.
The sensitivities of the weakly supervised methods are seen to depend nonlinearly on the signal cross section, because the amount of signal present in the data affects the training procedure and therefore the signal selection efficiency.

A version of the \cwola method similar to that of Ref. \cite{ATLAS_cwola}, using only the jet masses as input features, was also tested on these same signals.
Its performance was found not to be competitive with the methods employed in this search, failing to reach the $3 \sigma$ significance level for either signal model
in the cross section range considered.
This illustrates the significant gains provided by multivariate approaches employed in this analysis as compared to prior work.

Interpretability studies are also performed to verify that if a significant excess is seen by an anomaly detection method, its basic properties could be understood.
It is found that for discovery-level signal strengths, the substructure properties of the signal, such as the number of prongs and estimates of the masses of the daughter particles, could be determined by comparing the features of the excess events to those of backgrounds.
Such a characterization would allow follow-up studies, to verify its nature, including dedicated searches targeting signal models matching the properties of the excess.
Details of these studies are given in Appendix \ref{sec:sig_interp}.

The anomaly detection methods are also applied to the signal-depleted CR in data.
The fitting procedure is performed, and it is verified that the background model described the data well, and no method reported any significant excesses of events.

\section{Search results and interpretation}
\label{sec:results}
\begin{figure*}[htbp]
   \centering
   \includegraphics[width = 0.388\textwidth]{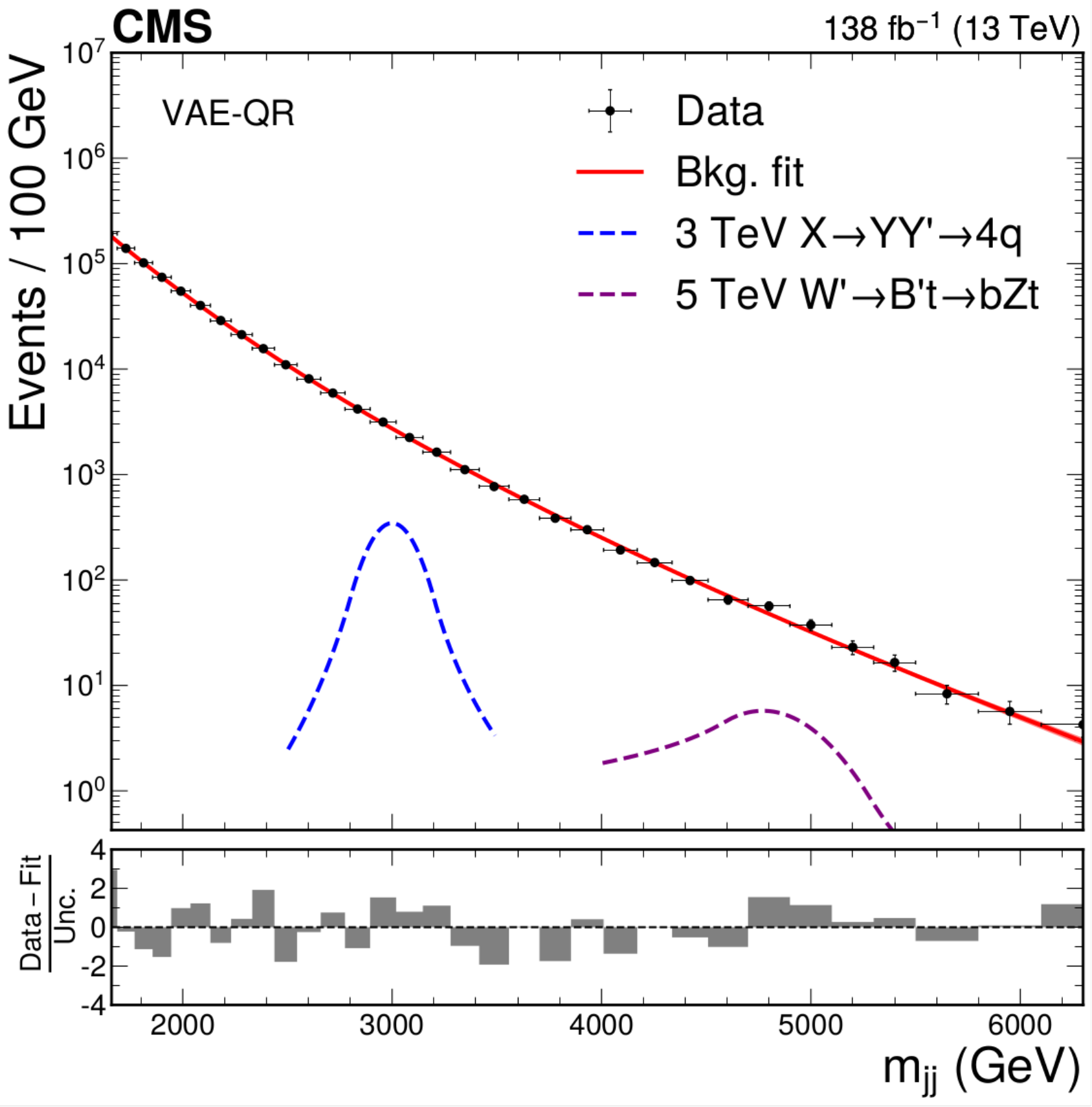}
   \includegraphics[width = 0.4\textwidth]{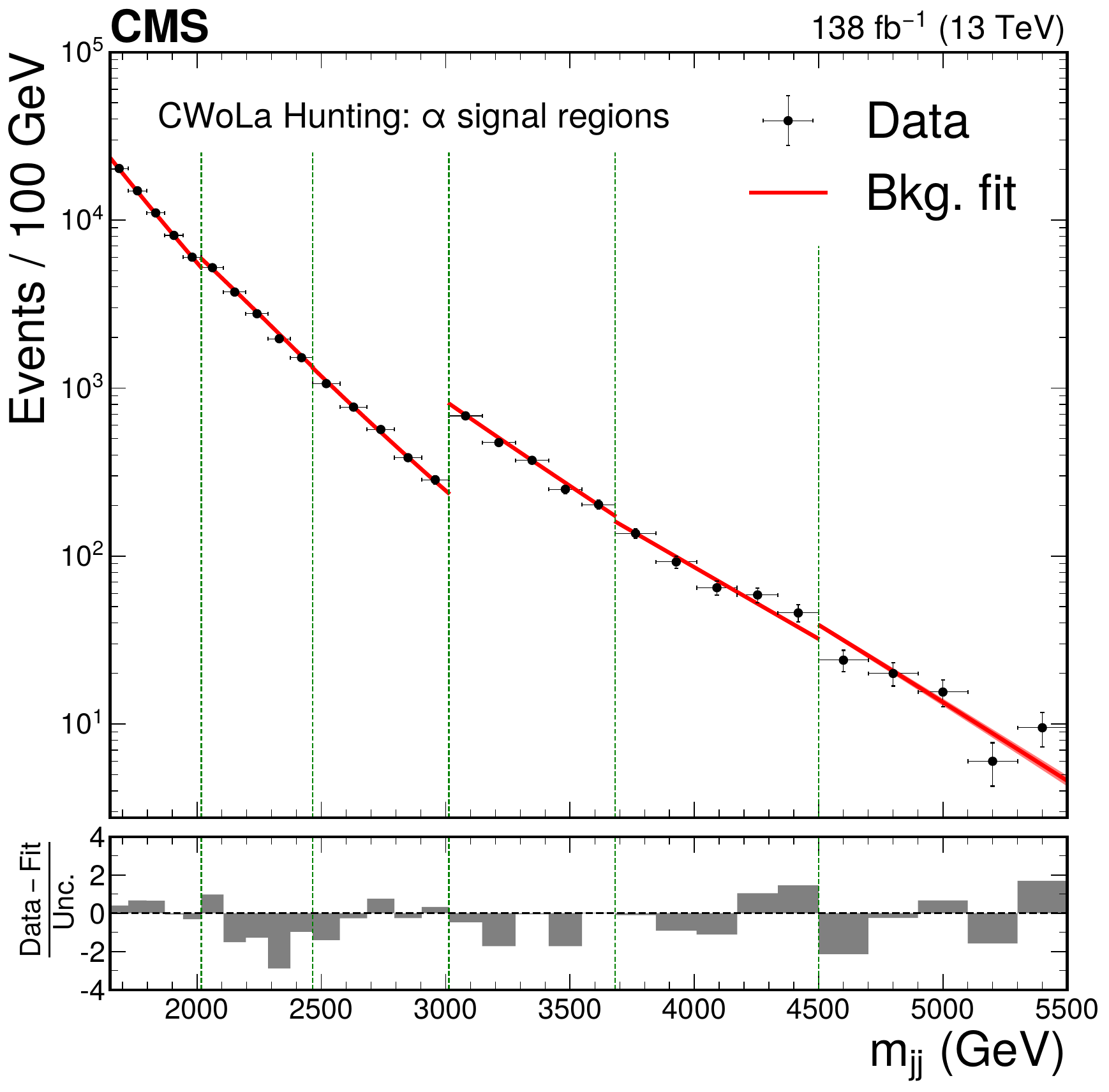} \\
   \includegraphics[width = 0.4\textwidth]{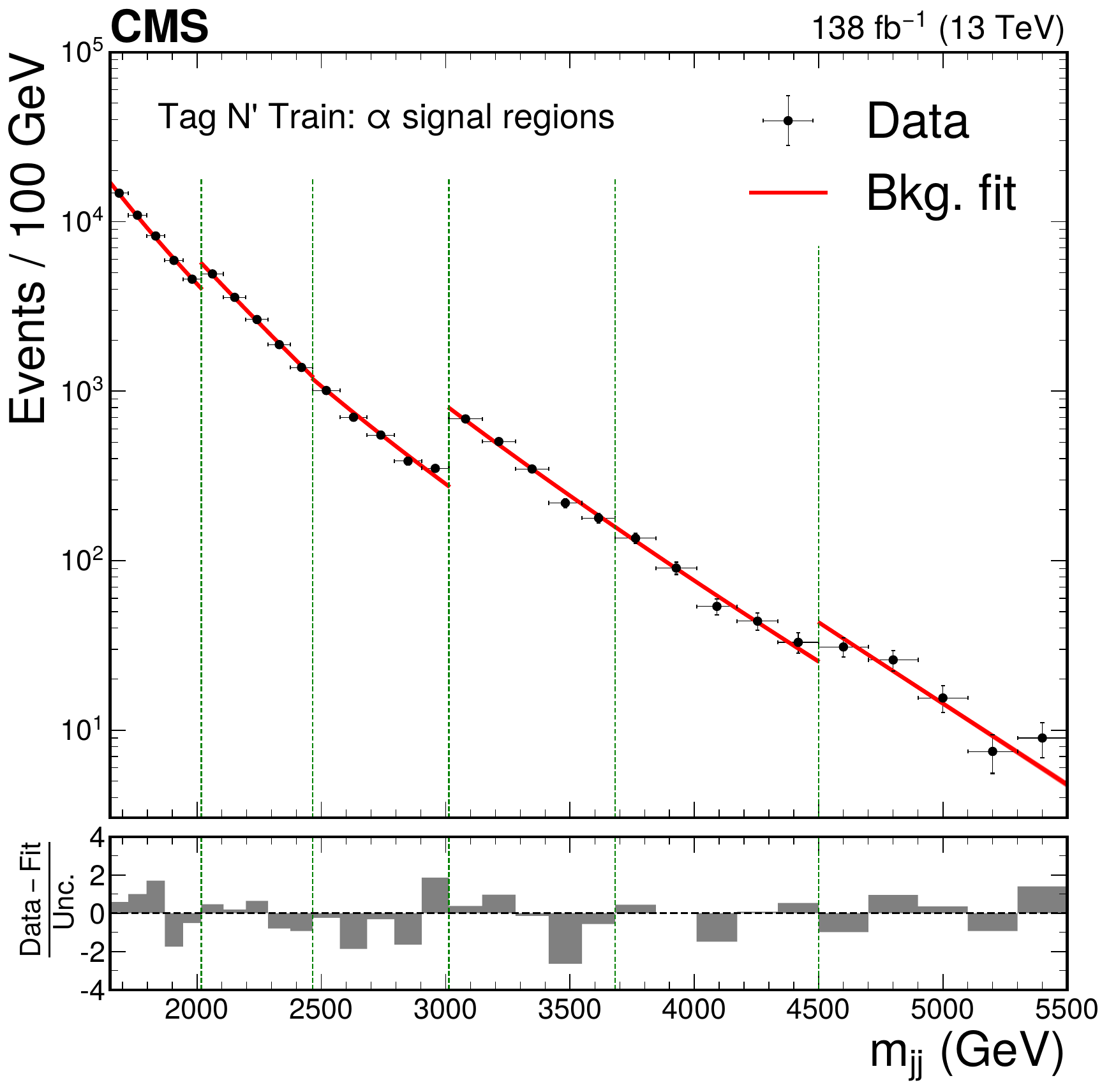}
   \includegraphics[width = 0.4\textwidth]{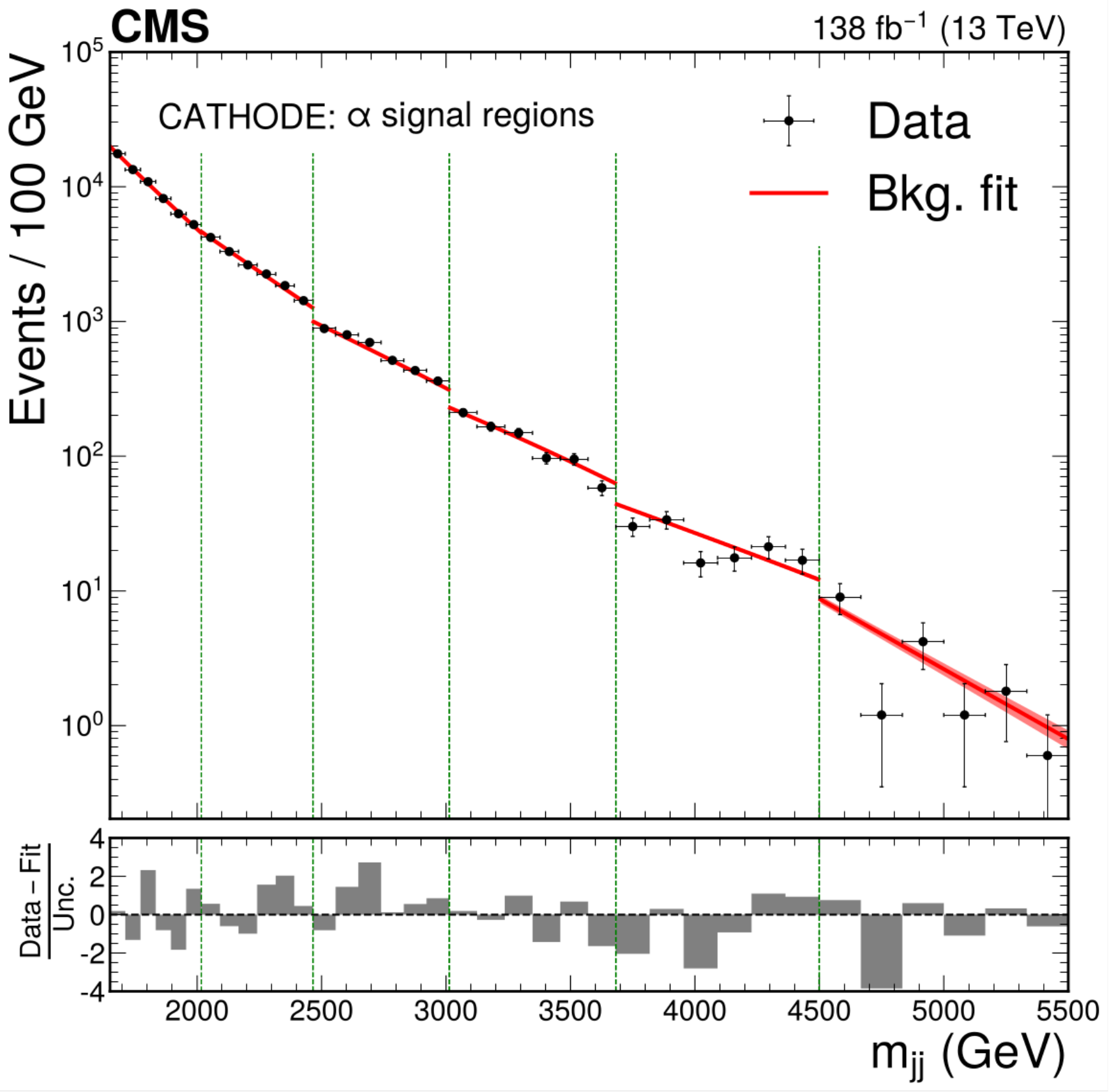} \\
   \includegraphics[width = 0.4\textwidth]{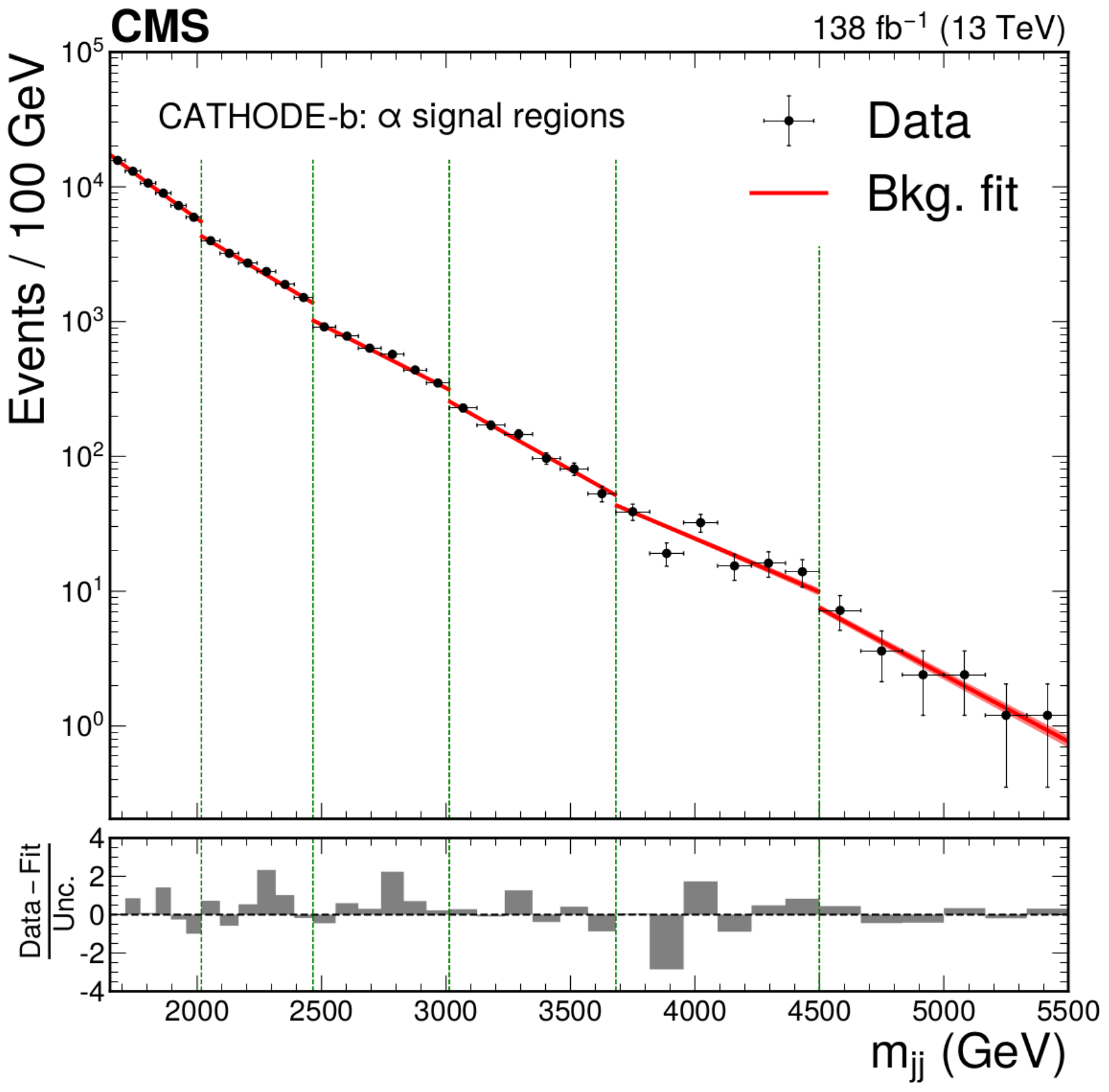}
   \includegraphics[width = 0.4\textwidth]{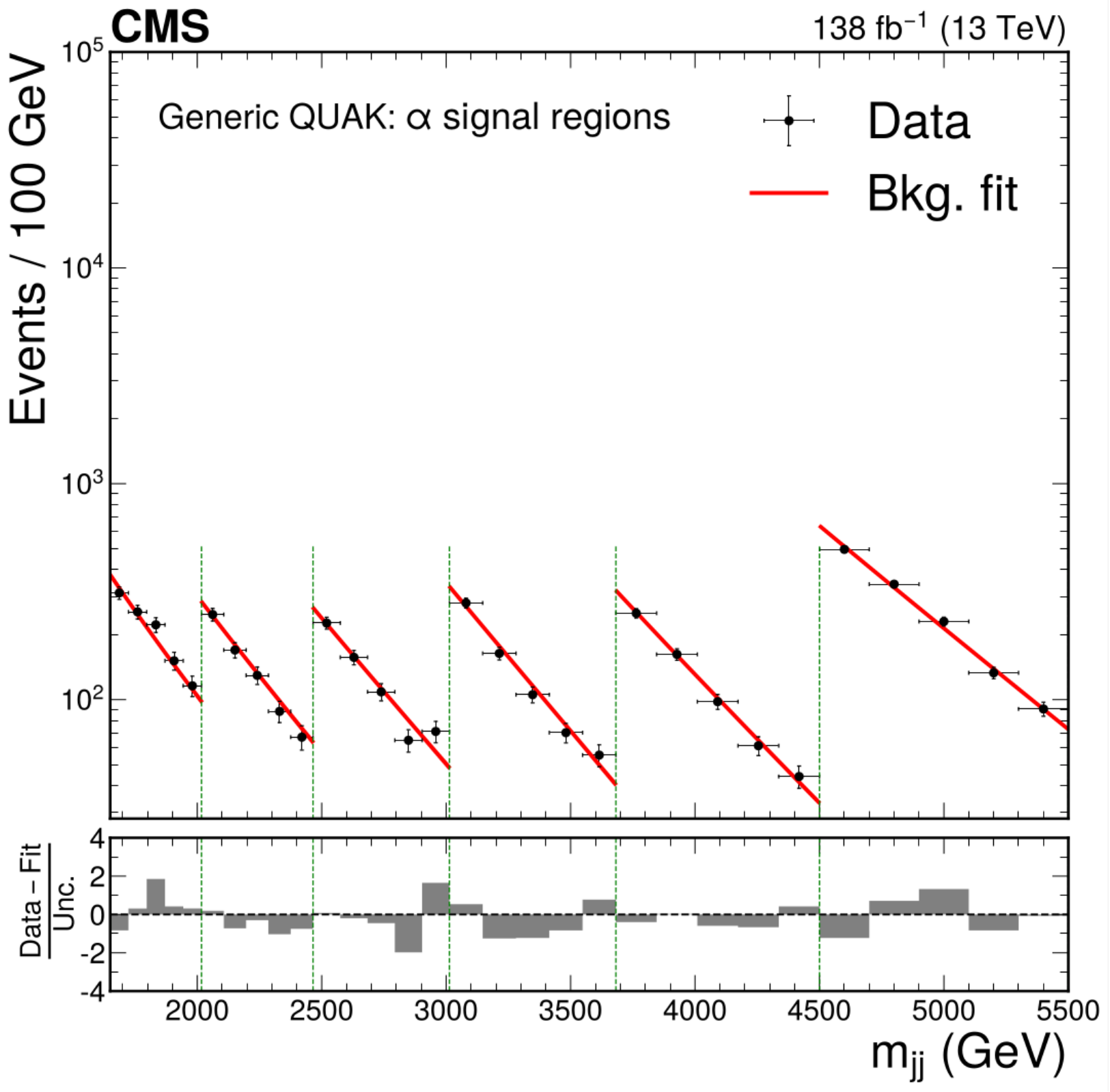} \\
   \caption{The dijet invariant mass spectrum and resulting background fit to the data for \vae~(upper left), \cwola~(upper right), \TNT~(middle left), \cathode~(middle right), \cathodeb~(lower left), and \quak~(lower right).
   The shapes of two benchmark signals are shown for the \vae method; the signal shapes for the other methods are similar.
   For all methods besides the \vae, separate selections are applied for different signal mass hypotheses and the resulting mass spectra are fit separately.
   The figures show the fitted and observed dijet mass distribution in the signal window of each selection, which results in a discontinuous distribution.
   The spectra in the $\alpha$ signal regions (indicated by the vertical dotted lines) are shown 
   for the weakly supervised methods and a similar selection of signal regions are shown for the \quak method.
}
   \label{fig:mjj1}
 \end{figure*}

\begin{figure*}[htbp]
   \centering
   \includegraphics[width = 0.4\textwidth]{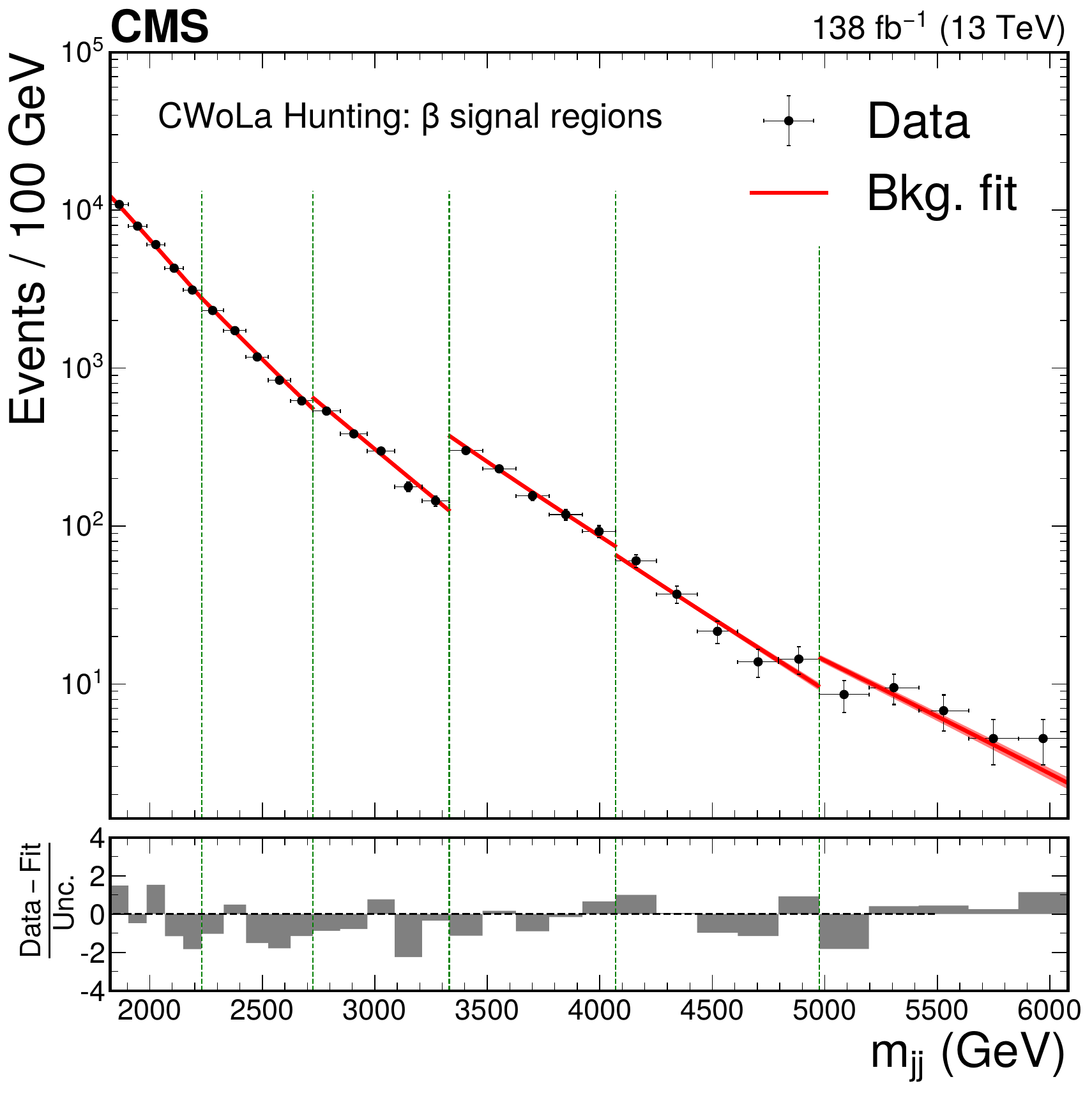} \\
   \includegraphics[width = 0.4\textwidth]{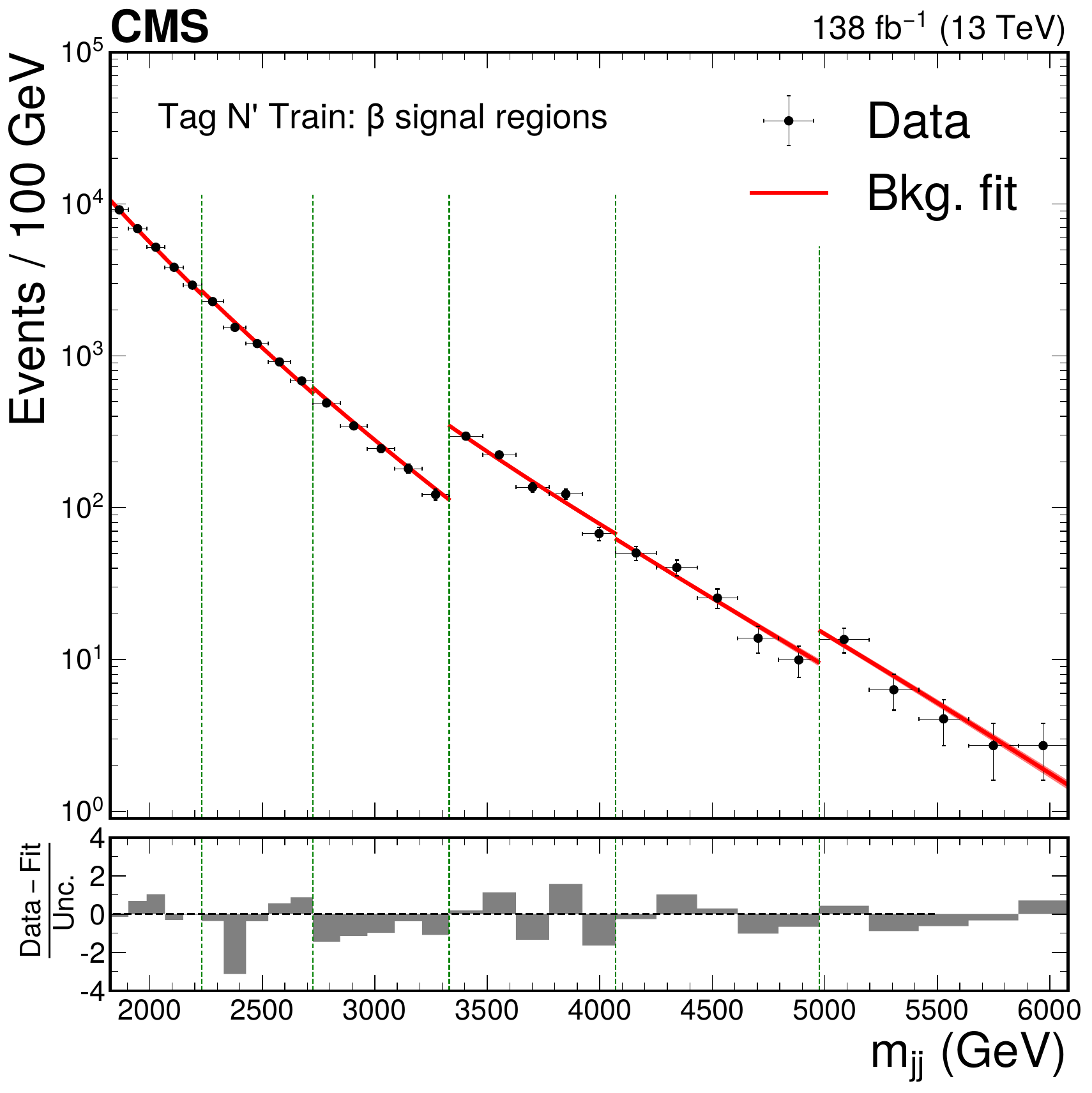}
   \includegraphics[width = 0.4\textwidth]{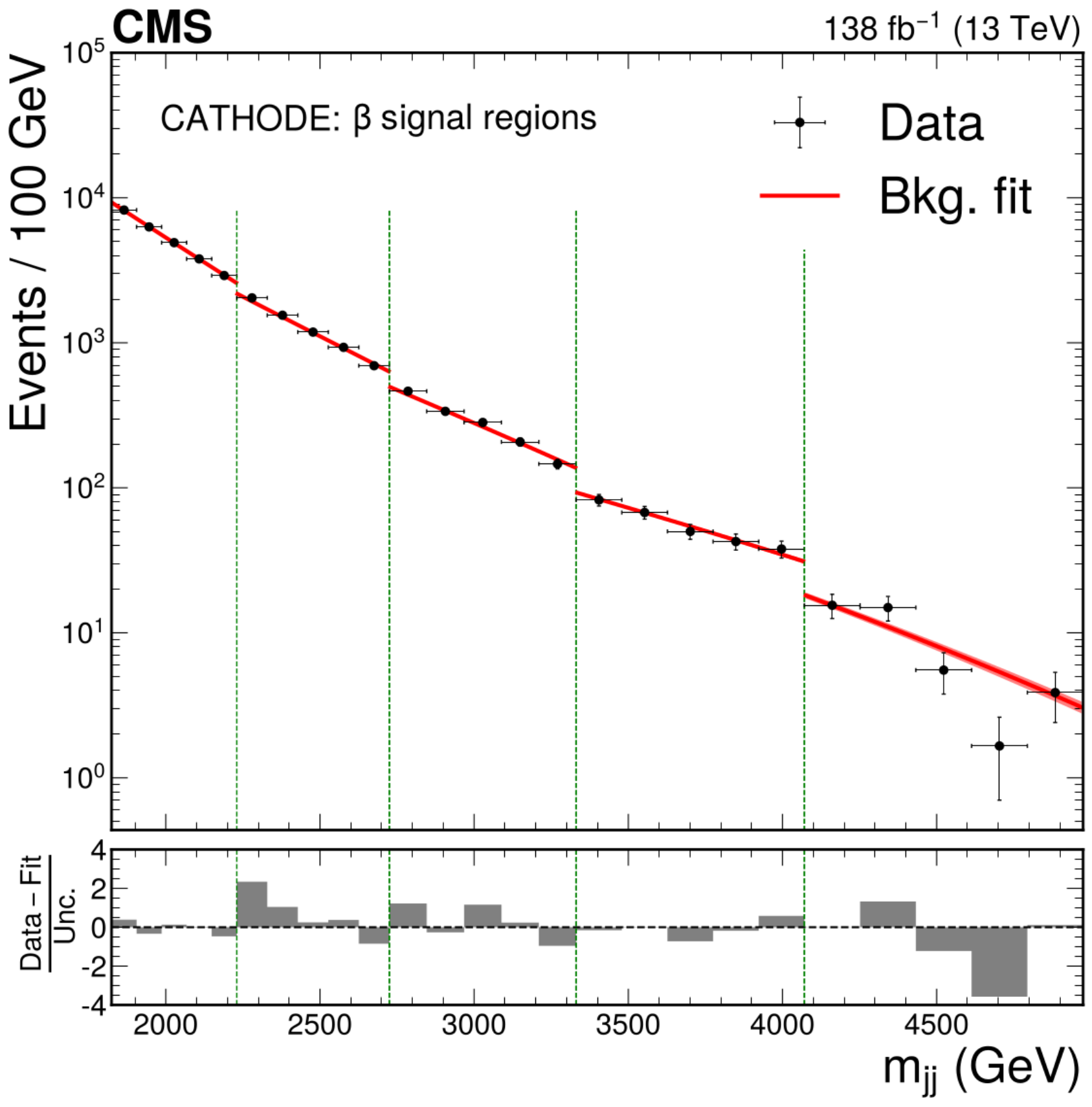} \\
   \includegraphics[width = 0.4\textwidth]{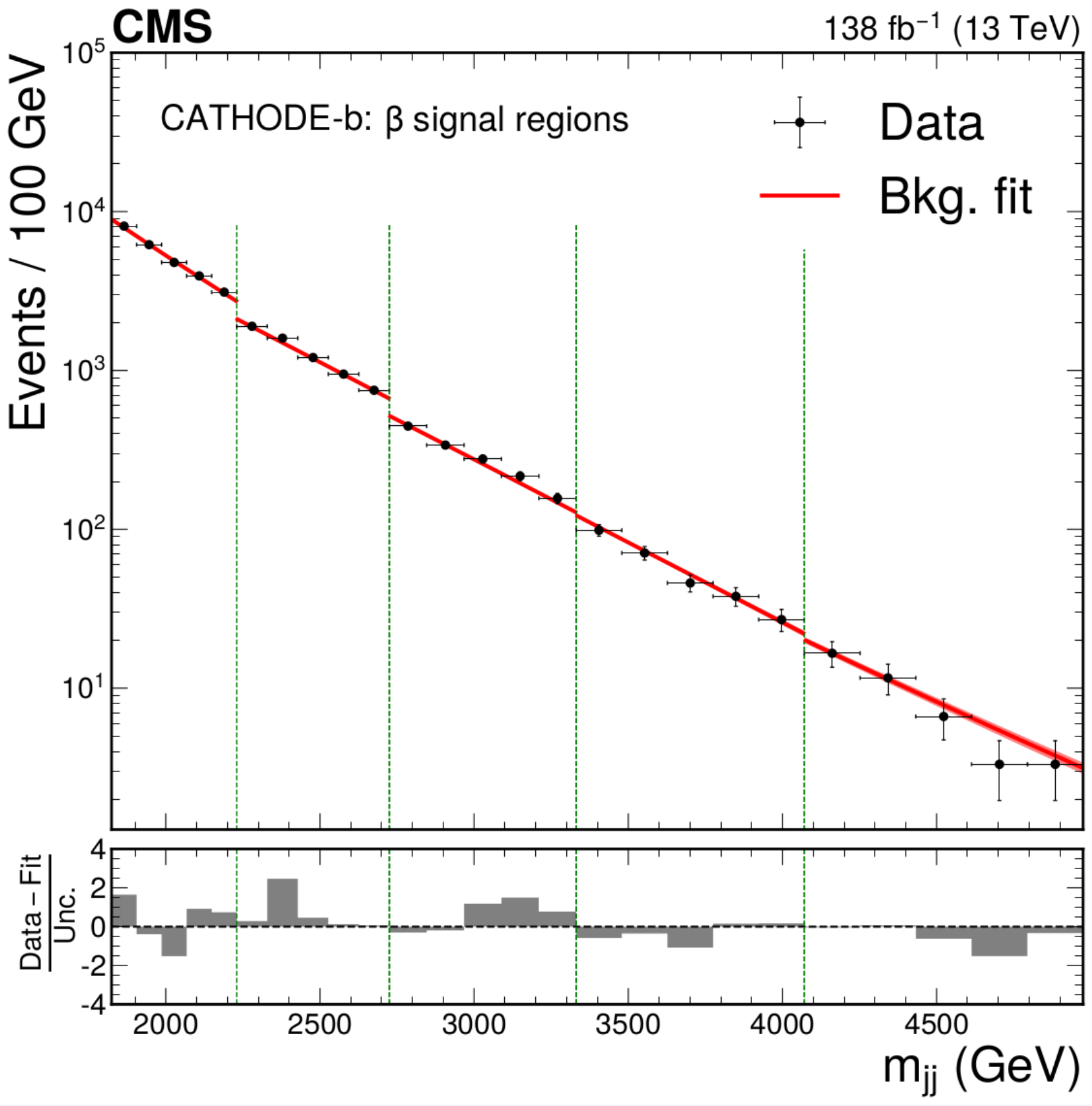}
   \includegraphics[width = 0.41\textwidth]{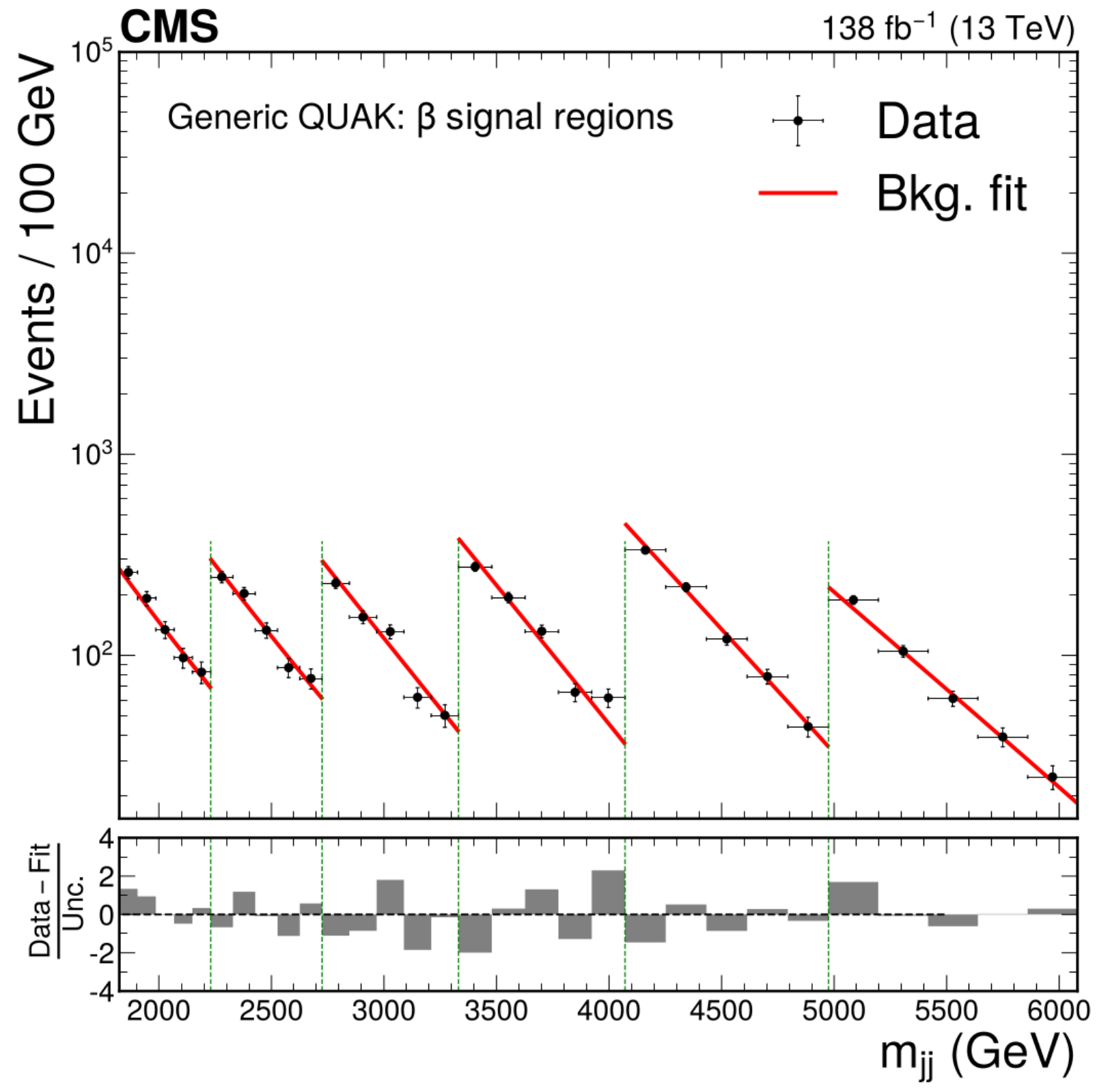}
   \caption{The dijet invariant mass spectrum and resulting background fit to the data for the $\beta$ signal regions (indicated by the vertical dotted lines) of \cwola~(upper), \TNT~(middle left), \cathode~(middle right), \cathodeb~(lower left), and a similar selection of signal regions of \quak~(lower right).
   Separate selections are applied for different signal mass hypotheses and the resulting mass spectra are fit separately.
   The figures therefore show the fitted and observed dijet mass distribution in the signal window of each selection, which results in a discontinuous distribution.
   The \cathode and \cathodeb methods are not used in the highest mass window of the $\beta$ signal regions due to the limited number of data events.
   They therefore have one fewer signal region shown than the other methods.
}
   \label{fig:mjj2}
 \end{figure*}

Following the validation in simulation and the data CR, the methods are applied to the data SR.
For the weakly supervised methods, the training and selection procedures are repeated multiple times to target each of the $\alpha$ and $\beta$ SRs.
For the \quak method, the selection is repeated targeting each $m_\mathrm{A}$ hypothesis.
The fitted dijet invariant mass spectra after a selection from each of the anomaly detection methods are shown in Figs.~\ref{fig:mjj1} and ~\ref{fig:mjj2}.
Good agreement between the data and background-only fits is observed for all methods.
The largest excesses of events seen by the \cathode, \cathodeb, \quak, and \vae methods had local significances of 2.2, 2.9, 2.6, and 2.3$\sigma$ at resonance masses of 2.3, 2.3, 4.7, and 4.9\TeV, respectively.
The \TNT and \cwola methods did not report any excesses of events larger than 1.5$\sigma$.

Having seen no significant excesses of events in the data, studies are performed to evaluate the sensitivity of the search procedure on a subset of the signal models.
These benchmark signal models cover several different combinations of the B and C particle substructure.
The considered models are \XtoYY, \Wp, \Wkk, and \GtoHH, which have 2+2, 3+3, 4+2, and 6+6 prongs in the substructure of the B and C particles, respectively.
The masses of the Y and Y' particles are set to 170 GeV, while the masses of the $\mathrm{B}'$, $\mathrm{R}$, and $\PH$ are set to 400\GeV.

For each method, the selection efficiency on the benchmark signal models is evaluated.
For the weakly supervised methods, the signal efficiency is known to vary based on the amount of signal present in the training data set, and therefore depends strongly on the signal cross section.
This dependence necessitated a special procedure to evaluate the efficiency and use it properly in deriving the exclusion limits.
The procedure involved injecting signals with various cross sections into the data, and each time performing the full analysis procedure, including the retraining of classifiers, to assess the signal efficiency at the injected cross section.
This procedure is described in detail in \applabel~\ref{sec:limits}.

\begin{figure*}[htbp]
   \centering
   \includegraphics[width = 0.7\textwidth]{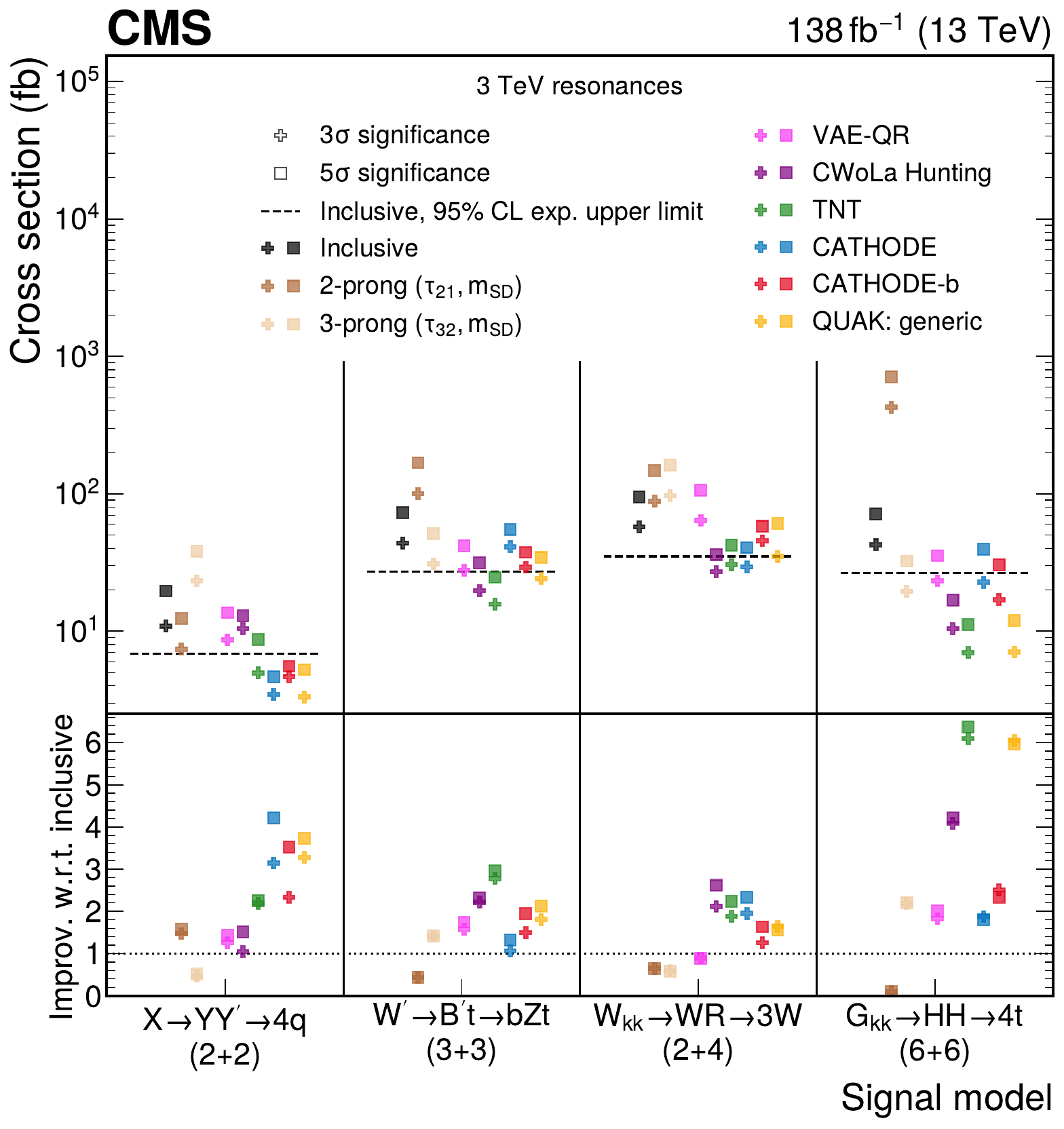}
   \caption{ The discovery sensitivity  for the process $\mathrm{A} \to \mathrm{BC}$, using the anomaly detection methods and a comparison to sensitivity of the inclusive search.
   In all signal processes, the mass of the heavy resonance is set to $m_\mathrm{A}=3\TeV$.
   For the BSM daughter particles, the masses of the \PY and \PYpr are set to 170\GeV, while the masses of the \PBpr,  $\mathrm{R}$, and  \PH are set to 400\GeV.
   In the upper panel, for each method, the cross section, which would have led to an expected $3\sigma$ ($5\sigma$) excess, is shown as a cross (square) marker.
   Sensitivities from six anomaly detection methods (six colors) are compared to an inclusive dijet search in which no substructure selection is made (black) and traditional substructure selections targeting 2-prong (dark brown) or 3-prong (tan) decays.
   The expected 95\% confidence level upper limits from the inclusive search are also shown in the upper panel as a dashed line.
   For all signal models at least one anomaly detection method is able to achieve an expected $5\sigma$ significance at a cross section at or below the upper limit of the inclusive search.
   Shown in the lower panel is the ratio of the cross section sensitivity from the inclusive search to the corresponding sensitivity for each method.}
   \label{fig:exp_signif}
 \end{figure*}

The discovery sensitivity of each method for the benchmark signals is then estimated.
For each method, the signal cross sections which would have led to an expected
$3\sigma$ and $5\sigma$ excess are determined.
The discovery sensitivities of the anomaly detection methods are compared to those of an inclusive search, and to traditional cutoff-based selections targeting 2-prong and 3-prong decays.
The results are shown in Fig. \ref{fig:exp_signif}.
It is found that the anomaly detection methods outperform the traditional cutoff-based selections for all of the benchmark signals considered, and significantly improve upon the discovery sensitivities of the inclusive search.
The best performing method varies across the different signals, illustrating the complementarity of the different approaches.
The largest improvements are seen for the 6+6 prong \GtoHH signal, in which anomaly detection methods are found to reduce the cross section needed for a $5\sigma$ discovery by a factor of 6.4 (2.9) as compared to the inclusive (3-prong) selection.
For every benchmark signal considered, at least one anomaly detection method is able to reach the $5\sigma$ level at a cross section below, or approximately equal to, the expected $2\sigma$ upper exclusion limit of the inclusive search.

Exclusion limits on the benchmark signal models are also derived for each anomaly detection method.
For the weakly supervised methods, the estimated signal efficiencies used in the limit setting are derived using the previously mentioned procedure, in which signals with various cross sections are injected into the data and the classifiers are retrained.
These efficiencies are used in conjunction with the observed data in the SR to derive exclusion limits.

\begin{figure*}[htbp]
   \centering
   \includegraphics[width = 0.49\textwidth]{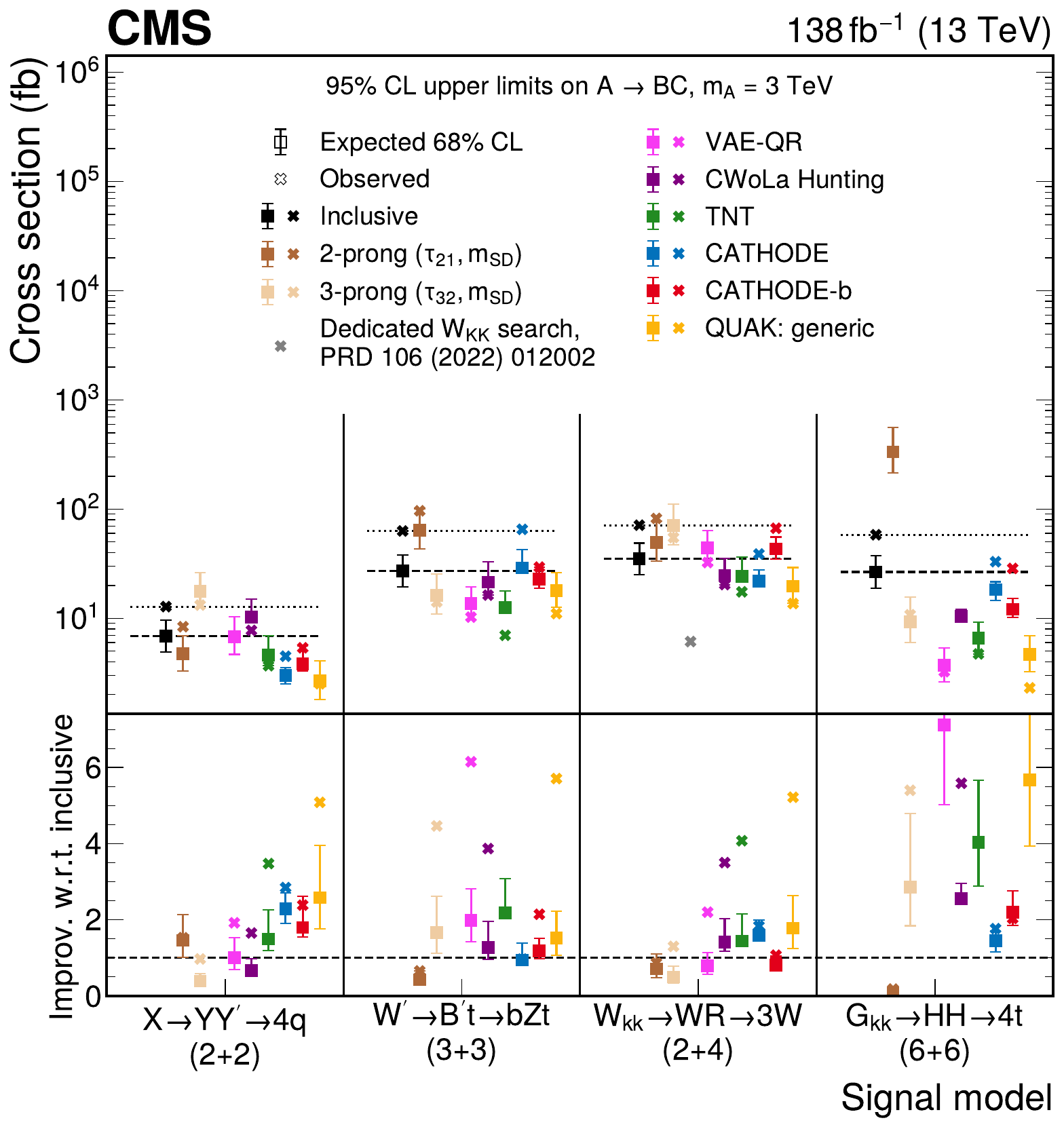}
   \includegraphics[width = 0.49\textwidth]{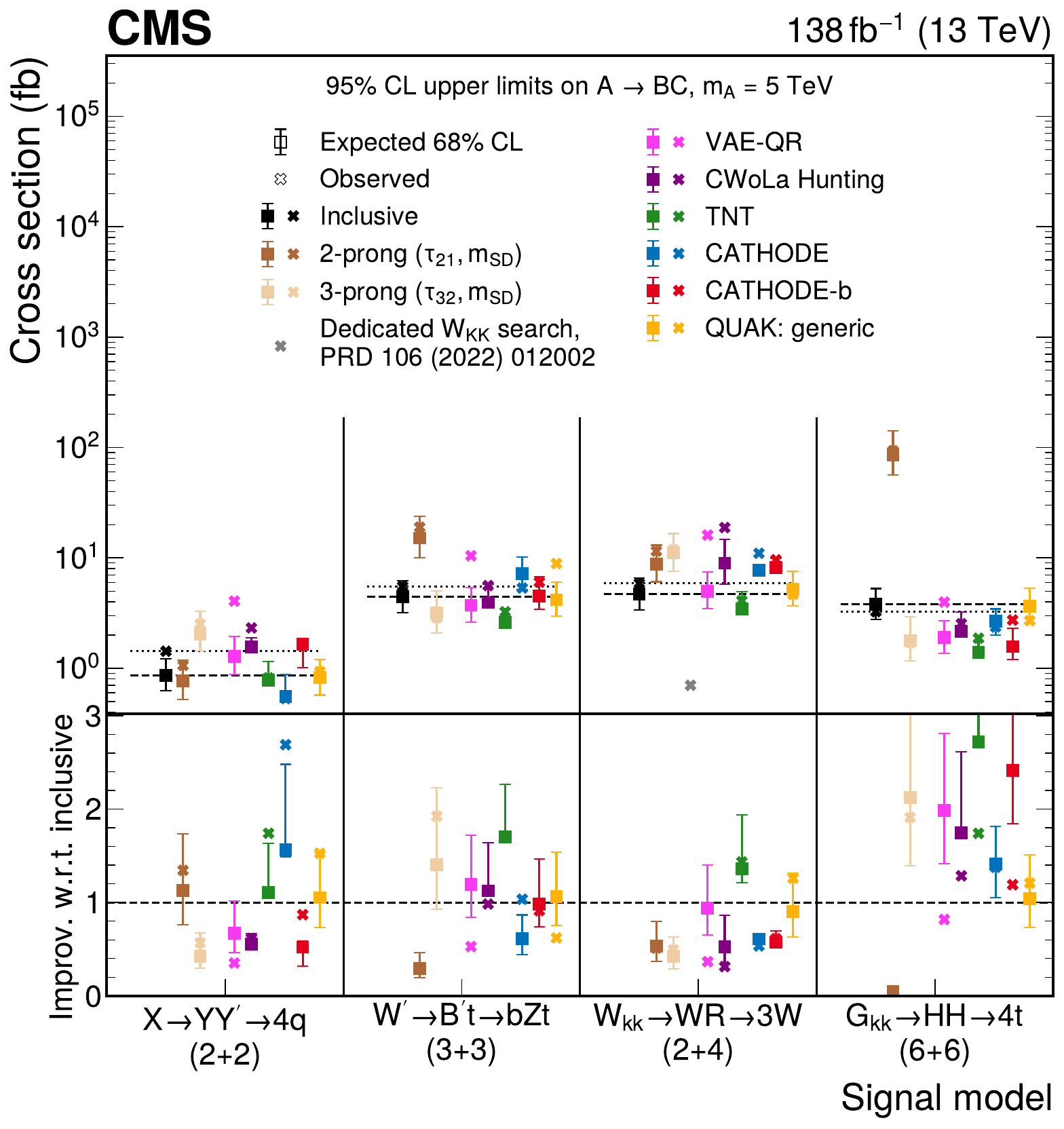}
   \caption{The upper limit at 95\% confidence level on the cross section for the process $\mathrm{A} \to \mathrm{BC}$, is shown for each search method applied to a variety of signal models.
   For a resonance mass $m_\mathrm{A}=3\TeV$ (left) and $m_\mathrm{A}=5\TeV$ (right), we show for each signal model (columns), and search method (all colors), the observed limits (crosses), expected limits (squares), and their 68\% expected central intervals (error bars).
   For the BSM daughter particles, the masses of the \PY and \PYpr are set to 170\GeV, while the masses of the \PBpr,  $\mathrm{R}$, and \PH are set to 400\GeV.
   Limits from the anomaly detection methods (six colors) are compared to those from an inclusive dijet search in which no substructure selection is made (black markers and horizontal lines),
 traditional substructure selections targeting 2-prong (dark brown) or 3-prong decays (tan), and the observed limit from a previous CMS search \cite{CMS_triboson_hadronic} for the \Wkk model in the all-hadronic channel (gray).}
   \label{fig:limits}
 \end{figure*}

These exclusion limits are compared to those from an inclusive dijet search, those from the traditional cutoff-based methods, and those from the previous CMS \Wkk search in the all-hadronic channel \cite{CMS_triboson_hadronic}, for benchmark signals with resonance masses of 3 and 5\TeV in Fig.~\ref{fig:limits}.
It is found that the expected limits of the anomaly detection methods improve upon those of the inclusive search and traditional cutoff-based approaches for all benchmark signals.
The largest gains are seen for the \GtoHH signal, where improvements up to a factor of 7.1 (2.5) are seen when compared to the inclusive (3-prong) selection.
Larger gains from the anomaly detection methods are seen for the 3\TeV signals than the 5\TeV signals, because at lower masses the anomaly detection methods employ tighter selections, which more effectively suppress the larger QCD backgrounds.
As expected, the previous CMS dedicated \Wkk search reports significantly more stringent limits on the \Wkk model than the anomaly detection methods.

Numerical limits on all considered signal models, including, but not limited to, the benchmark set, are summarized in Tables \ref{tab:3tev_extra_limits} and \ref{tab:5tev_extra_limits}.
It is found that the anomaly detection methods do not improve with respect to the inclusive search for some signals.
Signals in which only one of the two jets has a distinctive signature, such as the
 \Qstar signal or the \XtoYY mass points that feature very light daughter masses, are found to be difficult for the anomaly detection methods.
This is likely because for these signals one of the two jets originates from a single parton, or lacks sufficient jet substructure to be distinguished from a single parton, which makes signal discrimination more challenging.
For the \XtoYY signal, this occurs because light daughters are extremely boosted, making the 2-prong structure look similar to a single prong.
For example, for a 25\GeV daughter from a 3\TeV resonance, the typical separation ${\Delta R}$ between the two quarks is only $\sim$0.03.

It is important to mention that these benchmark signals are included in the signal prior of the generic \quak method.
In contrast, they are not used by the other methods, except in the procedure used to evaluate the signal efficiency.
It is found that when removing a given signal from the prior of the generic \quak method, the sensitivity to that signal degraded by $\sim$30\%.
The results from the other methods may therefore generalize better to untested signals than the results of the generic \quak search.

 \begin{table*}[ht!]
    \centering
    \topcaption{Limits on additional signal models and daughter mass combinations for a 3\TeV resonance mass. For each signal, the expected and observed 95\% CL upper limits on the signal cross section from the best performing anomaly detection method are reported.
    The expected limit from the anomaly detection method is also compared to the expected limit of the inclusive search to quantify the improvement.
    For some signals the anomaly detection methods do not improve with respect to the inclusive search. This is indicated by an improvement factor less than one.}
    \cmsTable{
    \begin{tabular}{l c c c c}
    \multicolumn{1}{p{3.2cm}}{ \centering { Model $\mathrm{A} \to \mathrm{BC}$ \\ ($m_\mathrm{A} = 3\TeV$)}} &
    \multicolumn{1}{p{3.2cm}}{ \centering {Daughter masses} \\ {(\GeVns)}} & {Method} &
    \multicolumn{1}{p{3.4cm}}{ \centering {Exp. (Obs.) limit \\ (fb)}} & \multicolumn{1}{p{3.4cm}}{ \centering {Improv. factor \\  w.r.t. inclusive }} \\
    \hline
        \Qstar & 25 & \cwola & 61.1 (30.1) & 0.3 \\
            & 80 & \cathode & 46.2 (76.3) & 0.4 \\
            & 170 & \cathode & 48.7 (86.3) & 0.4 \\
            & 400 & \cwola & 45.8 (24.3) & 0.5 \\\\

        \XtoYY & 25, 25 & \cathode & 7.4 (9.7) & 0.9 \\
            & 25, 80 & \cathode & 5.9 (8.2) & 1.2 \\
            & 25, 170 & \cathode & 8.3 (9.7) & 0.8 \\
            & 25, 400 & \vae & 13.6 (12.5) & 0.6 \\
            & 80, 80 & \cathode & 3.2 (4.3) & 2.1 \\
            & 80, 170 & \cathode & 4.5 (6.0) & 1.5 \\
            & 80, 400 & \cathode & 4.6 (6.0) & 1.6 \\
            & 170, 170 & \quak & 2.7 (2.5) & 2.6 \\
            & 170, 400 & \cathode & 4.3 (5.8) & 1.7 \\
            & 400, 400 & \vae & 2.1 (1.9) & 4.2 \\\\

        \Wp & 25 & \TNT & 22.6 (13.9) & 1.7 \\
            & 80 & \TNT & 18.2 (11.3) & 1.9 \\
            & 170 & \TNT & 12.2 (7.3) & 2.1 \\
            & 400 & \TNT & 12.5 (7.0) & 2.2 \\\\

        \Wkk & 170 & \TNT & 22.1 (15.2) & 1.6 \\
            & 400 & \quak & 19.7 (13.7) & 1.8 \\\\

        \Zp & 400 & \TNT & 39.1 (23.3) & 3.5 \\
        \GtoHH & 400 & \vae & 3.7 (3.2) & 7.1 \\
    \end{tabular}
    }
    \label{tab:3tev_extra_limits}
\end{table*}

\begin{table*}[ht!]
    \centering
    \topcaption{Limits on additional signal models and daughter mass combinations for a 5\TeV resonance mass. For each signal, the expected and observed 95\% CL upper limits on the signal cross section from the best performing anomaly detection method are reported.
    The expected limit from the anomaly detection method is also compared to the expected limit of the inclusive search to quantify the improvement.
    For some signals the anomaly detection methods do not improve with respect to the inclusive search. This is indicated by an improvement factor less than one.}
    \cmsTable{
    \begin{tabular}{l c c c c}
    \multicolumn{1}{p{3.2cm}}{ \centering { Model $\mathrm{A} \to \mathrm{BC}$ \\ ($m_\mathrm{A} = 5\TeV$)}} &
    \multicolumn{1}{p{3.2cm}}{ \centering {Daughter masses} \\ {(\GeVns)}} & {Method} &
    \multicolumn{1}{p{3.2cm}}{ \centering {Exp. (Obs.) limit \\ (fb)}} & \multicolumn{1}{p{3.2cm}}{ \centering {Improv. factor \\ w.r.t. inclusive }} \\
    \hline

        \Qstar & 25 & \quak & 3.5 (3.1) & 0.7 \\
               & 80 & \quak & 3.2 (2.8) & 0.8 \\
               & 170 & \quak & 3.3 (3.6) & 0.8 \\
               & 400 & \cathode & 3.7 (3.6) & 0.7 \\\\

        \XtoYY  & 25, 25 & \quak & 1.7 (1.6) & 0.5 \\
                & 25, 80 & \quak & 1.3 (1.3) & 0.7 \\
                & 25, 170 & \quak & 1.1 (1.1) & 0.8 \\
                & 25, 400 & \vae & 1.0 (3.4) & 0.9 \\
                & 80, 80 & \TNT & 1.1 (1.2) & 0.8 \\
                & 80, 170 & \quak & 0.9 (1.0) & 0.9 \\
                & 80, 400 & \vae & 0.9 (3.0) & 0.9 \\
                & 170, 170 & \cathode & 0.6 (0.5) & 1.6 \\
                & 170, 400 & \cathode & 0.7 (0.6) & 1.3 \\
                & 400, 400 & \cathode & 0.4 (0.3) & 2.4 \\\\

        \Wp     & 25 & \TNT & 3.6 (5.2) & 1.6 \\
                & 80 & \TNT & 3.5 (5.0) & 1.6 \\
                & 170 & \TNT & 2.5 (3.4) & 1.9 \\
                & 400 & \TNT & 2.6 (3.2) & 1.7 \\\\

        \Wkk    & 170 & \TNT & 4.4 (5.9) & 1.2 \\
                & 400 & \TNT & 3.4 (4.1) & 1.4 \\\\
        \GtoHH & 400 & \TNT & 1.4 (1.9) & 2.7 \\
    \end{tabular}
    }
    \label{tab:5tev_extra_limits}
\end{table*}

Except for \Wkk, none of the signals considered have been previously covered by dedicated searches in the mass range considered in this analysis, making the limits reported here the first of their kind.
This broad range of unique exploration is made possible by the combination of flexibility and sensitivity within the anomaly detection approach.
Though it does not achieve the same sensitivity to an individual model as a dedicated search, anomaly detection offers enhanced sensitivity to a much larger class of models.

Improvements in the upper limits and discovery potential, for the best performing anomaly detection method on each signal, are summarized in Table \ref{tab:limits_improvement}.
The \vae method is found to improve the exclusion limits more than the discovery potential.
This is due to the advantage of the multicategory approach, which is used only in the limit setting, when the relative yields of the considered signal model in each category are known, and is not used in the model-agnostic search.
In contrast, the weakly supervised methods are found to improve the discovery sensitivity more than they improved the limits, because their signal efficiency continues to improve with the higher signal cross sections necessary for discovery.
Overall, the anomaly detection methods are seen to improve the discovery sensitivity of all benchmark signals by at least a factor of 2.1 (2.6) compared to a dijet search with (without) the use of jet substructure information.
This illustrates the potential of anomaly detection methods to discover signals that may otherwise be missed, and highlights their complementarity to dijet searches that do use jet substructure information, those that do not, as well as dedicated searches.

 \begin{table*}[htbp!]
  \centering
    \topcaption{For the $\mathrm{A} \to \mathrm{BC}$ searches, the sensitivity improvement of the anomaly detection methods with respect to the best performing comparison method.
      The considered comparison methods are the inclusive search, 2-prong targeted selection, and 3-prong targeted selection.
  The fourth and fifth columns list, for each signal model, improvement factors on the exclusion limit for the best performing anomaly detection method for signals at masses of $m_\mathrm{A} = 3$ and 5\TeV, respectively.
  This is quantified as the ratio of the expected upper limit on the production cross section obtained by the anomaly detection method as compared to that of the inclusive search.
  The sixth column lists the improvement factor on the $5\sigma$ discovery potential for the best performing anomaly detection method for each signal at $m_\mathrm{A} = 3$\TeV.
  This is quantified as the ratio of the cross section that would have led to a $5\sigma$ excess for the comparison method as compared to that of the anomaly detection method.}
  \cmsTable{
    \begin{tabular}{l c c l l l}
     & & & \multicolumn{3}{c}{{Improvement factors in expected sensitivities}} \\
     {Prongs} & {Process} & {Comparison} & {95\% CL limit} & {95\% CL limit} & {5$\sigma$ discovery} \\
     & {$\mathrm{A} \to \mathrm{BC}$} & {method} & {$m_\mathrm{A} = 3$\TeV} & {$m_\mathrm{A} = 5$\TeV} & {$m_\mathrm{A} = 3$\TeV} \\
      \hline
     2+2 & \XtoYY & 2-prong $(\tau_{21},m_\mathrm{SD})$ &  1.8 (\quak)  & 1.4 (\cathode) &  2.9 (\cathode) \\
     3+3 & \Wp  & 3-prong $(\tau_{32},m_\mathrm{SD})$ &  1.3 (\TNT)   & 1.2 (\TNT) &  2.1 (\TNT) \\
     4+2 & \Wkk & Inclusive        &  1.5 (\cwola) & 1.4 (\TNT) &  2.6 (\cwola) \\
     6+6 & \GtoHH & 3-prong $(\tau_{32},m_\mathrm{SD})$ &  2.5 (\vae)  & 1.3 (\TNT) &  2.9 (\TNT) \\

  \end{tabular}
  }
  \label{tab:limits_improvement}
\end{table*}

\section{Summary}
\label{sec:summary}
To summarize, we have presented a model-agnostic search for new resonances in the dijet final state.
The search is based on \INTLUMI{}\fbinv of data collected at $\sqrt{s} = 13\TeV$ by the CMS experiment.
Five separate anomaly detection methods were employed to improve sensitivity to signals that produce jets with substructure distinct from that of QCD multijet events.
No significant excesses of events were observed by any of the methods.
The performance of the anomaly detection techniques was illustrated on a set of benchmark narrow-resonance signals covering a wide range of substructure signatures.
It was found that the anomaly detection methods improved the discovery sensitivity  and expected limits on the benchmark signals.
The anomaly detection methods were shown to enhance the sensitivity by larger factors, and on a much wider class of models, than traditional cutoff-based substructure selections, but fell short of the sensitivity of a dedicated model-specific search.

The performance of the anomaly detection methods on a diverse set of benchmark models demonstrates the sensitivity of the employed techniques to a wide class of dijet resonances that have substructure and fall within the considered mass range.
By construction, these approaches have sensitivity to an even broader class of models than the specific benchmarks studied.
The anomaly detection methods employed in this search represent a significant step forward in the search for new particles at the LHC in a model-agnostic fashion.
Further development and deployment of these techniques will play a crucial role in maximizing the discovery potential of LHC data.

\begin{acknowledgments}
We congratulate our colleagues in the CERN accelerator departments for the excellent performance of the LHC and thank the technical and administrative staffs at CERN and at other CMS institutes for their contributions to the success of the CMS effort. In addition, we gratefully acknowledge the computing centers and personnel of the Worldwide LHC Computing Grid and other centers for delivering so effectively the computing infrastructure essential to our analyses. Finally, we acknowledge the enduring support for the construction and operation of the LHC, the CMS detector, and the supporting computing infrastructure provided by the following funding agencies: SC (Armenia), BMBWF and FWF (Austria); FNRS and FWO (Belgium); CNPq, CAPES, FAPERJ, FAPERGS, and FAPESP (Brazil); MES and BNSF (Bulgaria); CERN; CAS, MoST, and NSFC (China); MINCIENCIAS (Colombia); MSES and CSF (Croatia); RIF (Cyprus); SENESCYT (Ecuador); ERC PRG, RVTT3 and MoER TK202 (Estonia); Academy of Finland, MEC, and HIP (Finland); CEA and CNRS/IN2P3 (France); SRNSF (Georgia); BMBF, DFG, and HGF (Germany); GSRI (Greece); NKFIH (Hungary); DAE and DST (India); IPM (Iran); SFI (Ireland); INFN (Italy); MSIP and NRF (Republic of Korea); MES (Latvia); LMTLT (Lithuania); MOE and UM (Malaysia); BUAP, CINVESTAV, CONACYT, LNS, SEP, and UASLP-FAI (Mexico); MOS (Montenegro); MBIE (New Zealand); PAEC (Pakistan); MES and NSC (Poland); FCT (Portugal); MESTD (Serbia); MCIN/AEI and PCTI (Spain); MOSTR (Sri Lanka); Swiss Funding Agencies (Switzerland); MST (Taipei); MHESI and NSTDA (Thailand); TUBITAK and TENMAK (Turkey); NASU (Ukraine); STFC (United Kingdom); DOE and NSF (USA).
  
\hyphenation{Rachada-pisek} Individuals have received support from the Marie-Curie program and the European Research Council and Horizon 2020 Grant, contract Nos.\ 675440, 724704, 752730, 758316, 765710, 824093, 101115353, 101002207, and COST Action CA16108 (European Union); the Leventis Foundation; the Alfred P.\ Sloan Foundation; the Alexander von Humboldt Foundation; the Science Committee, project no. 22rl-037 (Armenia); the Belgian Federal Science Policy Office; the Fonds pour la Formation \`a la Recherche dans l'Industrie et dans l'Agriculture (FRIA-Belgium); the F.R.S.-FNRS and FWO (Belgium) under the ``Excellence of Science -- EOS" -- be.h project n.\ 30820817; the Beijing Municipal Science \& Technology Commission, No. Z191100007219010 and Fundamental Research Funds for the Central Universities (China); the Ministry of Education, Youth and Sports (MEYS) of the Czech Republic; the Shota Rustaveli National Science Foundation, grant FR-22-985 (Georgia); the Deutsche Forschungsgemeinschaft (DFG), among others, under Germany's Excellence Strategy -- EXC 2121 ``Quantum Universe" -- 390833306, and under project number 400140256 - GRK2497; the Hellenic Foundation for Research and Innovation (HFRI), Project Number 2288 (Greece); the Hungarian Academy of Sciences, the New National Excellence Program - \'UNKP, the NKFIH research grants K 131991, K 133046, K 138136, K 143460, K 143477, K 146913, K 146914, K 147048, 2020-2.2.1-ED-2021-00181, TKP2021-NKTA-64, and 2021-4.1.2-NEMZ\_KI (Hungary); the Council of Science and Industrial Research, India; ICSC -- National Research Center for High Performance Computing, Big Data and Quantum Computing and FAIR -- Future Artificial Intelligence Research, funded by the NextGenerationEU program (Italy); the Latvian Council of Science; the Ministry of Education and Science, project no. 2022/WK/14, and the National Science Center, contracts Opus 2021/41/B/ST2/01369 and 2021/43/B/ST2/01552 (Poland); the Funda\c{c}\~ao para a Ci\^encia e a Tecnologia, grant CEECIND/01334/2018 (Portugal); the National Priorities Research Program by Qatar National Research Fund; MCIN/AEI/10.13039/501100011033, ERDF ``a way of making Europe", and the Programa Estatal de Fomento de la Investigaci{\'o}n Cient{\'i}fica y T{\'e}cnica de Excelencia Mar\'{\i}a de Maeztu, grant MDM-2017-0765 and Programa Severo Ochoa del Principado de Asturias (Spain); the Chulalongkorn Academic into Its 2nd Century Project Advancement Project, and the National Science, Research and Innovation Fund via the Program Management Unit for Human Resources \& Institutional Development, Research and Innovation, grant B39G670016 (Thailand); the Kavli Foundation; the Nvidia Corporation; the SuperMicro Corporation; the Welch Foundation, contract C-1845; and the Weston Havens Foundation (USA).
\end{acknowledgments}

\bibliography{auto_generated}

\numberwithin{figure}{section}
\appendix
\section{Details of anomaly detection methods}
\label{sec:method_details}

This section provides technical details of the anomaly detection methods employed in this analysis. 

\subsection{Unsupervised method details}
\label{subsec:vae_details}
The VAE used in the \vae method is based on a one-dimensional convolutional neural network \cite{CNN}  using the 100 highest $\pt$ individual PF constituents of the jet.
The jet is represented as a $100{\times}3$ matrix. 
For jets with less than 100 particles, the remaining entries in the matrix were filled with zeros (zero padding).  
The size of the maximally compressed representation (latent space) of the network is 12. 
During training, jets in the CR were selectively sampled so that their distribution matches the kinematics of jets in the SR.
The total loss function minimized during training consists of the sum of the reconstruction loss, computed as the permutation-invariant Chamfer distance~\cite{fan2016point} between the input and the output jet, and the typical Kullback--Leibler regularization term of a VAE \cite{kingma2014autoencoding}, computed from the mean ($\mu$) and variance ($\sigma$) vectors in the latent space. 
The latter term is minimized when the distribution of latent representations is close to that of a multidimensional Gaussian distribution. 
The two terms in the loss function force the model to find a good compromise between representing the different modes found in the data and having a latent space with a regular structure.
After training, the loss computed on each jet is used as the anomaly score metric, where high values correspond to signal-like events.
The event-level discriminator is built using the minimum loss of the two jets, L1 and L2.

\subsection{Weakly supervised method details}
\label{subsec:weaksup_details}
In the \cwola~method, fully connected feed-forward neural networks were used for the classifiers applied to each jet.
The distribution of anomaly scores for each classifier were used to convert each score of each jet into a percentile.
The event-level anomaly score is then defined as the maximum of the percentile score for the two jets.
A selection threshold on the event-level anomaly score is defined based on its efficiency in the sidebands of the targeted SR. 
Because the QCD background decreases at higher dijet masses, a higher selection efficiency is used: 
efficiencies of 1, 3, and 5\% were used for SRs with \mjj centers below 3100, between 3100 and 4500, and above 4500 \GeV, respectively.

\begin{figure*}
    \centering
    \includegraphics[width = 0.95\textwidth]{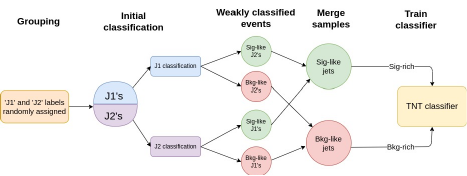}
    \caption{A flowchart outlining how the samples for the weakly supervised training were constructed in the \TNT method. 
    The two jets in the dijet candidate were randomly assigned labels J1 and J2.
    For each event, the J1 (J2) jet is placed into either a signal-like or background-like sample based on the autoencoder scores evaluated on the J2 (J1) jet. 
    The samples of signal-like (background-like) J1's and signal-like (background-like) J2's were merged together to construct a single sample of signal-like (background-like) jets. 
    The \TNT classifier is then trained to distinguish between these two samples. 
}
    \label{fig:TNT_diagram}
\end{figure*}

For \TNT, in order to construct the jet images used by the autoencoder, we follow Ref.~\cite{Macaluso:2018tck} and apply preprocessing to the jet constituents before pixelating.
We center the image on the \pt centroid of the jet constituents, and then divide the image into 32$\times$32 pixels. 
The image covers a region of $\eta$-$\phi$ space with width $\Delta \eta = \Delta \phi = 1.2$ centered on the jet axis.
 In order to reduce dependence on the \pt of the jet, each image is
normalized so that the sum of all the pixel intensities is equal to 1.
The autoencoder is constructed out of a series of convolutional layers and fully connected layers, which compress the input 32$\times$32 image into a latent space of size 6.
The architecture is then mirrored to decode the latent space back to the size of the original image.
A diagram of the \TNT training algorithm is shown in Fig.~\ref{fig:TNT_diagram}.

The selection of anomalous events is identical to that of the \cwola~method, 
except that the final anomaly score is defined as the multiplication of the percentile score of the two jets rather than the maximum.

The normalizing flows used for \cathode~were trained using 15 MADE blocks~\cite{germain2015made}, each made of a single hidden layer of 128 nodes.
The loss function is the negative log likelihood of the transformed samples under a standard Gaussian distribution.
The flows were trained in data using all events outside of the SR of interest.
After training, the flows were used to sample events in the SR, generating four times as many events as there were in the actual data in this region, as was done in Ref.~\cite{Hallin:2021wme}.
The \mjj distribution used for sampling is taken from a kernel density estimation fit \cite{kde1,kde2} of the data \mjj distribution in the SR.

After sampling, a fully connected feed-forward neural network classifier using three hidden layers of 64 nodes each and ReLU activation \cite{relu} is trained to distinguish between the synthetic events, discussed in Section \ref{subsec:weak_sup}, and actual events from the SR.
The binary cross-entropy \cite{cross_entropy} loss function is used, with class weights to compensate for the imbalance introduced by oversampling. 
The anomaly metric is then the average score from this classifier over the 10 best epochs, as quantified by the validation loss.

The sub-dominant \ttbar, $\PQt\PW$, single \PQt production, {\PZ}+jets, and {\PW}+jets backgrounds, will include jets that contain either a boosted top quark, 
or a boosted W or Z boson, and these jets are clearly distinct from those of the dominant QCD multijet background.
The effect of this on the performance of the \cwola and \TNT algorithms is tested using the simulated pseudo-data set
A version of the simulated pseudo-data set without these backgrounds is constructed, and the same signal injection test described in Section \ref{sec:validation}
is performed.
The performance of the two algorithms using this altered pseudo-data set is found to be very similar to the performance on the original pseudo-data set,
an indication these backgrounds have a minor effect on anomaly detection performance.

\subsection{Semisupervised method details}
\label{subsec:quak_details}
The \quak~method uses density estimators, trained on dedicated signal and background simulations, to construct a multidimensional space with several ``signal-like" axes and one ``background-like" axis. 
Each density estimator is implemented as a six-layer normalizing flow using autoregressive rational quadratic spline transformations~\cite{durkan2019splineflows}. 
The background flow is trained on a mixture of QCD, $\ttbar$, and V+jets events. 
The six  signal flows were trained on combinations of signal samples pooled according to the B and C particle masses $(m_\mathrm{B},m_\mathrm{C})$:
$(80,80)$, $(80,170)$, $(80,400)$, $(170,170)$, $(170,400)$, and $(400,400)$\GeV. 
Each flow learns the multidimensional probability distribution of its training sample.
Once trained, the flow can be used to evaluate the likelihood of any new sample under this probability distribution.  
Each event in data is assigned a ``loss" along each flow axis, defined as the negative log likelihood of the event evaluated by the corresponding flow. 
The losses were decorrelated from \mjj to prevent sculpting of dijet spectra, transformed to have a Gaussian-like shape using the Box--Cox transformation \cite{boxcox}, and normalized to have an average value of 0 and standard deviation of 1. 
A single signal score is extracted by taking the ``signed L5 norm" of the density scores from the six signal axes.
The signed L5 norm is defined as $L_5^{\sgn}(\mathbf{x}) = \sgn(y)\abs{y}^{1/5}$, where $y = \sum_{i=0}^5 \sgn(x_i)\abs{x_i}^5$ and $\mathbf{x}$ is a vector of the six signal losses. 
This is chosen to emphasize very signal-like or very anomalous events, which should have large negative or positive losses, respectively, along a particular axis or axes. The $L5$ norm was chosen over the more familiar $L2$ norm by scanning over exponents in the range 2--10 and observing improved performance that leveled off near 5. 

Events populate a 2D \quak space, with positions determined by the background loss and combined signal losses. 
The precise distribution of background events in this space is determined from control samples in data by making a template from events with \mjj within the sideband intervals.
Because there is no straightforward single anomaly score from a 2D distribution, a binned approach is used to determine the most anomalous events. 
First, 50\% of events were removed, based on a selection on the background-like probability.
The final selection is determined based on the targeted signal resonance mass $m_\mathrm{A}$.
A 2D histogram representation of the QUAK space with 500 bins is constructed from events within the sidebands defined in Section \ref{subsec:quak}.
The SR events were put into a separate histogram with the same binning.
Events from the SR were selected from bins corresponding to the least populated, second-least populated, the third-least populated bins, (and so on) from the sideband histogram.
The procedure terminates when at least 1500 events or at least $500  ([{m_\mathrm{A}}/\TeVns] - 3)^3$ events from the SR have been selected. 
All events that fall into the bins selected in the above procedure were selected for inclusion in the final fit.
Studies in simulation and the data sideband show that the \quak selection has a higher efficiency for \ttbar events than for QCD multijet events,
enhancing the relative contribution of this background.
However, because the contribution of \ttbar background prior to selection is quite small, and \ttbar events do not form a resonance peak in the dijet mass,
this behavior does not hinder the extraction of a signal.

\subsection{Cross validation procedure}
\label{subsec:cross_val}
Since the weakly supervised methods and the QR technique of the \vae use data from the SR to train, a multifold cross-validation scheme is used.
The scheme ensures all data events in the SR were used for the final search, while avoiding the same data being used in both the training and evaluation of a model.

For the training of the QR, the data were randomly split into $k=20$ equal subsamples (folds) and the QR network is trained separately for each fold.
For the selection of anomalous events in each fold, the average of the QR networks trained in the other nineteen folds, smoothed by a third-order polynomial fit is used. 
The events selected from these 20 folds were then combined for the final search.
It was verified that this procedure does not artificially diminish the size of localized excesses should they be present in the data. 

For the weakly supervised methods, the data were split into five equal folds. 
Four of these folds were used to train and validate a classifier which is then used to select the most anomalous events in the remaining fold. 
The process is then repeated four times, varying which folds were used for training and selection.
The selected events from all five folds were merged together before the fit to the dijet mass spectrum.
In addition, for increased stability of the training performance, an ensemble of four models is used to define each anomaly discriminant. 
The ensemble is constructed by varying which of the four folds were used for training and which for validation.
The final anomaly discriminant used in the fifth fold is taken as the mean of the anomaly scores of the four different models.

\section{Signal interpretation studies}
\label{sec:sig_interp}
If one or more of the anomaly detection methods were to find a statistically significant excess, it would be important to discern its basic properties so that follow-up studies could be undertaken.
This type of study would validate that the properties of the excess match a realistic new particle, rather than arising from a subtle instrumental or reconstruction-related artifact picked up by the anomaly detection method.
If the basic substructure properties of an excess are determined, a dedicated search targeting a signal model with a matching signature could be performed to verify the nature of the excess. 
Based on studies in simulation, two types of analysis were found to be effective to characterize its nature.

The first type of study is to plot the observables of the events in the region of the excess, which have the highest anomaly score.
These distributions of observables are then compared to those of all events in the
region of the excess, without any anomaly score cutoff. 
Differences in these two distributions demonstrate the distinctive observables of the anomalous events compared to standard events.
It should be noted that because the weakly supervised methods will learn to identify the specific
characteristics of the signal, they will preferentially select background events that have observables
similar to that of the signal.
This helps in identifying what observables the anomaly detection algorithm has learned
are signal like, but means these distributions cannot be used for a precision extraction of signal properties, because the background is significantly biased.

An additional type of study is used to directly interpret the anomaly detection model itself, by testing which input observables are most important to the network giving an event a high anomaly score.
This task requires special strategies for the interpretation of an unknown excess, where one does not know the true origin of any event.
We use an interpretability technique called the permutation feature importance \cite{perm_feat_importance} which is suitable for situations without labeled examples.
For each observable used as an input for the network, its permutation score is computed as the average change in the anomaly score, when the value of that observable is randomly 
replaced with the value obtained from a different event.
We assess the permutation score on the set of events with the highest anomaly scores, to determine which of their properties are most crucial for the anomaly classification.
For each high-anomaly-score event, we swap one individual observable at a time with the values obtained from 100 randomly selected events.
We then compute the average anomaly score variation  across these 100 perturbations. 
This is then averaged across all events from the high-anomaly-score sample to determine the permutation score of that observable.
This procedure is repeated to asses the permutation score of all observables. 
We find the permutation scores of unimportant observables to be small, but not exactly zero, while the scores of important observables are generally larger.
We note that this method is not as useful in interpreting the \vae results, because for that method the input observables are 4-momenta of individual PF candidates, which are difficult 
to connect to the physics properties of the signal.

\begin{figure*}[!h]
  \centering
  \includegraphics[width = 0.32\textwidth]{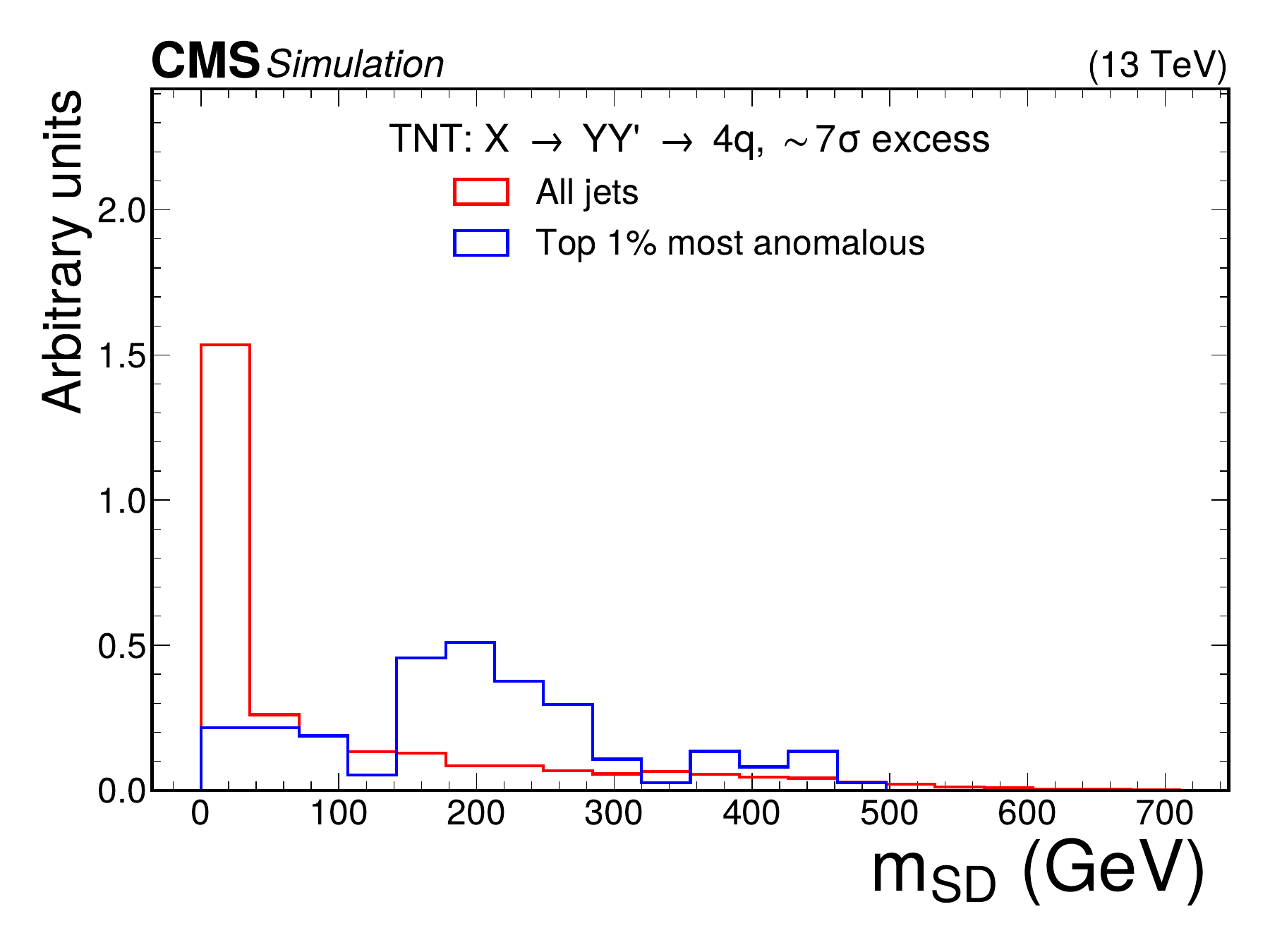}
  \includegraphics[width = 0.32\textwidth]{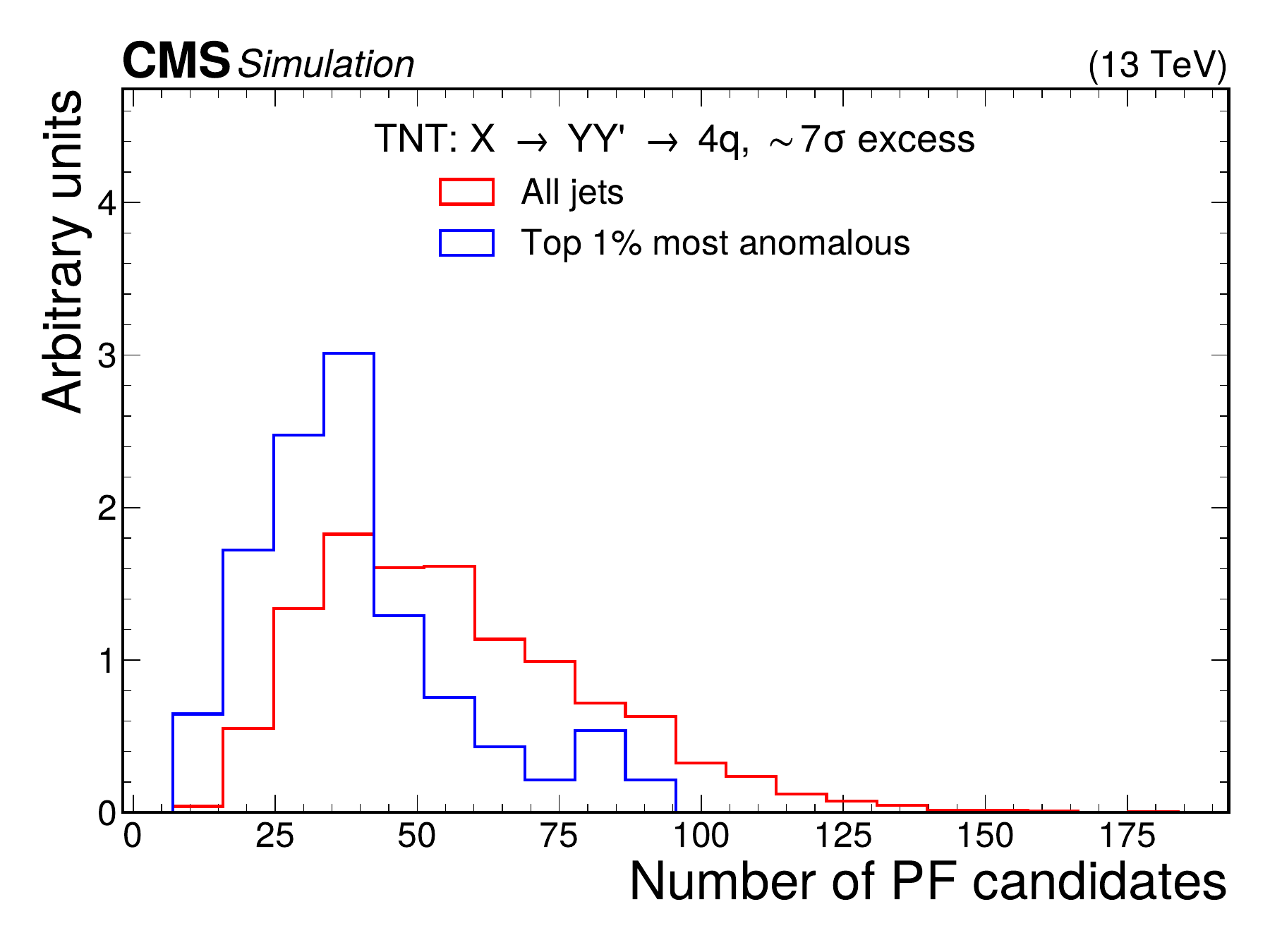}
  \includegraphics[width = 0.32\textwidth]{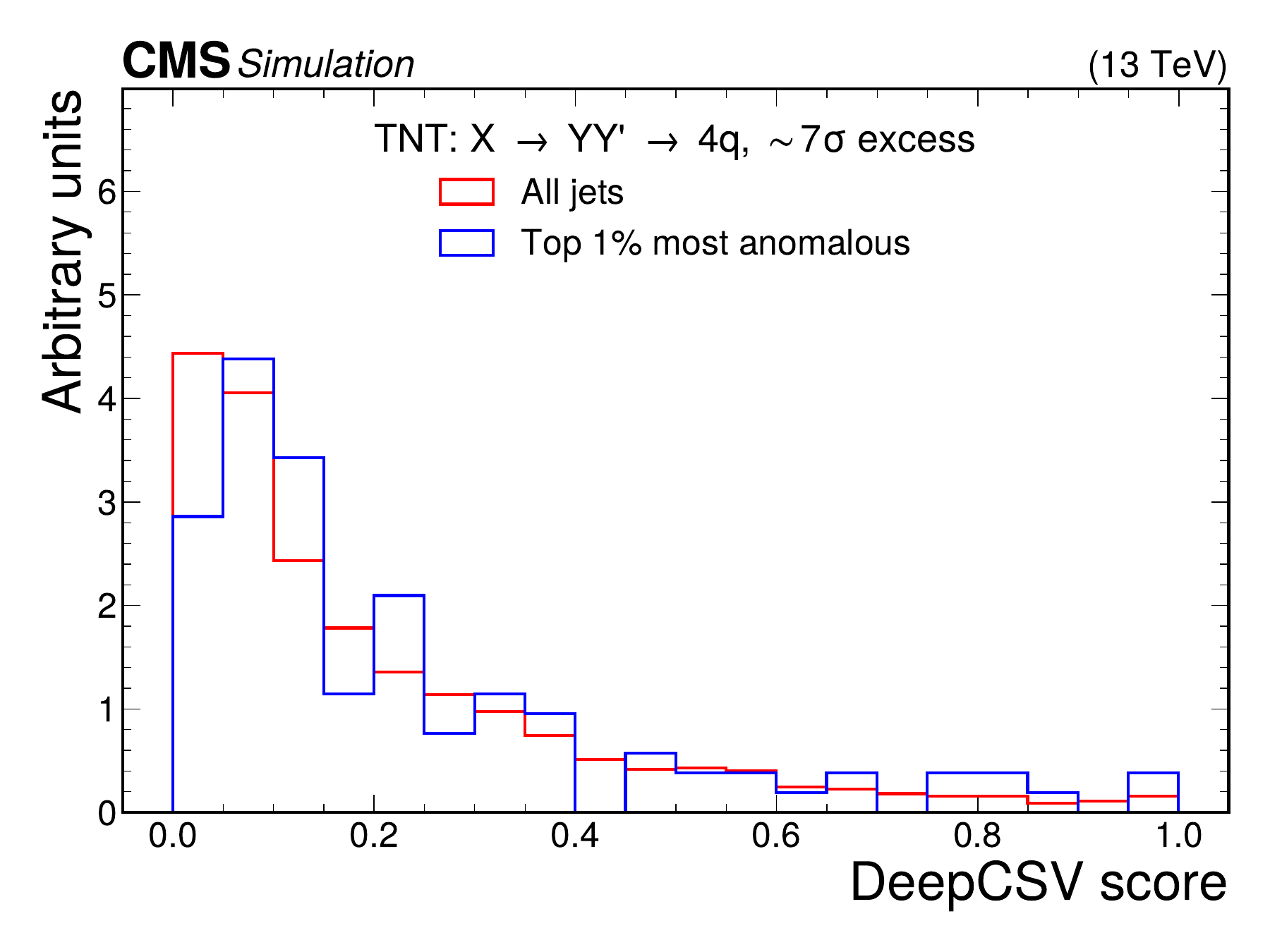}\\
  \includegraphics[width = 0.32\textwidth]{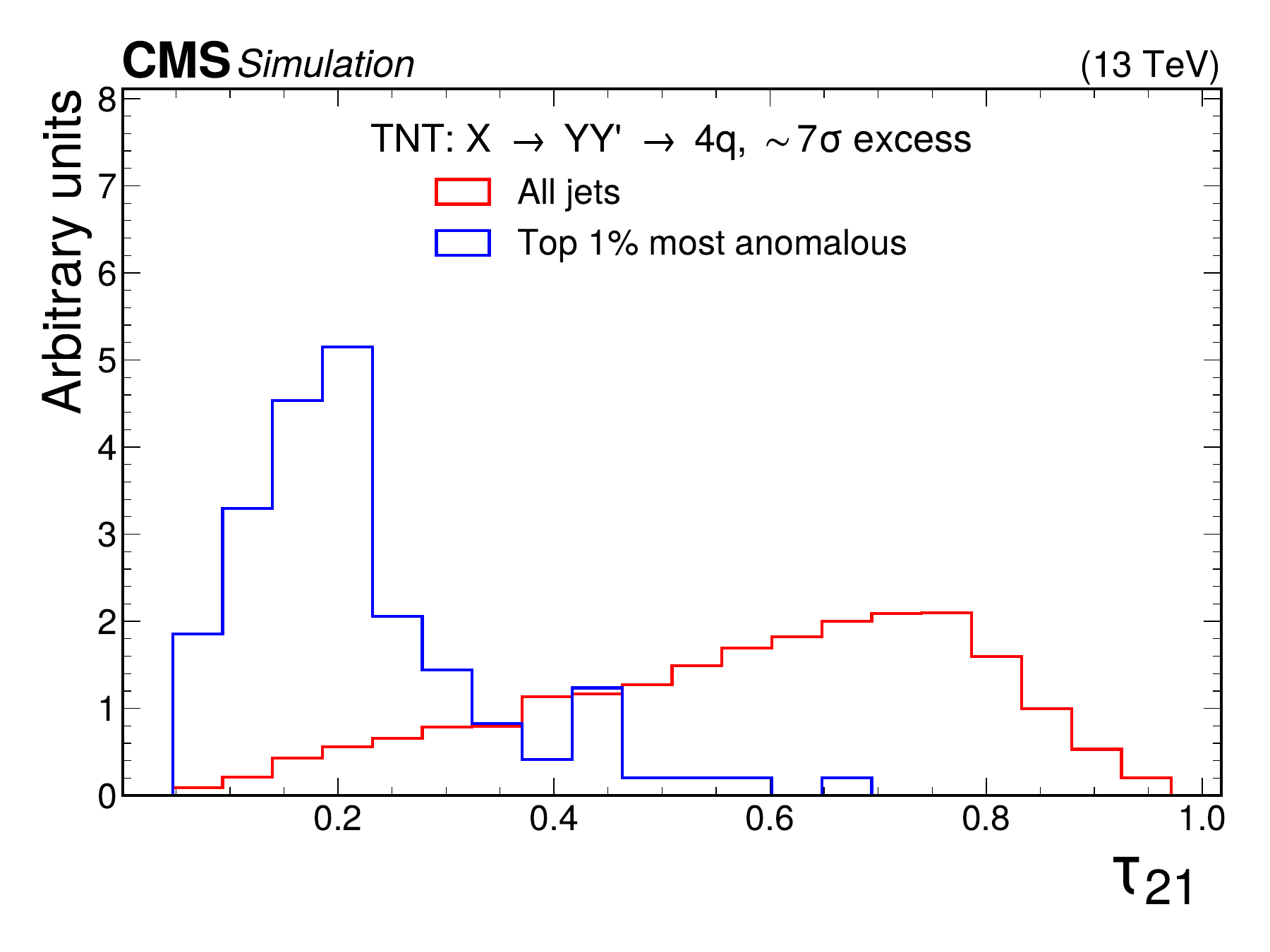}
  \includegraphics[width = 0.32\textwidth]{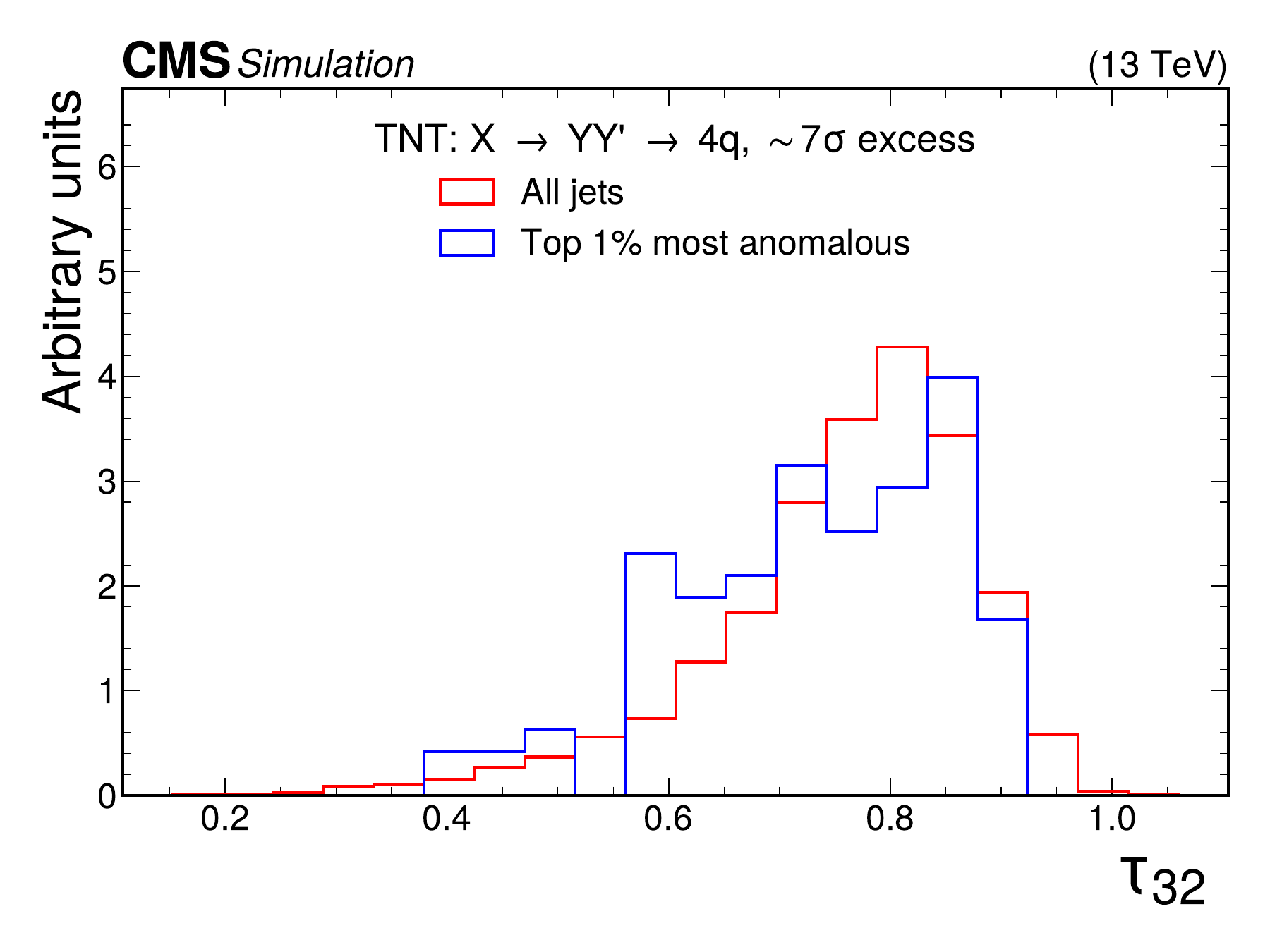}
  \includegraphics[width = 0.32\textwidth]{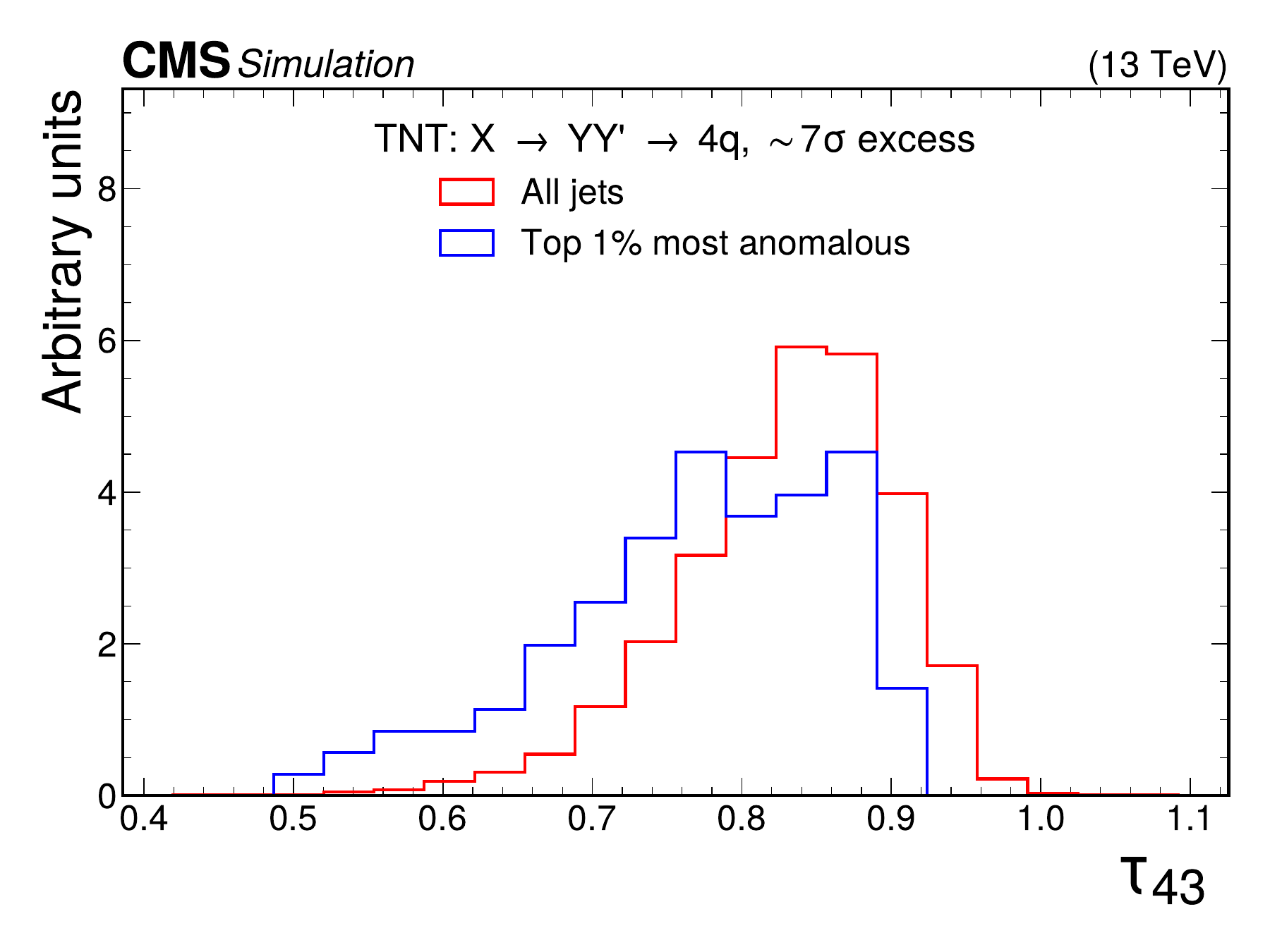}
  \caption{Excess interpretation example for the \TNT method trained on a simulated sample, with an injection an \XtoYY signal with a cross section of 24 fb, $M_\PX$ = 3 \TeV, and $M_{\PY/\PYpr} = 170 \GeV$.
    The plots compare the properties of the jets with the highest anomaly score (blue) as compared to those for all jets in the region of the excess (red).
    The two-pronged nature of the anomaly is evident from the low $\tau_{21}$ scores, and the approximate mass of the $\PY$ and $\PYpr$ resonance can be seen as a peak in the jet mass ($m_\mathrm{SD}$) distribution.
}
  \label{fig:interp_XYY}
\end{figure*}

\begin{figure*}[!h]
  \centering
  \includegraphics[width = 0.32\textwidth]{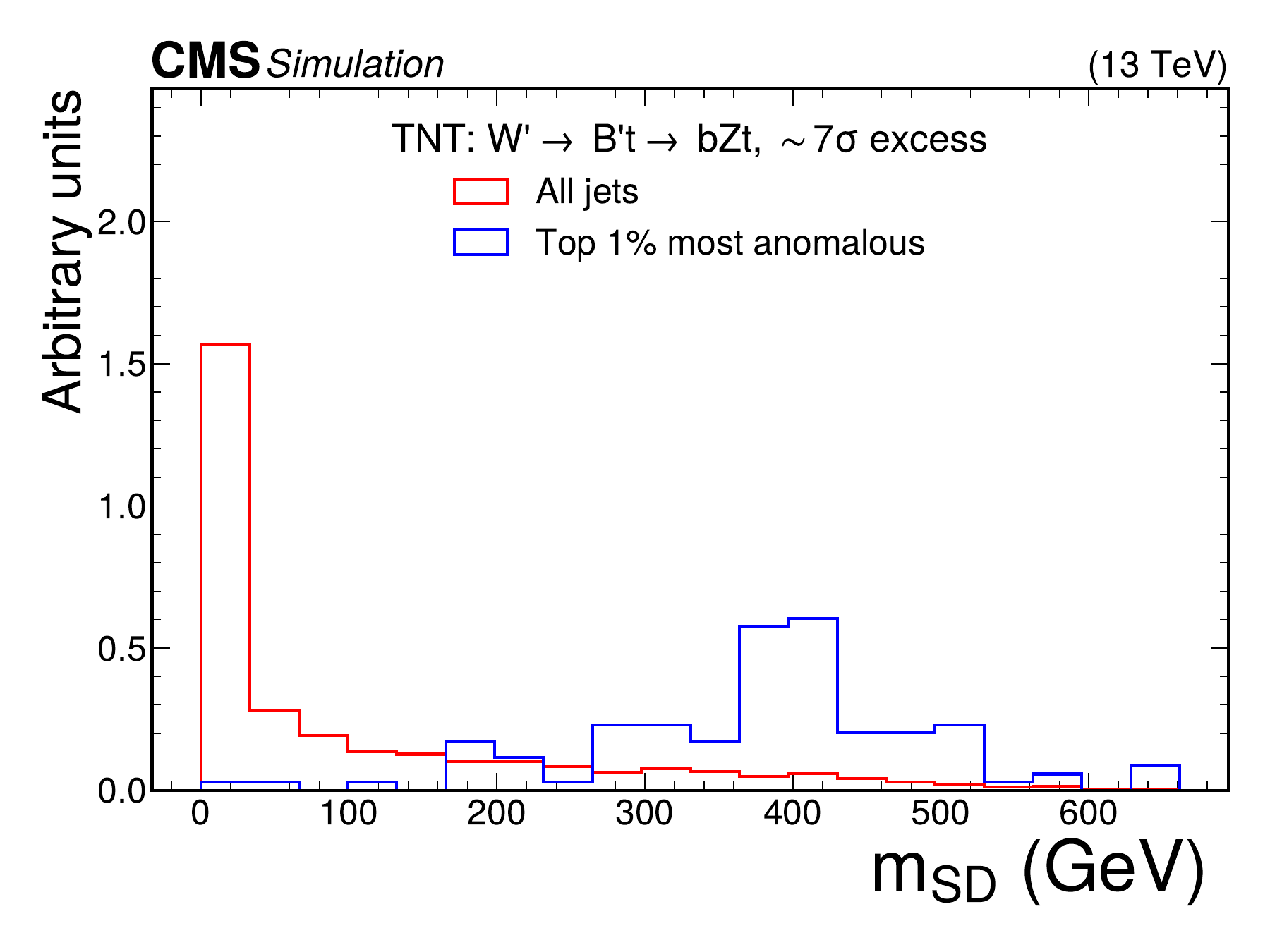}
  \includegraphics[width = 0.32\textwidth]{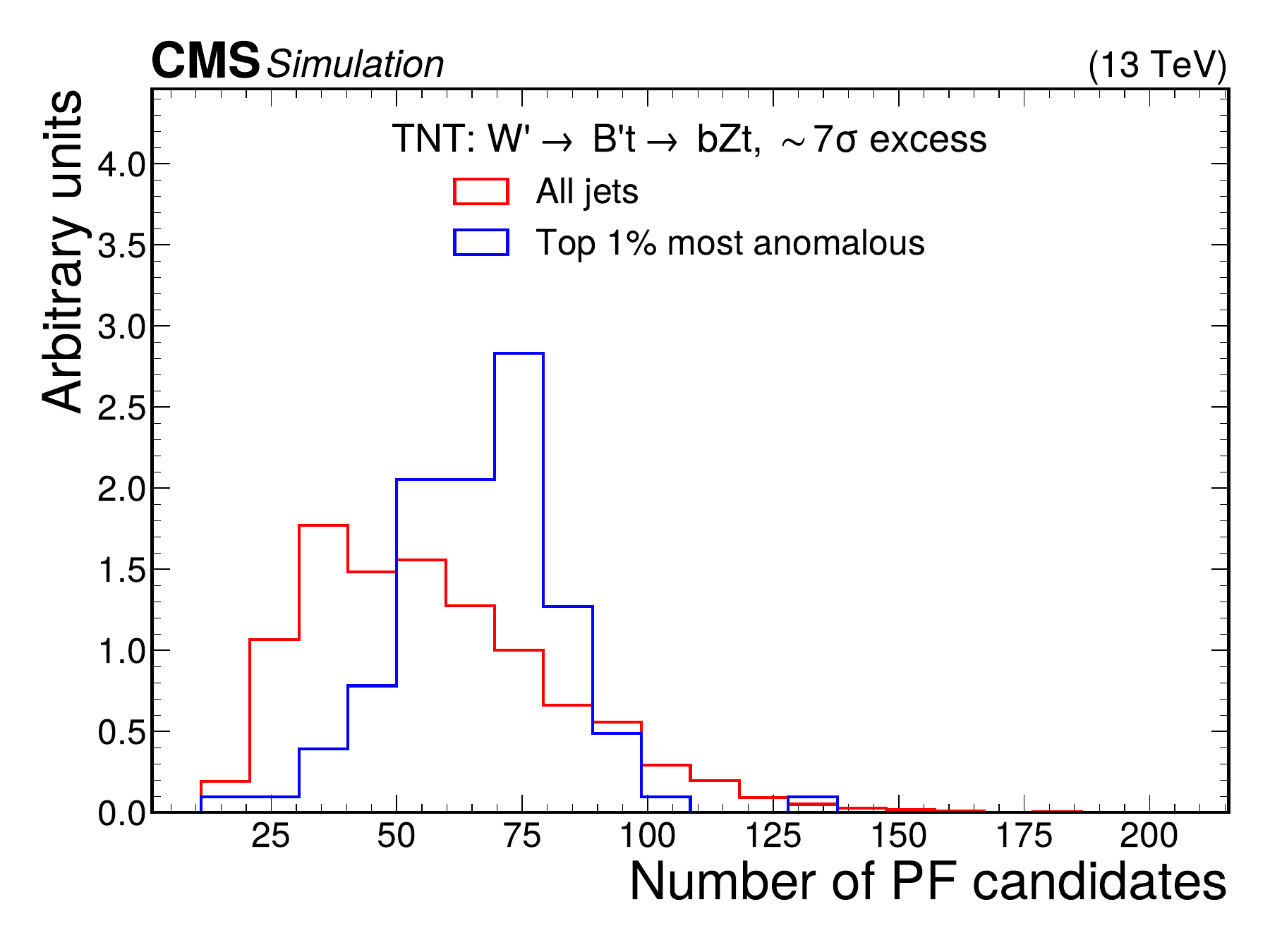}
  \includegraphics[width = 0.32\textwidth]{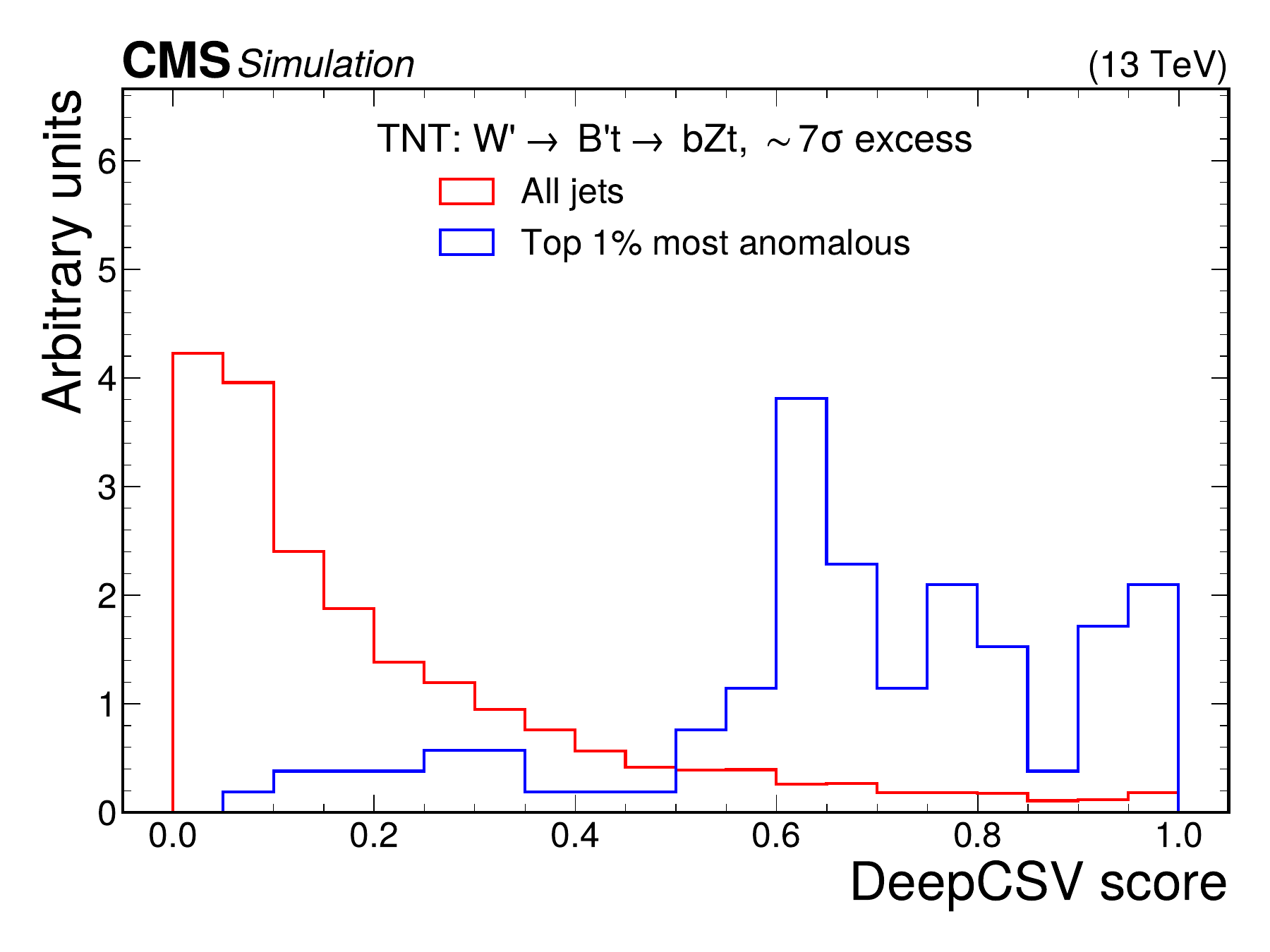}\\
  \includegraphics[width = 0.32\textwidth]{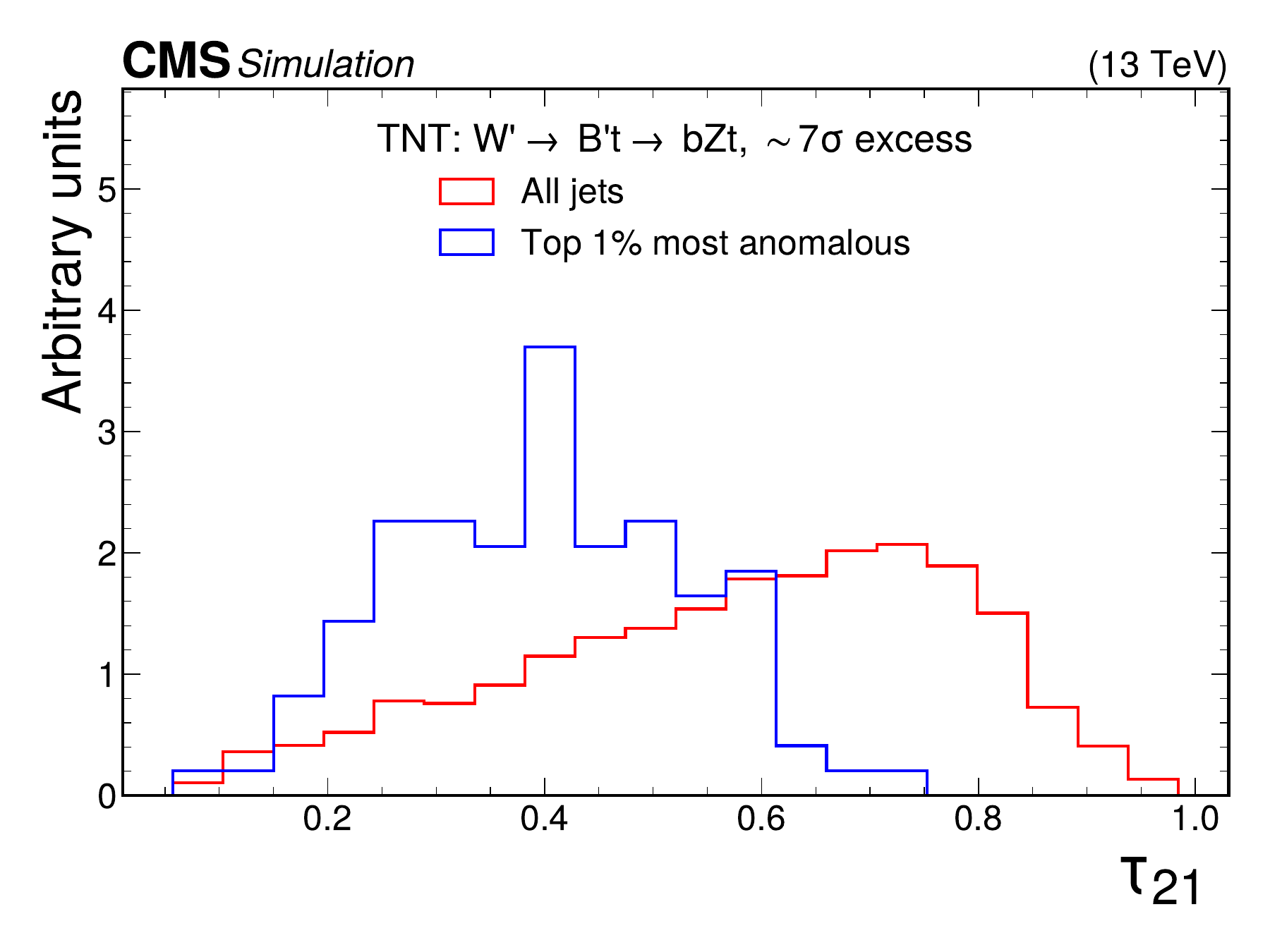}
  \includegraphics[width = 0.32\textwidth]{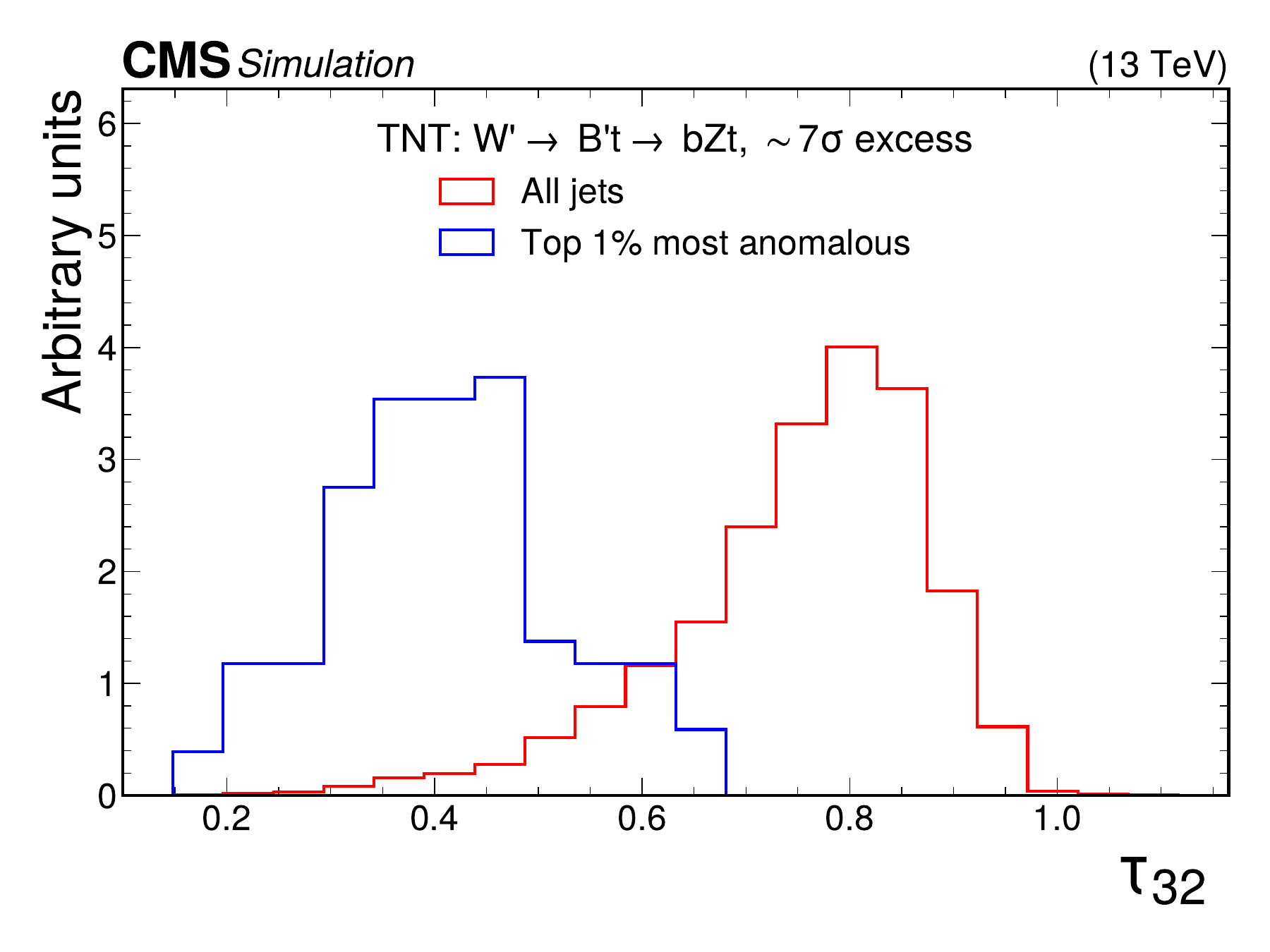}
  \includegraphics[width = 0.32\textwidth]{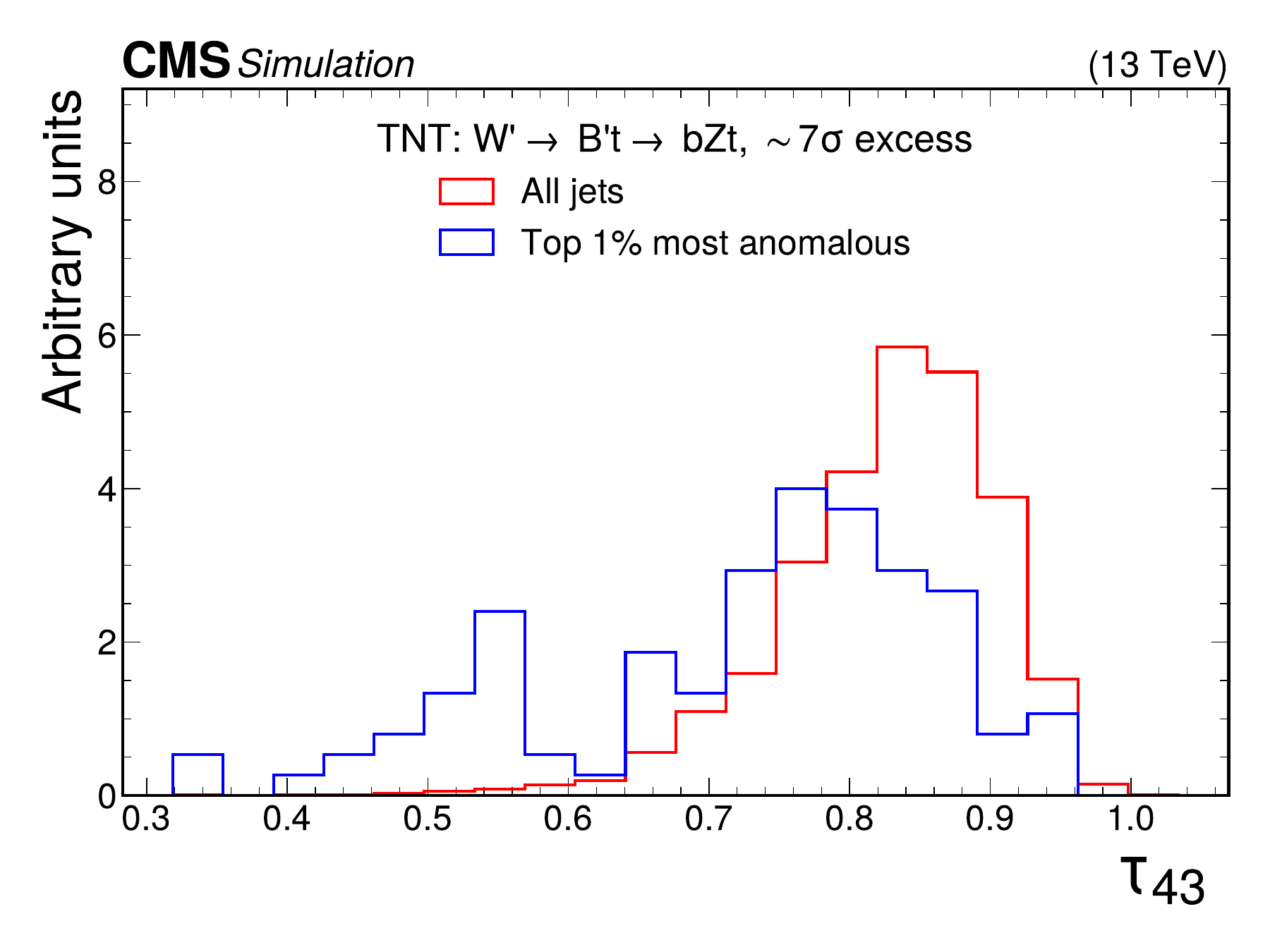}
  \caption{Excess interpretation example for the \TNT method trained on a simulated sample, with an injection of a \Wp signal with a cross section of 97 fb, $M_{\PWpr} = 3 \TeV$, and $M_{\PBpr} = 400 \GeV$.
    The plots compare the properties of the jets with the highest anomaly score (blue) as compared to those for all jets in the region of the excess (red).
    The three-pronged nature of the signal is clear from the low $\tau_{32}$ scores, the presence of b tags from the high \textsc{DeepCSV} score, and peaks in the jet mass ($m_\mathrm{SD}$) at 170 \GeV and 400 \GeV
    indicate the top quark and $B'$ resonances.
  }
  \label{fig:interp_Wp}
\end{figure*}

\begin{figure*}[!h]
  \centering
  \includegraphics[width = 0.4\textwidth]{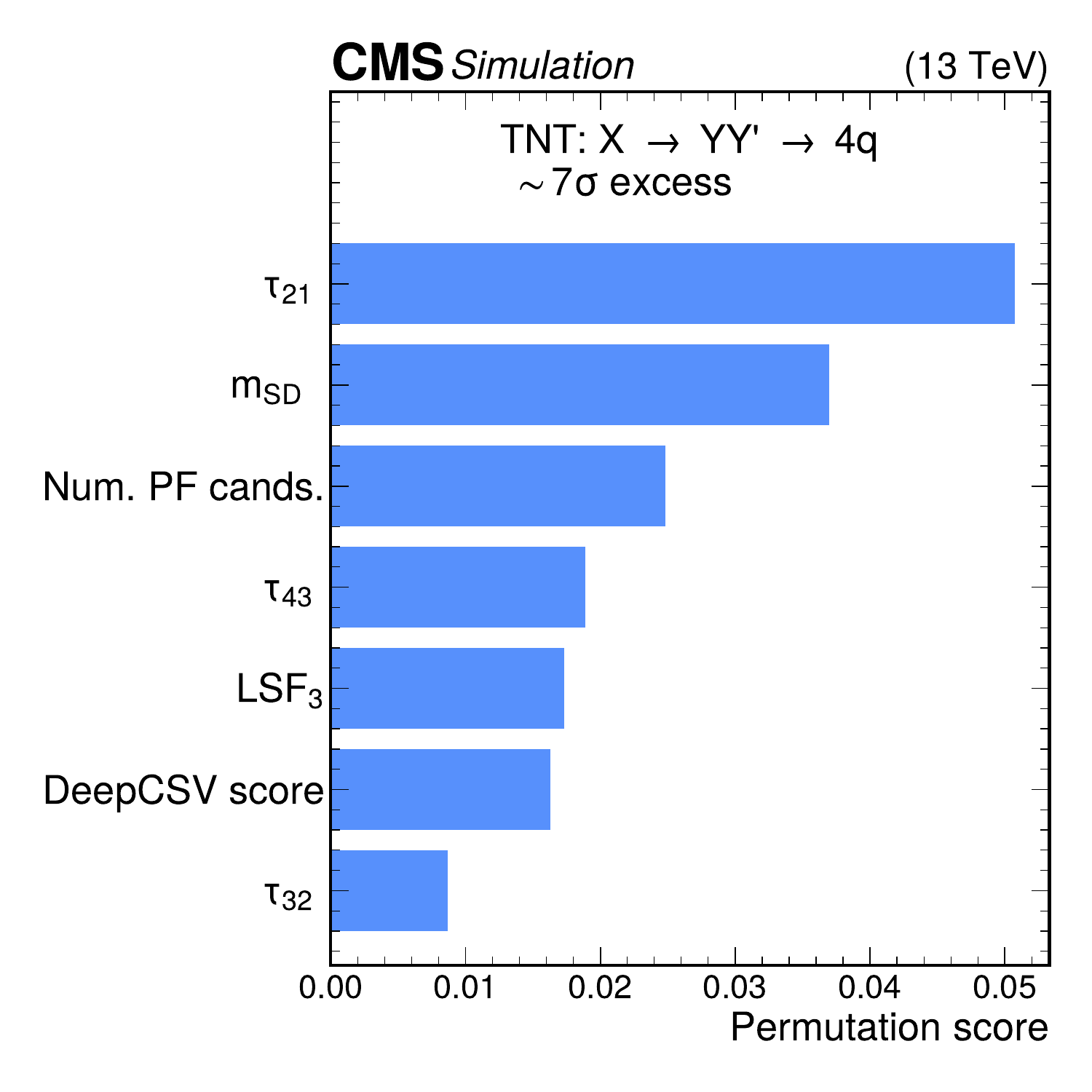}
  \includegraphics[width = 0.4\textwidth]{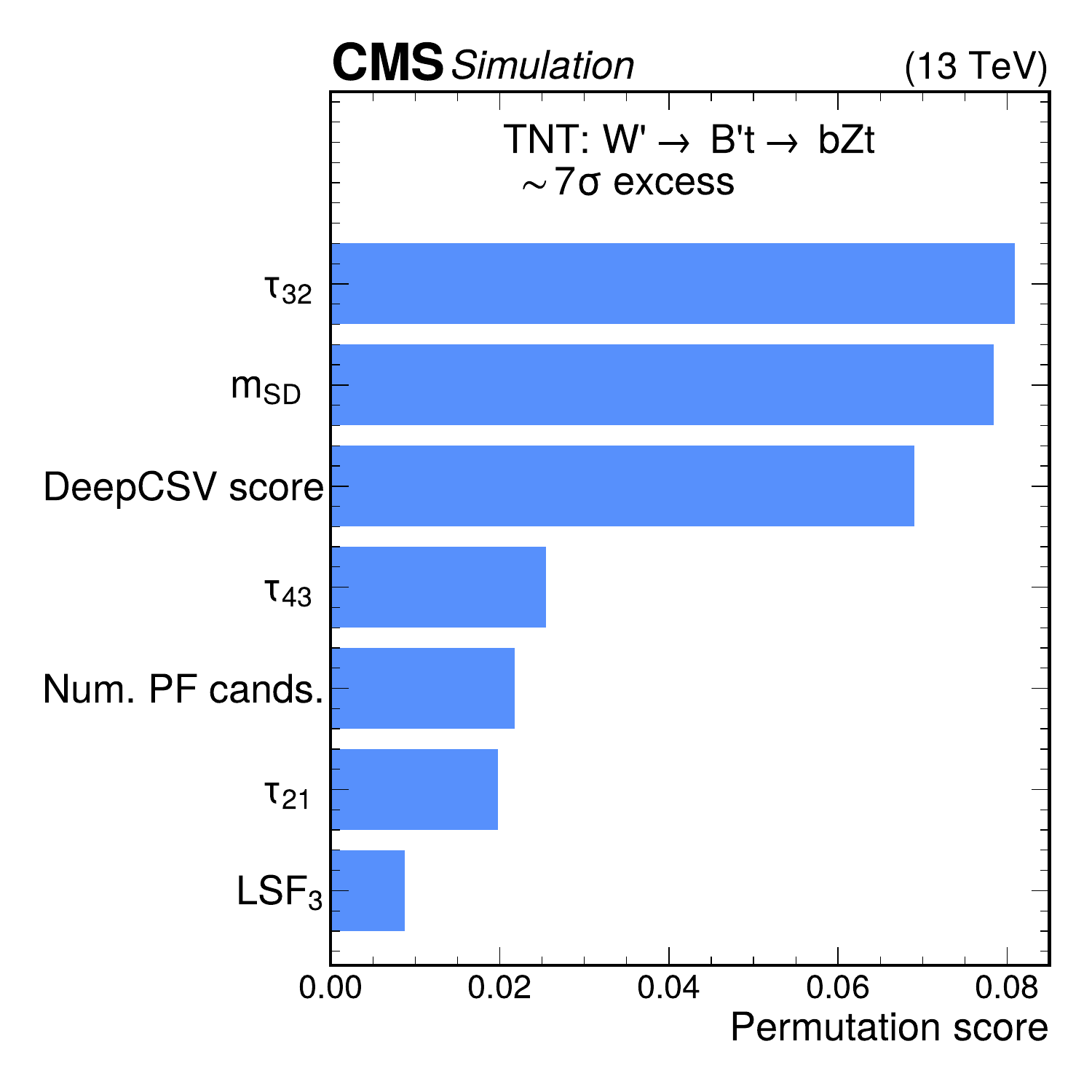}
  \caption{Excess interpretation examples for the \TNT method trained on a simulated sample, with an injection of an \XtoYY signal with a cross section of 24 fb (left) and a \Wp signal with a cross section of 97 fb (right).
    The sensitivity of the anomaly score to the different input observables is assessed to aid in the determination of the properties of the excess.
    For the \XtoYY injection, the $\tau_{21}$ and jet masses are seen to be the most important observables.
    For the \Wp injection, the $\tau_{32}$, jet masses, and b tagging score are seen to be the most important observables.
  }
  \label{fig:interp_scores}
\end{figure*}

The combination of these two types of studies makes it possible to identify characteristics of the signal in the case of an excess. 
A demonstration of the first type of study applied to the \TNT method on the \XtoYY and \Wp signals are shown in Figs. \ref{fig:interp_XYY} and \ref{fig:interp_Wp}, respectively. 
A demonstration of the second type of study is shown in Fig. \ref{fig:interp_scores}. 
The signal injections from the performance studies of Section \ref{sec:validation} were used to test the interpretability of the excesses that were found.
For the \XtoYY signal, the two pronged substructure is clear from the low $\tau_{21}$ score of the anomalous jets, and the rough value of the $Y$ mass can be identified from the peak in jet mass at 170 GeV.
For the \Wp signal, the three pronged substructure can be discerned from the low $\tau_{32}$ score.
A high b tagging score indicates that both the B and C particle decays produce b quarks, and 
the top quark mass (170 \GeV) and $B'$ mass (400 \GeV) can be seen from the peaks in the jet mass distribution.

\section{Limit-setting procedure}
\label{sec:limits}

For the weakly supervised methods, special procedures need to be taken in the limit setting to account for nonstandard dependencies.
The procedure begins with a fit to the observed \mjj spectrum using the signal shape from the desired signal model. 
An upper limit on the signal yield, obtained from one such fit, corresponds to a number of signal events that
needs to be converted to a cross section by accounting for the luminosity,
acceptance, and efficiency.
For a traditional analysis with a fixed selection, the efficiency is a constant
number that depends only on the signal model.
However, in weakly supervised methods, the event selection is adjusted dynamically
based on the presence of signal in data.
This causes the signal efficiency to depend on several factors,
including the total size of the training data set, the amount of signal in the
training data set, and the distinctiveness of the signal as compared to the background.
The amount of signal in the training data set, in particular, is directly related to
the unknown signal cross section $\sigma_\text{sig}$ that one wishes to set a limit on.
Therefore, when setting the limit, the efficiency $\epsilon$ needs to be
understood as depending on $\sigma_\text{sig}$. This leads to the
following relation between the expected signal yield $N_\text{sig}$ in the
SR and the cross section:
\begin{equation}
  N_\text{sig}(\sigma_\text{sig}) = \sigma_\text{sig} \; \mathcal{L} \; A \; \epsilon(\sigma_\text{sig}),
    \label{eq:limits}
\end{equation}
where $\mathcal{L}$ is the integrated luminosity, $A$ is the acceptance, and
$\epsilon(\sigma_\text{sig})$ is the efficiency as a function of the cross section.
Given an upper limit $N_\text{UL}$ on the signal yield, the corresponding limit
on the cross section is obtained by finding the $\sigma_\text{sig}$ such that
$N_\text{sig}(\sigma_\text{sig}) = N_\text{UL}$ in the expression above.
This is straightforward for the \vae and \quak methods that do not use weak
supervision and have a fixed signal efficiency, $\epsilon(\sigma_\text{sig}) = \epsilon_0$,
as in this case Eq.~\eqref{eq:limits} can be inverted explicitly.
For \cwola, \TNT, and \cathode, the efficiency function is not known \textit{apriori} and
the inversion needs to be done numerically.
This Appendix describes the procedure used to evaluate $\epsilon(\sigma_\text{sig})$ and
obtain the limit on the cross section.

\begin{figure*}
    \centering
    \includegraphics[width=0.7\textwidth]{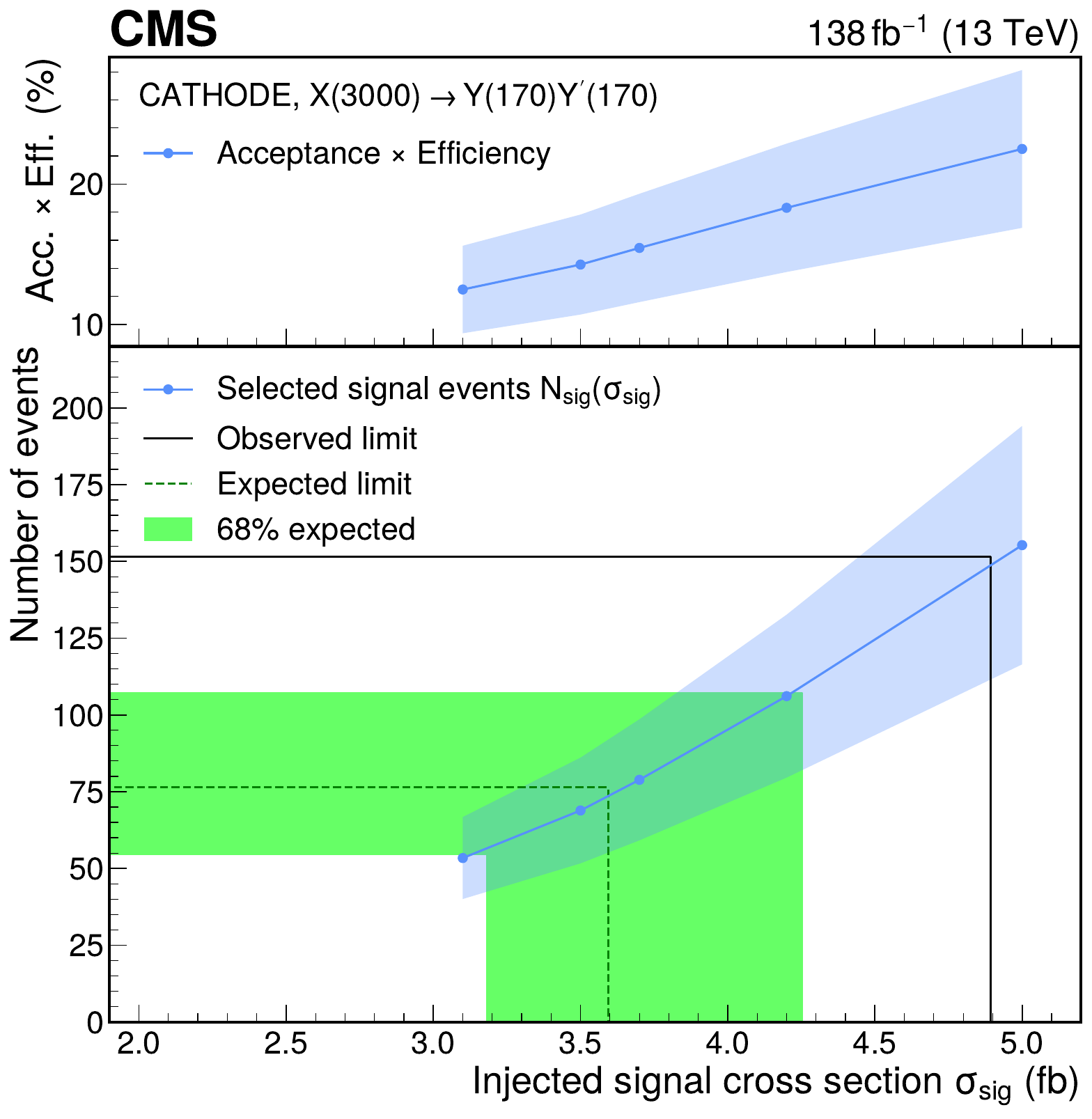}
    \caption{
        Diagram of the limit-setting procedure for the \XtoYY signal at 3\TeV
        with the \cathode method.
        The upper panel shows the estimated signal acceptance times efficiency as a
        function of the cross section injected in data.
        The shaded region in the upper panel shows the total statistical and systematic uncertainty in the efficiency.
        The resulting $N_\text{sig}(\sigma_\text{sig})$ curve and its corresponding uncertainty band from the efficiency
        are shown in blue in the lower panel.
        The expected and observed limits on the number of signal events are
        shown as a horizontal solid black line and green dashed lines,
        respectively, and connected to the corresponding limits on the cross
        section (vertical lines).
        The 68\% confidence level band around the expected limit is displayed similarly.
    }
    \label{fig:limit_diagram}
\end{figure*}

The dependence of the signal efficiency on $\sigma_\text{sig}$ can be understood as follows.
If the signal cross section is very low, such that there is no signal present in
the SR, the training procedure produces an essentially random classifier.
Therefore, events selected as having high anomaly scores fall into a random phase space region of the input variables.
Typically, this region is not signal-enriched and the selection on the anomaly score
has a very low signal efficiency, lower or similar to that of background (1--5\%).
If, on the other hand, there is a large amount of signal present in the data,
the weakly supervised training procedure produces a classifier that
correctly identifies events from the signal phase space as the most anomalous.
Many signal events therefore pass the anomaly score selection and the signal
efficiency is significantly higher (20--50\%).
For intermediate signal strengths, the signal-finding performance of the
classifier smoothly transitions between these two extremes, resulting in an
efficiency function similar to the one shown in the upper panel of Fig.~\ref{fig:limit_diagram}.

The efficiency functions of the weakly supervised methods were calibrated by
injecting each signal into the data at various cross sections, performing the
training procedure, and evaluating the efficiency using the full signal sample.
To account for the statistical fluctuations in the sampling, each injection was
repeated five times using different random signal events.
The average efficiency of these five runs was used as the estimate of the signal
efficiency at that cross section, and their envelope was taken as an uncertainty
in the efficiency due to random variations in the sampling and training
performance. For both the expected and observed upper limit, a binary search was used to solve Eq.~\eqref{eq:limits} for the corresponding
signal cross section, $N_\text{UL} = N_\text{sig}(\sigma_\text{UL})$. 
The stopping condition of the search procedure required both sides of
the equation to agree within 10\%. The results of the search for one mass point
of the  \XtoYY signal is shown in the lower panel of Fig.~\ref{fig:limit_diagram}.
It should be emphasized that, due to the dependence on the signal cross section, the phase space of events selected by the classifiers trained on the data itself, differs from the phase space selected by the classifiers trained with injected signal used in this calibration.

The trained networks resulting from the injection of signal events in data can be affected
by significant systematic uncertainties.
The impact of each source on the efficiency was assessed using signal samples
varied up and down by one standard deviation.
Retrainings were performed for all sources affecting the input distributions of
the neural networks significantly.
The dominant source arises from the correction to signal modeling based on the
Lund plane~\cite{Lund_PAS}, followed by the stochasticity of signal sampling for
the injection.
Other modeling and experimental uncertainties were found to be small.
All sources were added in quadrature, and the final limit was obtained by
repeating the fit to data, with a log-normal multiplicative nuisance parameter
inserted in Eq.~\eqref{eq:limits}.

Injecting signal events directly into the data ensures that the background
distribution is modeled accurately, removing the need for generating large
SM event samples and estimating the associated uncertainties.
The calibration procedure is appropriate when no signal is present in the data.
In the opposite case, the estimation of the signal efficiencies would be biased by signal events present in the data, which could result in limits with improper coverage.
This bias was studied in simulation, and its magnitude and direction
were found to depend on which signal was present in the data and in what amount.
If the features of the signal already present in the data matched those of the signal being injected, the procedure would overestimate the signal efficiency at a given cross section.
Conversely, the presence of a signal with features distinct from the signal
being injected, would in some cases result in an underestimated signal
efficiency.
It was found that for signals in data which produce excesses of ${\sim}3$ standard deviations or less, the amount of bias in the signal efficiency estimate is less than ${\sim}20\%$, and the resulting exclusion limits have proper coverage.
Having seen no excesses above this level in the search, the injection procedure was determined to be valid for estimating the signal efficiency and producing exclusion limits.

\section{Study of potential fit bias}
\label{sec:fit_bias}

We test for potential bias in the extraction of a signal, due to the chosen parameterization of the background, by comparing against an additional functional form.
Specifically, we compare to the alternate functional form given by:
\begin{equation}
  \frac{\rd{N}}{\rd{\mjj}} = \frac{P_0} {(x)^{P_{1}}} \; \exp(-P_2 x  -P_3 (x)^2 - P_4 (x)^3).
\label{eq:alt_dijet1}
\end{equation}

We perform a test for potential bias, using the background-only version of the simulated pseudo-data set described in Section \ref{sec:validation}. 
For each anomaly detection method, we perform the full analysis procedure to select the final spectrum of dijet masses.
These distributions are then fit using the alternate form of Eq. \ref{eq:alt_dijet1}. 
Pseudo-data dijet mass distributions of varying signal strength are then generated based on the fitted parameterization and signal templates. 
These pseudo-data distributions are then fit with the background parameterization used in the search (Eq. \ref{eq:dijet_fn}) plus a signal with varying signal strength parameter. 
The bias in the extracted signal strength is computed as the average difference between the estimated signal strength parameter and the true injected signal strength, divided by
the uncertainty in the estimated signal strength, across the different pseudo-experiments.
This test is performed for three different injected signal strengths: zero injected signal, 
an injection strength resulting in approximately $2\sigma$ statistical significance of a signal, 
and an injection strength resulting in approximately $5\sigma$ statistical significance of a signal.
This procedure is performed scanning over all the signal masses considered in the search. 

The computed biases, for the different chosen signal strengths and resonance masses, were found to be close 
to zero for all anomaly detection methods.
Most biases were less than $0.5\sigma$ and no scenarios were found to produce biases of larger than $1 \sigma$.
The amount of bias due to the background parameterization choice was therefore deemed acceptable. 
\cleardoublepage \section{The CMS Collaboration \label{app:collab}}\begin{sloppypar}\hyphenpenalty=5000\widowpenalty=500\clubpenalty=5000\input{EXO-22-026-public-authorlist.tex}\end{sloppypar}
\end{document}

%% file: EXO-22-026-public-authorlist.tex
\cmsinstitute{Yerevan Physics Institute, Yerevan, Armenia}
{\tolerance=6000
V.~Chekhovsky, A.~Hayrapetyan, V.~Makarenko\cmsorcid{0000-0002-8406-8605}, A.~Tumasyan\cmsAuthorMark{1}\cmsorcid{0009-0000-0684-6742}
\par}
\cmsinstitute{Institut f\"{u}r Hochenergiephysik, Vienna, Austria}
{\tolerance=6000
W.~Adam\cmsorcid{0000-0001-9099-4341}, J.W.~Andrejkovic, L.~Benato\cmsorcid{0000-0001-5135-7489}, T.~Bergauer\cmsorcid{0000-0002-5786-0293}, S.~Chatterjee\cmsorcid{0000-0003-2660-0349}, K.~Damanakis\cmsorcid{0000-0001-5389-2872}, M.~Dragicevic\cmsorcid{0000-0003-1967-6783}, P.S.~Hussain\cmsorcid{0000-0002-4825-5278}, M.~Jeitler\cmsAuthorMark{2}\cmsorcid{0000-0002-5141-9560}, N.~Krammer\cmsorcid{0000-0002-0548-0985}, A.~Li\cmsorcid{0000-0002-4547-116X}, D.~Liko\cmsorcid{0000-0002-3380-473X}, I.~Mikulec\cmsorcid{0000-0003-0385-2746}, J.~Schieck\cmsAuthorMark{2}\cmsorcid{0000-0002-1058-8093}, R.~Sch\"{o}fbeck\cmsAuthorMark{2}\cmsorcid{0000-0002-2332-8784}, D.~Schwarz\cmsorcid{0000-0002-3821-7331}, M.~Sonawane\cmsorcid{0000-0003-0510-7010}, W.~Waltenberger\cmsorcid{0000-0002-6215-7228}, C.-E.~Wulz\cmsAuthorMark{2}\cmsorcid{0000-0001-9226-5812}
% ...
\par}
\vskip\cmsinstskip
\dag:~Deceased\\
$^{1}$Also at Yerevan State University, Yerevan, Armenia\\
$^{2}$Also at TU Wien, Vienna, Austria\\
% ...
